\documentclass[aps,prx,twocolumn,superscriptaddress]{revtex4-2}
\usepackage{amsmath}
\usepackage{amsfonts}
\usepackage{amssymb}
\usepackage{array}
\usepackage{bm}
\usepackage{booktabs}
\usepackage{braket}
\usepackage{cancel}
\usepackage{dcolumn}
\usepackage{graphicx}
\usepackage{hyperref}
\usepackage{mathtools}
\usepackage{multirow}
\usepackage{siunitx}
\usepackage{spreadtab}
\usepackage[caption=false]{subfig}
\usepackage[dvipsnames]{xcolor}
\usepackage{tabularx} 
\usepackage[noabbrev]{cleveref}
\usepackage{threeparttablex}

\usepackage{pax} 
\hypersetup{
breaklinks=true,   
colorlinks=true,    
linkcolor=blue,
citecolor=blue,
urlcolor=blue,
}
\graphicspath{{./figs/}}		

\usepackage{pdfpages} 
\usepackage{pgffor} 

\makeatletter
\AtBeginDocument{\let\LS@rot\@undefined}
\makeatother

\def\supplementfilename{SI.pdf}

\pdfximage{\supplementfilename}
\def\numbersupplementpages{\the\pdflastximagepages}

\newif\ifarXiv
\arXivtrue 

\begin{document}

\title{Elastic Constants and Bending Rigidities from Long-Wavelength Perturbation Expansions}

\author{Changpeng Lin}
\email{changpeng.lin@epfl.ch}
\affiliation{%
Theory and Simulation of Materials (THEOS), and National Centre for Computational Design and Discovery of Novel Materials (MARVEL), \'Ecole Polytechnique F\'ed\'erale de Lausanne, 1015 Lausanne, Switzerland
}%
\author{Samuel Ponc\'e}
\affiliation{%
European Theoretical Spectroscopy Facility, Institute of Condensed Matter and Nanosciences, Universit\'e catholique de Louvain, Chemin des \'Etoiles 8, B-1348 Louvain-la-Neuve, Belgium
}%
\affiliation{%
WEL Research Institute, Avenue Pasteur 6, 1300 Wavre, Belgium
}%
\author{Francesco Macheda}
\affiliation{Dipartimento di Fisica, Università di Roma La Sapienza, Piazzale Aldo Moro 5, 00185 Roma, Italy}
\affiliation{Istituto Italiano di Tecnologia, Graphene Labs, Via Morego 30, 16163 Genova, Italy}%
\author{Francesco Mauri}
\affiliation{Dipartimento di Fisica, Università di Roma La Sapienza, Piazzale Aldo Moro 5, 00185 Roma, Italy}
\affiliation{Istituto Italiano di Tecnologia, Graphene Labs, Via Morego 30, 16163 Genova, Italy}%
\author{Nicola Marzari}
\affiliation{%
Theory and Simulation of Materials (THEOS), and National Centre for Computational Design and Discovery of Novel Materials (MARVEL), \'Ecole Polytechnique F\'ed\'erale de Lausanne, 1015 Lausanne, Switzerland
}%
\affiliation{%
Laboratory for Materials Simulations, Paul Scherrer Institut, 5232
Villigen PSI, Switzerland
}%

\date{\today}

\begin{abstract}
Mechanical and elastic properties of materials are among the most fundamental quantities for many engineering and industrial applications.
Here, we present a formulation that is efficient and accurate for calculating the elastic and bending rigidity tensors of crystalline solids, leveraging interatomic force constants and long-wavelength perturbation theory.
Crucially, in the long-wavelength limit, lattice vibrations induce macroscopic electric fields which further couple with the propagation of elastic waves, and a separate treatment on the long-range electrostatic interactions is thereby required to obtain elastic properties under the appropriate electrical boundary conditions.
A cluster expansion of the charge density response and dielectric screening function in the long-wavelength limit has been developed to efficiently extract multipole and dielectric tensors of arbitrarily high order.
We implement the proposed method in a first-principles framework and perform extensive validations on silicon, NaCl, GaAs and rhombohedral BaTiO$_3$ as well as monolayer graphene, hexagonal BN, MoS$_2$ and InSe, obtaining good to excellent agreement with other theoretical approaches and experimental measurements.
Notably, we establish that multipolar interactions up to at least octupoles are necessary to obtain the accurate short-circuit elastic tensor of bulk materials, while higher orders beyond octupole interactions are required to converge the bending rigidity tensor of 2D crystals.
The present approach greatly simplifies the calculations of bending rigidities and will enable the automated characterization of the mechanical properties of novel functional materials.
\end{abstract}

\maketitle

\section{Introduction}

Mechanical properties dictate almost every aspect of daily, industrial, engineering, and scientific sectors.
A system can be mechanically deformed by applying external loads, which originates from the microscopic nature of interatomic bonding forming it.
In 1678, Hooke first obtained his empirical law to describe the force needed to linearly extend or compress a mechanical spring from its equilibrium distance~\cite{hooke1678}, which states the relationship $\sigma=E\varepsilon$ in a stress-strain language with the proportional factor $E$ being the modulus of elasticity.
The complete theory of linear elasticity can be elegantly described by the generalized Hooke's law, where the constant of elastic modulus is replaced by a fourth-rank elastic tensor~\cite{nye1985}.
With this elastic tensor, macroscopic mechanical properties, such as bulk, shear, and Young's moduli, Poisson's ratio, and Lam\'e's parameters, can be derived as well~\cite{voigt1928,reuss1929,hill1952}.
Importantly, thermal transport and thermomechanical properties are intimately linked with elastic tensors and moduli, which enables their use in screening materials for specific applications.
For example, elastic constants and the associated sound velocities are key descriptors for identifying materials with high thermal conductivity for efficient thermal management~\cite{Jia2017,Yan2022}.
One can also have a rapid but approximate screening of candidate materials for high-performance thermoelectrics and thermal barrier coatings using the extended Debye model and minimum thermal conductivity~\cite{cahill1992,clarke2003,toher2014}.
Furthermore, mechanical properties play a critical role in the performance of lithium-ion and solid-state batteries~\cite{Stallard2022,Kalnaus2023}.
Particularly, the design of cathode materials with high mechanical strength is essential for enhancing battery lifespan and maintaining energy capacity, as charging and discharging cycles cause dimensional changes that can lead to fracture of cathodes.
In addition, elastic properties are crucial in the determination of the ductile-to-brittle behavior of solids~\cite{pugh1954,pettifor1992,greaves2011,gschneidner2003,niu2012}, elastic-energy-storage-based programmable soft machines~\cite{pal2020}, molecular crystal engineering~\cite{saha2018,spackman2022}, and the analysis of structure and composition of planets in geophysics~\cite{anderson1968,karki2001,oganov2001}.
Therefore, it is of significant importance to explore mechanical and elastic properties of novel materials for a wide range of applications.

Two-dimensional (2D) materials, particularly since the first exfoliation of atomically thin graphene in 2004~\cite{novoselov2004,novoselov2005}, have been drawing sustained interest owing to their intriguing physical properties, such as tunable electronic band gap~\cite{drummond2012,kim2015}, superior thermal conductivity~\cite{cai2010,chen2012}, ultrahigh Young's modulus~\cite{lee2008}, quantum (anomalous) Hall effect~\cite{goerbig2011,chang2023}, and the emergence of excitonic~\cite{varsano2020,jia2022} and topological insulators~\cite{ren2016}.
Over the past decade, both computational design~\cite{choudhary2017,mounet2018,haastrup2018,gjerding2021} and experimental synthesis~\cite{butler2013} have greatly enriched the family of 2D materials.
Thanks to their reduced dimensionality, 2D materials can also exhibit unique and anomalous mechanical properties, including extraordinary strength~\cite{lee2008}, outstanding flexibility~\cite{fasolino2007,kim2015-2}, negative Poisson’s ratio~\cite{jiang2014,du2016}, and highly anisotropic elastic moduli~\cite{li2020}.
As materials become atomically thin, they can be easily corrugated by mechanical and temperature stimuli, where bending stiffness or rigidity is an essential physical parameter.
For example, ripples (the thermal fluctuations of membrane height) are found ubiquitous as an intrinsic feature in ultrathin 2D crystals like graphene~\cite{meyer2007,bao2009,lui2009}, which play a crucial role in the thermodynamic stability of free-standing samples~\cite{fasolino2007} and the negative in-plane thermal expansion at low temperatures~\cite{mounet2005}.
Besides, the out-of-plane bending properties are also of great significance for the synthesis and preparation of 2D materials~\cite{coleman2011}, which eventually drives their applications in highly flexible electronic devices~\cite{gao2019}.
However, due to the challenge in performing \emph{in-situ} nanomechanics to measure mechanical properties of 2D materials~\cite{li2018}, only a few experimental data are available~\cite{akinwande2017}.
Especially for bending rigidities, the measurements were only performed for graphene and a few 2D transition metal dichalcogenides, including MoS$_2$, WS$_2$, and WSe$_2$~\cite{zhao2015,han2020,yu2021}.
Hence, developing precise and reliable computational tools for predicting the bending rigidities of 2D materials is highly desirable.

First-principles calculations based on density functional theory (DFT) have been the workhorse to provide accurate mechanical properties of crystalline solids at relatively low cost~\cite{de2015,choudhary2018,dal2016}.
There exist mainly two approaches to determine the elastic tensor of crystalline solids: finite differences~\cite{steinle1999,le2001,le2002,dal2016} and linear response method within density functional perturbation theory (DFPT)~\cite{baroni1987,gonze1997,gonze1997a,baroni2001,hamann2005,wu2005}.
In the finite-difference method, several small deformations (strains) are applied to the equilibrium lattice configuration, and the stress tensors of strained structures are obtained thanks to the introduction of the quantum theory of stress~\cite{Nielsen1985}, from which the elastic constants can be calculated according to the stress-strain relation.
Since elastic constants are the second-order derivatives of strain energy density with respect to strain~\cite{nye1985,born1954}, an alternative approach is to compute the variation of total energy as a function of strain.
The number of strains to be applied in these methods depends on the crystal system~\cite{nye1985}, where the Laue class for the centrosymmetric point groups is used to analyze the independent components of elastic tensors.
While the energy-strain method requires more distorted configurations than the stress-based one, the latter is more sensitive to the convergence criteria and requires stringent computational parameters~\cite{le2001,le2002}.
Both finite-difference schemes require a convergence study on the magnitude of the applied strain, increasing the total number of ground-state calculations for strained systems by several times.
Additionally, the structural optimization under external strains becomes computationally expensive for low-symmetry crystals and requires a high-precision calculation to identify the correct minimum~\cite{sluiter1998}.
By contrast, DFPT adopts the analytic differentiation of the stress under the strain perturbation in a linear-response region and solves self-consistently the Sternheimer equation~\cite{baroni1987,gonze1997,gonze1997a,baroni2001,hamann2005,wu2005}.
For a given set of DFT parameters, DFPT calculations circumvent these difficulties and generally provide numerically more accurate elastic tensors than finite differences~\cite{janssen2016}.
Unfortunately, the implementation of DFPT is complex and depends on the specific exchange-correlation functional and pseudopotential classes.

Unlike elastic constants, bending rigidities are challenging to  calculate because introducing curvature generally induces a large deformation to atomic structures and small curvatures require a large supercell with many atoms~\cite{passerone1999}.
The calculations of bending properties of monolayers have thus far relied on the combination of continuum medium models with atomistic simulations using empirical interatomic potentials~\cite{arroyo2004,gao2014,davini2017,zhang2011}, lacking predictive abilities.
Furthermore, scaling down to nanometers, the usage of continuum medium models is also questionable due to the breakdown of the continuum approximation~\cite{zhang2011,verma2016}.
When it comes to first-principles calculations of bending properties,  a few approaches~\cite{koskinen2010,koskinen2010-2,kit2011,wei2013,banerjee2016,ghosh2019} have been proposed to evaluate the bending rigidities of 2D materials, mostly limited to the mean curvature modulus.
One method is to compute the elastic configurational energy due to curvature effect, which can be derived based on the Helfrich Hamiltonian of a membrane:
$\mathcal{U}=AD^\mathrm{P}r^{-2}/2$~\cite{helfrich1973,lipowsky1991} with
$D^\mathrm{P}$ and $A$ being the principal (mean curvature) bending rigidity and the area of atoms per chemical formula.
In this case, one needs to obtain the elastic energy $\mathcal{U}$ of nanotubular configurations at different radii $r$ and perform a fit to extract $D^\mathrm{P}$~\cite{wei2013,lai2016}.
The Gaussian curvature modulus $D^\mathrm{G}$ is also accessible through this method, where a series of spheroidal fullerene-like structures with different radii are built and DFT calculations are performed to obtain the corresponding elastic energy due to Gaussian curvature: $\mathcal{U}=A(2D^\mathrm{P}+D^\mathrm{G})r^{-2}$~\cite{wei2013}.
The main drawback of such a method is the difficulty to study nanotubes with a small curvature, where each configuration can contain hundreds of atoms, while the simulation with a large curvature may cause prominent local distortions due to significant in-plane strains and the induced relaxation effects~\cite{liu2014}.
Koskinen and Kit~\cite{koskinen2010,koskinen2010-2,kit2011} introduced an approach to model the spherical membrane system using revised periodic boundary conditions (PBCs) and Bloch's theorem which are compatible with spherical symmetry.
However, it is not straightforward to adopt such revised PBCs in plane-wave DFT codes, and their applications are limited to calculations in the density-functional tight-binding formalism.
Besides, cyclic DFT~\cite{banerjee2016,ghosh2019} (that directly use the cyclic symmetries in the Kohn-Sham equations) has been proposed to study cylindrical systems and bending deformations.
The advantage of cyclic DFT is that the calculations are carried out within one unit cell with the correct cyclic symmetries and PBCs, which is computationally efficient.
The developed computational framework has been applied to study the low-curvature bending rigidities of 44 monolayer materials~\cite{kumar2020}, and a caveat is that one should verify the selected range of small curvatures yields the correct low-curvature results.
Recently, Shirazian and Sauer~\cite{shirazian2022} presented another first-principles approach to calculate the bending rigidities of 2D materials, by adding directly the curvature perturbation to atomic structures.
They chose the bending deformed configurations that minimize membrane deformations, meanwhile preserving PBCs, and bending rigidities are then obtained by comparing the calculated bending energies to two classical models from structural mechanics.
Although the mentioned \emph{ab initio} approaches have been applied to study the bending deformation of 2D materials,
none of them are tensorial, giving only the mean curvature bending rigidities and Gaussian modulus which limit the study of elastic anisotropy.

Here we present an efficient and accurate approach for calculating simultaneously the elastic and bending rigidity tensors of crystalline solids based on Huang's atomic theory of elasticity~\cite{huang1949,huang1950,born1954}, valid equivalently in any dimension.
Huang~\cite{huang1950} derived the expression for elastic stiffness tensor from a long-wavelength equation for lattice vibrations, and we extend here his method to compute also the bending rigidity tensor via a higher-order long-wavelength perturbation expansion.
Crucially, we provide practical guidelines for the implementation in the framework of modern first-principles simulations, and finally demonstrated its accuracy by validating extensively against the results available in the literature.
We find that good agreements can only be achieved with the subtraction of long-range multipolar electrostatics, ensuring convergence of all the perturbative summations and the correct electric boundary conditions.
Therefore, we develop a novel cluster expansion approach which enables an automated, accurate, and efficient extraction of multipole and dielectric tensors up to an arbitrarily high order, based on the charge density and dielectric responses in the long-wavelength limit.
Besides, our high-order extension of Huang's theory gives access to the complete fourth-rank bending rigidity tensor, which is significant for anisotropic nanoplates, otherwise inaccessible from existing methods.
This enables a direct evaluation of Gaussian curvature modulus without extra cost, and our simulations also reveal the pivotal role of lattice relaxation in 2D materials during bending deformation, establishing a fundamental platform for designing future flexible electronics.

This work is organized as follows.
In Sec.~\ref{atomic_theory}, we first recall the classical continuum treatment of linear elasticity, equations of motion for elastic waves and the electromechanical coupling in dielectric medium.
Then, Huang's results on the long-wavelength acoustic vibrations and his expression for elastic stiffness tensor are derived, with the exact treatment of long-range electrostatic interactions that can be applied in the current DFT codes to correctly accounts for the electrical boundary conditions of elastic constants.
In Sec.~\ref{theory_bending_rgd}, using a higher-order perturbation expansion, we extend the Huang's method to describe the long-wavelength dynamics of flexural acoustic (ZA) modes in 2D materials and obtain the corresponding expression for calculating the full bending rigidity tensor.
In Sec.~\ref{results}, we demonstrate the accuracy and wide applicability of our method by showing example calculations on well-studied bulk and 2D materials, and compare thoroughly with both the existing experimental and theoretical results.
These include silicon (Si), sodium chloride (NaCl), gallium arsenide (GaAs), and rhombohedral barium titanate (BaTiO$_3$) for bulk solids, and monolayer graphene, molybdenum disulfide (MoS$_2$), hexagonal boron nitride (\emph{h}-BN), and indium selenide (InSe) as representative 2D materials.
Finally, in Sec.~\ref{conclusions}, we conclude and briefly summarize our main results in this study.

\section{First-principles theory of elasticity}\label{atomic_theory}

When a solid is homogeneously deformed within the elastic limit, the resulting second-rank stress tensor $\sigma_{\alpha\gamma}$ in the system can be described by the generalized Hooke's law~\cite{nye1985,born1954}:
\begin{equation}
\sigma_{\alpha\gamma}=C_{\alpha\gamma,\beta\delta}\varepsilon_{\beta\delta}, \label{eq:hooke_law}
\end{equation}
where $C_{\alpha\gamma,\beta\delta}$ is the fourth-rank elastic stiffness tensor, and $\varepsilon_{\beta\delta}$ denotes the second-rank strain tensor, with the subscripts $\alpha$, $\beta$, $\gamma$, and $\delta$ labeling the Cartesian components.
Except when explicitly mentioned, we assume Einstein summation on repeated atomic positions and Cartesian directions.
Apart from the induced stress field, the lattice deformation also gives rise to a change in the total energy $\mathcal{E}$~\cite{nye1985,born1954}:
\begin{equation}\label{eq:energy_taylor}
\mathcal{E}(V,\boldsymbol{\varepsilon})-\mathcal{E}(V_0)= \frac{V}{2}C_{\alpha\gamma,\beta\delta}\varepsilon_{\alpha\gamma}\varepsilon_{\beta\delta}+\mathcal{O}(\varepsilon^3),
\end{equation}
in which $V$ and $V_0$ are the strained and equilibrium volumes of the system, respectively.
The linear term of Eq.~\eqref{eq:energy_taylor} is not present when the system is initially at the equilibrium, and this gives the usual definition of the strain energy density function $\mathcal{U}(\boldsymbol{\varepsilon})=\frac{1}{2}C_{\alpha\gamma,\beta\delta}\varepsilon_{\alpha\gamma}\varepsilon_{\beta\delta}$ in a quadratic form.
Since the stress-strain tensors are symmetric and the elastic constant is the second-order derivative of strain energy density with respect to strain components $C_{\alpha\gamma,\beta\delta}=\partial^2 \mathcal{U}(\boldsymbol{\varepsilon})/\partial\varepsilon_{\alpha\gamma}\partial\varepsilon_{\beta\delta}$, these conditions imply that the elastic tensor obeys the symmetry relationships
\begin{equation}\label{eq:elastic_sym_relation}
C_{\alpha\gamma,\beta\delta}=C_{\gamma\alpha,\beta\delta}=C_{\alpha\gamma,\delta\beta}=C_{\beta\delta,\alpha\gamma}.
\end{equation}
As a result, the number of independent components in the fourth-rank elastic tensor decreases from 81 to 21 and can be further reduced, depending on the point group symmetry of crystals.
By calculating the tensor inverse of $C_{\alpha\gamma,\beta\delta}$, one can introduce the elastic compliance tensor $S_{\alpha\gamma,\beta\delta}$~\cite{mavko2020}:
\begin{equation}\label{eq:intro_compliance}
C_{\alpha\gamma,\beta\delta}S_{\beta\delta,\lambda\mu}=\frac{1}{2}(\delta_{\alpha\lambda}\delta_{\gamma\mu}+\delta_{\alpha\mu}\delta_{\gamma\lambda}),
\end{equation}
where $\delta_{\alpha\lambda}$ denotes the Kronecker delta.
The same elastic tensor symmetry Eq.~\eqref{eq:elastic_sym_relation} also holds for $S_{\alpha\gamma,\beta\delta}$, and the stress-strain relation can be then written as
\begin{equation}\label{eq:hooke_law_compliance}
\varepsilon_{\alpha\gamma}=S_{\alpha\gamma,\beta\delta}\sigma_{\beta\delta}.
\end{equation}
The equations presented so far and the equations of motion for elastic waves introduced in the following Sec.~\ref{strain_eq_motion} constitute the basic continuum description of elasticity in solids.
While this work focuses on zero-temperature results, finite-temperature effects can be taken into account using a free energy formalism~\cite{monacelli2021}.

\subsection{Equations of motion for elastic waves}\label{strain_eq_motion}

In continuum mechanics, when a material is subjected to a homogeneous elastic deformation, the deformation parameter $\tilde{\varepsilon}_{\alpha\beta}$ as a second-rank tensor can be defined as
\begin{equation}\label{eq:deform_para}
\tilde{\varepsilon}_{\alpha\beta}\equiv\frac{\partial u_\alpha}{\partial \tau_\beta},
\end{equation}
where $\mathbf{u}$ is the rigid displacement of a material point, and $\boldsymbol{\tau}$ is its initial position.
It is noted that the deformation parameter $\tilde{\varepsilon}_{\alpha\beta}$ itself is not a symmetric tensor, as it consists of a symmetric strain with an antisymmetric pure rotation~\cite{nye1985}.
We indicate with a tilde the quantities that are not symmetric and the strain tensor can be thus obtained by removing such a rotational component as
\begin{equation}\label{eq:strain_tensor}
\varepsilon_{\alpha\beta}=\frac{\tilde{\varepsilon}_{\alpha\beta}+\tilde{\varepsilon}_{\beta\alpha}}{2}=\frac{1}{2}\bigg(\frac{\partial u_\alpha}{\partial\tau_\beta}+\frac{\partial u_\beta}{\partial\tau_\alpha}\bigg),
\end{equation}
which also stems from the symmetry of elastic tensor in Eq.~\eqref{eq:elastic_sym_relation}, and the strain energy density $\mathcal{U}(\boldsymbol{\varepsilon})$ is only a function of the symmetrized tensors~\cite{born1954}.
Although inhomogeneity may occur in special cases such as polycrystalline samples or those with defects, Eq.~\eqref{eq:deform_para} is still valid locally, and the homogeneous assumption can be made in a small neighborhood around any lattice position.

To derive the equations of motion for stress fields in the dynamic theory, we use the divergence theorem for stress tensor, which states that the divergence of the stress tensor is equal to the force per unit volume.
Accordingly, the equations of motion for stress fields can be written as
\begin{equation}\label{eq:eq_of_motion}
\rho\frac{\partial^2 u_\alpha}{\partial t^2} = \frac{\partial \sigma_{\alpha\gamma}}{\partial \tau_\gamma},
\end{equation}
where $\rho$ is the mass density of the crystal, $t$ is the time variable.
Applying the generalized Hooke's law and substituting the strain tensor by Eq.~\eqref{eq:strain_tensor} gives
\begin{equation}\label{eq:eq_of_motion2}
\rho\frac{\partial^2 u_\alpha}{\partial t^2} = C_{\alpha\gamma,\beta\delta}\frac{\partial^2 u_\beta}{\partial \tau_\gamma\partial \tau_\delta},
\end{equation}
in which the symmetry $C_{\alpha\gamma,\beta\delta}=C_{\alpha\gamma,\delta\beta}$ has been used.
Finally, using a plane-wave ansatz $\mathbf{u}(\boldsymbol{\tau},t)=\mathbf{\overline{U}}\exp{(i\mathbf{q}\cdot\boldsymbol{\tau}-i\omega t)}$, the equations of motion for displacement fields or elastic waves are solved as
\begin{equation}\label{eq:wave_eq_of_motion}
\rho\omega^2\overline{U}_\alpha= C_{\alpha\gamma,\beta\delta}q_\gamma q_\delta\overline{U}_\beta,
\end{equation}
where $\mathbf{\overline{U}}$ is a vector of displacement amplitude (all overline quantities are amplitudes), $\mathbf{q}$ is the wave vector, $\omega$ is the frequency of the elastic wave, and $i$ is the imaginary unit.

However, the spontaneous macroscopic electric fields induced by atomic vibrations can couple with stress fields, contributing additionally to the stress tensor through the piezoelectric effects.
For any solid where piezoelectricity is allowed by its group symmetry, such electromechanical couplings between the dielectric polarization and homogeneous deformation are linked by the
constitutive equations~\cite{born1954}:
\begin{align}
\sigma_{\alpha\gamma}&=C_{\alpha\gamma,\beta\delta}\varepsilon_{\beta\delta}-e_{\beta,\alpha\gamma}E_{\beta}, \label{eq:stress_tensor_piezo} \\
P_{\alpha}&=e_{\alpha,\beta\gamma}\varepsilon_{\beta\gamma}+\epsilon_{\alpha\beta}E_{\beta}, \label{eq:polarization}
\end{align}
where $E_{\beta}$ and $P_{\alpha}$ are the macroscopic electric and dielectric displacement fields along the corresponding Cartesian direction,  $\epsilon_{\alpha\beta}$ is the dielectric permittivity tensor and $e_{\alpha,\beta\delta}$ is the third-rank piezoelectric tensor.
The second term in Eq.~\eqref{eq:stress_tensor_piezo} and the first term in Eq.~\eqref{eq:polarization} represent the piezoelectric stress and polarization due to the electric and strain fields, respectively.
With the additional piezoelectric contribution to the stress tensor, the equations of motion Eq.~\eqref{eq:eq_of_motion2} thus become
\begin{equation}\label{eq:eq_of_motion_piezo2}
\rho\frac{\partial^2 u_\alpha}{\partial t^2} = C_{\alpha\gamma,\beta\delta}\frac{\partial^2 u_\beta}{\partial \tau_\gamma\partial \tau_\delta} - e_{\beta,\alpha\gamma}\frac{\partial E_{\beta}}{\partial \tau_{\gamma}}.
\end{equation}
Using a similar wave ansatz for the electric field $\mathbf{E}(\boldsymbol{\tau},t)=\mathbf{\overline{E}}\exp{(i\mathbf{q}\cdot\boldsymbol{\tau}-i\omega t)}$ and dielectric displacement field $\mathbf{P}=\overline{\mathbf{P}}\exp{(i\mathbf{q}\cdot\boldsymbol{\tau}-i\omega t)}$, we arrive at the propagation of electromechanical couplings as
\begin{align}
\rho\omega^2\overline{U}_\alpha&= C_{\alpha\gamma,\beta\delta}q_\gamma q_\delta\overline{U}_\beta+i e_{\beta,\alpha\gamma}q_\gamma\overline{E}_{\beta}, \label{eq:wave_eq_of_motion_piezo} \\
\overline{P}_{\alpha}&=i e_{\alpha,\beta\gamma}q_\gamma\overline{U}_\beta+\epsilon_{\alpha\beta}\overline{E}_{\beta},\label{eq:eq_polarization}
\end{align}
where $\mathbf{\overline{E}}$ and $\overline{\mathbf{P}}$ are the vectors of electric field and dielectric displacement amplitudes, respectively.
The coupling between the long-wavelength elastic waves and the macroscopic electric fields shows that the interatomic forces in any type of crystals are inherently long-range, which leads to a divergent lattice summation for an infinite crystal.
In particular for infrared-active solids, the long-wavelength longitudinal optical (LO) phonons have a non-analytic behavior in the vicinity of the Brillouin zone center and should be treated separately~\cite{gonze1997,baroni2001}.
In this sense, the definition of elastic tensor actually depends on the electrical boundary conditions, and a similar separate treatment for the long-range electrostatic interactions will also be needed for the calculations of elastic constants.
The $C_{\alpha\gamma,\beta\delta}$ discussed so far are the ``short-circuit" ones, which are under the boundary condition of zero electric field and can be obtained by removing the macroscopic long-range Coulomb interactions.
In contrast, the ``open-circuit" elastic constants defined under the boundary condition of zero electric displacement field can be obtained by considering the additional contributions from the macroscopic electric field effects~\cite{wu2005,royo2020},
\begin{equation}\label{eq:open_circuit_C}
C^{\hat{\mathbf{q}}}_{\alpha\gamma,\beta\delta}=C_{\alpha\gamma,\beta\delta}+4\pi\frac{(\hat{\mathbf{q}}\cdot\mathbf{e})_{\alpha\gamma}(\hat{\mathbf{q}}\cdot\mathbf{e})_{\beta\delta}}{\hat{\mathbf{q}}\cdot\boldsymbol{\epsilon}\cdot\hat{\mathbf{q}}},
\end{equation}
where the second term is the direction-dependent piezoelectric contribution to elastic tensor with the unit vector $\hat{\mathbf{q}}$ pointing in the direction of the open-circuit and zero electric displacement field.
The separate treatment of the long-range multipolar interactions to obtain the elastic tensor under different electrical boundary conditions is addressed in Sec.~\ref{multipole}.

\subsection{Long-wavelength acoustic vibrations}\label{long-wavelength}

The secular equation for lattice dynamics from Newton's second law of motion reads~\cite{dove1993,born1954}
\begin{equation}\label{eq:eigeneq}
 M_\kappa\omega^2_\nu(\mathbf{q})U_{\nu,\kappa\alpha}(\mathbf{q}) = \Phi_{\kappa\alpha,\kappa'\beta}(\mathbf{q})U_{\nu,\kappa'\beta}(\mathbf{q}),
\end{equation}
where $M_\kappa$ is the atomic mass of the $\kappa$th atom within the unit cell and $\Phi_{\kappa\alpha,\kappa'\beta}(\mathbf{q})$ is the so-called dynamical matrix.
Such a lattice-dynamical problem is solved by diagonalizing the mass-scaled dynamical matrix $\Phi_{\kappa\alpha,\kappa'\beta}(\mathbf{q})/\sqrt{M_\kappa M_{\kappa'}}$.
If there is $n_\mathrm{a}$ atoms in the unit cell, one can obtain $3n_\mathrm{a}$ phonon modes labeled by the momentum $\mathbf{q}$ and mode index $\nu$, with the corresponding phonon frequencies $\omega_\nu(\mathbf{q})$ and mass-weighted eigenvectors $\sqrt{M_\kappa}U_{\nu,\kappa\alpha}(\mathbf{q})$.
Hereafter, the band index $\nu$ is dropped for clarity.
With the knowledge of the second-order interatomic force constants (IFCs) in real space, the dynamical matrix at an arbitrary $\mathbf{q}$ in reciprocal space can be obtained via a Fourier transform:
\begin{equation}\label{eq:dyn_mat}
\Phi_{\kappa\alpha,\kappa'\beta}(\mathbf{q}) = \sum_{l}\Phi^{l}_{\kappa\alpha,\kappa'\beta}{\rm e}^{-i \mathbf{q}\cdot\boldsymbol{\tau}^{l}_{\kappa\kappa'}},
\end{equation}
where $\boldsymbol{\tau}^{l}_{\kappa\kappa'}\equiv\boldsymbol{\tau}_{l\kappa'}-\boldsymbol{\tau}_{0\kappa}$ is introduced with $\boldsymbol{\tau}_{l\kappa}$ denoting the equilibrium position of the $\kappa$th atom within the unit cell $\mathbf{R}_l$.
Thanks to the translational invariance, the second-order IFCs only depend on the relative position between the two atoms.
As a result, it is sufficient to consider only the second-order IFCs $\Phi^{l}_{\kappa\alpha,\kappa'\beta}$ that involve the atom $\kappa'$ in the unit cell $l$ and the other atom $\kappa$ in the reference unit cell $0$.
We note that the real-space IFCs discussed here include macroscopic long-range effects that can be removed to obtain the short-circuit elastic constants, see Sec.~\ref{multipole}.
In addition, these IFCs need to satisfy the translational and rotational invariances and the equilibrium conditions for vanishing external stress, as summarized in Appendix~\ref{inv_conds}.

To derive the long-wavelength equations for acoustic vibrations, we follow the same perturbation approach adopted by Born and Huang~\cite{born1954}.
We perform a Taylor expansion of the dynamical matrices $\Phi_{\kappa\alpha,\kappa'\beta}(\mathbf{q})$, mode eigenvectors $U_{\kappa\alpha}(\mathbf{q})$, and the corresponding frequencies $\omega(\mathbf{q})$ in terms of the momentum $\mathbf{q}$ to the second order as:
\begin{align}
\Phi_{\kappa\alpha,\kappa'\beta}(\mathbf{q}) \simeq & \Phi^{(0)}_{\kappa\alpha,\kappa'\beta} \!+\! i q_\gamma\Phi^{(1),\gamma}_{\kappa\alpha,\kappa'\beta} \!+\! \frac{q_\gamma q_\delta}{2} \Phi^{(2),\gamma\delta}_{\kappa\alpha,\kappa'\beta} \label{eq:dynmat_expand} \\
U_{\kappa\alpha}(\mathbf{q}) \simeq & U^{(0)}_{\kappa\alpha}+ i U^{(1)}_{\kappa\alpha}(\mathbf{q})+U^{(2)}_{\kappa\alpha}(\mathbf{q}) \label{eq:eigenfunc_expand} \\
\omega(\mathbf{q}) \simeq & \omega^{(1)}(\mathbf{q}),\label{eq:freq_expand}
\end{align}
which gives the acoustic phonon dispersions up to a linear term and related elastic properties.
It should be noted that the imaginary unit $i$ is separated from the derivatives of those complex physical quantities to make the perturbation series all real.
Also, there is no zeroth-order term in Eq.~\eqref{eq:freq_expand}, because the frequencies of acoustic modes vanish at the Brillouin zone center ($\mathbf{q=0}$), imposed by the acoustic sum rules for translational invariance.
By taking the derivative of Eq.~\eqref{eq:dyn_mat} with respect to $\mathbf{q}$, we obtain the detailed perturbation factors for dynamical matrix as
\begin{align}
\Phi^{(0)}_{\kappa\alpha,\kappa'\beta} &= \sum_{l} \Phi^{l}_{\kappa\alpha,\kappa'\beta},\label{eq:D0} \\
\Phi^{(1),\gamma}_{\kappa\alpha,\kappa'\beta} &= -\sum_{l}\Phi^{l}_{\kappa\alpha,\kappa'\beta}\tau^{l}_{\kappa\kappa'\gamma}, \label{eq:D1} \\
\Phi^{(2),\gamma\delta}_{\kappa\alpha,\kappa'\beta} &=  -\sum_{l}\Phi^{l}_{\kappa\alpha,\kappa'\beta} \tau^{l}_{\kappa\kappa'\gamma}\tau^{l}_{\kappa\kappa'\delta}, \label{eq:D2}
\end{align}
where it should be noted that the Einstein summation is not used here.
By substituting Eqs.~\eqref{eq:dynmat_expand} to~\eqref{eq:freq_expand} into the secular equation, Eq.~\eqref{eq:eigeneq}, keeping only the resulting terms up the second order in $\mathbf{q}$, and equating the terms involving the same order in $\mathbf{q}$, one obtains the following three linear equations:
\begin{align}
0 =& \Phi^{(0)}_{\kappa\alpha,\kappa'\beta}U^{(0)}_{\kappa'\beta}, \label{eq:eq_0th} \\
0 =& q_\gamma\Phi^{(1),\gamma}_{\kappa\alpha,\kappa'\beta}U^{(0)}_{\kappa'\beta} \!+\! \Phi^{(0)}_{\kappa\alpha,\kappa'\beta} U^{(1)}_{\kappa'\beta}(\mathbf{q}), \label{eq:eq_1st} \\
\!M_\kappa [\omega^{(1)}\!(\mathbf{q})]^2 U^{(0)}_{\kappa\alpha} =&  \frac{q_\gamma q_\delta}{2}\Phi^{(2),\gamma\delta}_{\kappa\alpha,\kappa'\beta}U^{(0)}_{\kappa'\beta}
 \nonumber \\
-  q_\gamma \Phi^{(1),\gamma}_{\kappa\alpha,\kappa'\beta} &  U^{(1)}_{\kappa'\beta}(\mathbf{q}) + \Phi^{(0)}_{\kappa\alpha,\kappa'\beta} U^{(2)}_{\kappa'\beta}(\mathbf{q}), \label{eq:eq_2nd}
\end{align}
which represent the zeroth- to second-order long-wavelength equations, respectively.
These are a set of coupled equations with the perturbation series of phonon frequencies $\omega(\mathbf{q}) $ and eigenvectors $U_{\kappa\alpha}(\mathbf{q})$ as the unknowns, which must be solved iteratively~\cite{born1954}.
The derivation of their solutions is presented in Appendix~\ref{solvability}.

When seeking the solutions of the second-order equation, its solvability condition in Eq.~\eqref{eq:solvability_2nd} gives rise to the sound propagation equation and can be rewritten as
\begin{equation}\label{eq:eq_acoustic_wave}
M [\omega^{(1)}(\mathbf{q})]^2 \overline{U}_\alpha  = q_\gamma q_\delta T_{\alpha\beta,\gamma\delta}\overline{U}_\beta,
\end{equation}
where $M=\sum_{\kappa}M_\kappa$ is the total atomic mass of unit cell, $\overline{U}_\alpha$ is the polarization vector related to the displacements induced by the zeroth-order acoustic waves, and $T_{\alpha\beta,\gamma\delta}$ can be expressed as
\begin{equation}\label{eq:tensor_T}
T_{\alpha\beta,\gamma\delta}=T^{\mathrm{CI}}_{\alpha\beta,\gamma\delta}+\frac{1}{2}\Big(T^{\rm LM}_{\alpha\gamma,\beta\delta}+T^{\rm LM}_{\alpha\delta,\beta\gamma}\Big).
\end{equation}
In the above equation, $T^{\mathrm{CI}}_{\alpha\beta,\gamma\delta}$ is the clamped-ion contribution to elastic deformation, representing the anisotropic components of the stress tensor~\cite{born1954,lin2022,leibfried1961}, while $T^{\rm LM}_{\alpha\gamma,\beta\delta}$ is the contribution associated with the relaxation of internal ionic coordinates in response to the external stress fields~\cite{born1954}.
These two tensors are defined using Eqs.~\eqref{eq:strain_gradient_frozen} and \eqref{eq:strain_gradient_relax}, which yields
\begin{align}
T^{\mathrm{CI}}_{\alpha\beta,\gamma\delta} = \sum_\kappa T^{\mathrm{CI},\kappa}_{\alpha\beta,\gamma\delta} &= \frac{1}{2}\sum_{\kappa\kappa'}\Phi^{(2),\gamma\delta}_{\kappa\alpha,\kappa'\beta},\label{eq:bracket} \\
T^{\mathrm{LM}}_{\alpha\gamma,\beta\delta} = \sum_\kappa T^{\mathrm{LM},\kappa}_{\alpha\gamma,\beta\delta}&=-\Lambda^{\kappa}_{\lambda,\alpha\gamma}\Upsilon^{\kappa}_{\lambda,\beta\delta} \nonumber \\
&=-\Lambda^{\kappa}_{\lambda,\alpha\gamma}\Gamma_{\kappa\lambda,\kappa'\mu}\Lambda^{\kappa'}_{\mu,\beta\delta}, \label{eq:curly}
\end{align}
where we have used the definitions given by Eq.~\eqref{eq:Gamma_orthogonal}, \eqref{eq:strain_force_response}, and \eqref{eq:displace_response} for $\boldsymbol{\Gamma}$, $\boldsymbol{\Lambda}$, and $\boldsymbol{\Upsilon}$ which act as the inverse of the zone-center IFC matrix, internal-strain force-response tensor and internal-strain displacement-response tensor, respectively.
The antisymmetry relation $\Phi^{(1),\gamma}_{\kappa\alpha,\kappa'\beta}=-\Phi^{(1),\gamma}_{\kappa'\beta,\kappa\alpha}$ is also used to obtain Eq.~\eqref{eq:curly}.
In addition, $T^{\mathrm{CI}}_{\alpha\beta,\gamma\delta}$ and $T^{\rm LM}_{\alpha\gamma,\beta\delta}$ follow the symmetry relations:
\begin{align}\label{eq:T_CI_sym}
T^{\mathrm{CI}}_{\alpha\beta,\gamma\delta} =& T^{\mathrm{CI}}_{\beta\alpha,\gamma\delta} = T^{\mathrm{CI}}_{\alpha\beta,\delta\gamma}=T^{\mathrm{CI}}_{\gamma\delta,\alpha\beta}, \\
T^{\rm LM}_{\alpha\gamma,\beta\delta} =& T^{\rm LM}_{\gamma\alpha,\beta\delta}=T^{\rm LM}_{\alpha\gamma,\delta\beta}=T^{\rm LM}_{\beta\delta,\alpha\gamma},\label{eq:T_LM_sym}
\end{align}
where Eq.~\eqref{eq:T_CI_sym} comes from its definition as a symmetrized tensor by removing surface effects~\cite{lin2022} and the equilibrium conditions Eq.~\eqref{eq:eq_conds}, and Eq.~\eqref{eq:T_LM_sym} is a result of the
rotational invariance Eq.~\eqref{eq:ri_2nd} and the fact that $\Gamma_{\kappa\alpha,\kappa'\beta}$ is symmetric.
The long-wavelength Eq.~\eqref{eq:eq_acoustic_wave} for sound wave propagation is intimately linked with the equations of motion for elastic waves Eq.~\eqref{eq:wave_eq_of_motion}, and it directly determines the first-order linear dispersion for acoustic modes around the Brillouin zone center,
\begin{equation}\label{eq:linear_disp}
[\omega^{(1)}(\mathbf{q})]^2=q_\gamma q_\delta\frac{T_{\alpha\beta,\gamma\delta}}{M}
\overline{U}_\alpha\overline{U}_\beta,
\end{equation}
where $\overline{U}_\alpha$ can be chosen to be a real and orthonormal vector that diagonalizes the zone-center dynamical matrix.
By introducing $\mathbf{q}\equiv q\hat{\mathbf{q}}$, Eq.~\eqref{eq:eq_acoustic_wave} can be further recast into the Christoffel's equation for elastic waves \cite{fedorov2013}:
\begin{equation}\label{eq:christoffel}
\Big[M v^2 \delta_{\alpha\beta} - \hat{q}_\gamma \hat{q}_\delta T_{\alpha\beta,\gamma\delta}\Big]\overline{U}_\beta=0,
\end{equation}
with the phase velocity $v=\omega^{(1)}(\mathbf{q})/|\mathbf{q}|$.
Such an eigenvalue equation is central to the theory of elastic waves in crystals and can be solved by diagonalizing the positive-definite tensor $M^{-1}\hat{q}_\gamma \hat{q}_\delta T_{\alpha\beta,\gamma\delta}$, which gives rise to the sound propagation velocity in crystals and the corresponding displacement vector along the $\hat{\mathbf{q}}$ direction.
Finally, we show in Appendix~\ref{relations_longwave_elastic} that the coupled long-wavelength Eqs.~\eqref{eq:eq_0th}--\eqref{eq:eq_2nd} are directly linked to the homogeneous elastic deformation.

\subsection{Huang formula for elastic stiffness tensor}\label{Huang_formula}

The expression for the short-circuit elastic stiffness tensor can be obtained by comparing the long-wavelength equation for acoustic vibrations with the macroscopic equation for elastic waves.
By comparing Eqs.~\eqref{eq:wave_eq_of_motion} and \eqref{eq:eq_acoustic_wave} divided by the unit cell volume $\Omega$, we have
\begin{equation}\label{eq:eq_acoustic_wave_elcons}
\rho [\omega^{(1)}(\mathbf{q})]^2 \overline{U}_\alpha  = \frac{1}{\Omega}  T_{\alpha\beta,\gamma\delta} q_\gamma q_\delta \overline{U}_\beta,
\end{equation}
with $\rho=M/\Omega$, which implies that the following identity holds:
\begin{equation}\label{eq:elastic_identity_q}
 C_{\alpha\gamma,\beta\delta}q_\gamma q_\delta=\frac{1}{\Omega}  T_{\alpha\beta,\gamma\delta} q_\gamma q_\delta.
\end{equation}
Since this identity is valid for any small value of $\mathbf{q}$, we can further have
\begin{equation}\label{eq:elastic_identity}
C_{\alpha\gamma,\beta\delta}+C_{\alpha\delta,\beta\gamma}=\frac{1}{\Omega}\Big(2T^{\mathrm{CI}}_{\alpha\beta,\gamma\delta} + T^{\rm LM}_{\alpha\gamma,\beta\delta}+T^{\rm LM}_{\alpha\delta,\beta\gamma}\Big),
\end{equation}
where the symmetry of $T^{\mathrm{CI}}_{\alpha\beta,\gamma\delta}$ is applied, and the unsymmetrized elastic tensor~\cite{stengel2013} can be represented as
\begin{equation}\label{eq:elastic_tensor_unsym}
\tilde{C}_{\alpha\beta,\gamma\delta}=\frac{1}{\Omega}\Big[T^{\mathrm{CI}}_{\alpha\beta,\gamma\delta} + \frac{1}{2}\Big(T^{\rm LM}_{\alpha\gamma,\beta\delta}+T^{\rm LM}_{\alpha\delta,\beta\gamma}\Big)\Big].
\end{equation}
This degree of freedom in defining the elastic tensor stems from using a reciprocal-space formulation as in Eq.~\eqref{eq:wave_eq_of_motion} or \eqref{eq:eq_acoustic_wave} from the lattice-dynamical theory, which is often referred to as \emph{dynamic} elastic tensor in contrast to the \emph{static} one defined in real space~\cite{stengel2016,divincenzo1986}.
When considering the symmetry relations of $C_{\alpha\gamma,\beta\delta}$, $T^{\mathrm{CI}}_{\alpha\beta,\gamma\delta}$ and $T^{\rm LM}_{\alpha\gamma,\beta\delta}$ given by Eqs.~\eqref{eq:elastic_sym_relation}, \eqref{eq:T_CI_sym}, and \eqref{eq:T_LM_sym}, respectively, the symmetric elastic tensor is defined as~\cite{born1954,huang1950}
\begin{equation}\label{eq:elastic_tensor}
C_{\alpha\gamma,\beta\delta}=\frac{1}{\Omega}\Big(T^{\mathrm{CI}}_{\alpha\beta,\gamma\delta} + T^{\mathrm{CI}}_{\beta\gamma,\alpha\delta} - T^{\mathrm{CI}}_{\beta\delta,\alpha\gamma} + T^{\rm LM}_{\alpha\gamma,\beta\delta}\Big),
\end{equation}
which is the Huang formula for the elastic stiffness tensor and is the only solution to Eq.~\eqref{eq:elastic_identity}.
It is important to emphasize that the elastic tensor just defined is under short-circuit boundary conditions, where the macroscopic electrostatics must be correctly handled (see Sec.~\ref{multipole} for details).
Thus, the elastic constants of crystals can be fully determined from the two tensors, $T^{\mathrm{CI}}_{\alpha\beta,\gamma\delta}$ and $T^{\rm LM}_{\alpha\gamma,\beta\delta}$.

Due to mechanical stability, the results of $T^{\mathrm{CI}}_{\alpha\beta,\gamma\delta} + T^{\mathrm{CI}}_{\beta\gamma,\alpha\delta} - T^{\mathrm{CI}}_{\beta\delta,\alpha\gamma}$ must be a positive-definite matrix, which always increases the total energy of crystals under elastic deformations.
By contrast, $T^{\rm LM}_{\alpha\gamma,\beta\delta}$ is in general negative as the relaxation of internal ionic degrees of freedom lowers the total energy, if there is more than one atom in the unit cell without enough site symmetries (e.g. inversion).
We note that the long-wavelength equation for acoustic modes and the macroscopic equation of motion for elastic waves are equivalent only if the system is initially free of stress
such that the equilibrium conditions $T^{\mathrm{CI}}_{\alpha\beta,\gamma\delta}=T^{\mathrm{CI}}_{\gamma\delta,\alpha\beta}$ hold [see Eq.~\eqref{eq:eq_conds}].
Nevertheless, the equilibrium conditions for vanishing external stress can still hold under isotropic and hydrostatic pressures, $\sigma_{\alpha\beta}=-p\delta_{\alpha\beta}$, where $p$ is a constant.
This means that the Huang formula for elastic constants can be still applied to crystals under any symmetry-preserving hydrostatic pressure.

\subsection{Separation of long-range electrostatic interactions}\label{multipole}

In semiconductors and insulators, lattice vibrations will generate macroscopic fields described by a long-range electrostatic potential~\cite{Pick1970,stengel2013}, which leads to a slow spatial decay of real-space IFCs with increasing interatomic distances.
To guarantee the real-space locality and smoothness of the IFCs for an accurate Fourier interpolation of the dynamical matrix in Eq.~\eqref{eq:dyn_mat} near the Brillouin zone center,
the dynamical matrix can be decomposed into a short-ranged ($\mathcal{S}$) and a long-range ($\mathcal{L}$) parts as~\cite{baroni2001,gonze1997}
\begin{equation}\label{eq:ifc_range_separation}
\Phi(\mathbf{q})=\Phi^{\mathcal{S}}(\mathbf{q})+\Phi^{\mathcal{L}}(\mathbf{q}),
\end{equation}
where the latter is the non-analytic electrostatic contribution to lattice dynamics.
For bulk materials, the long-range part is given by~\cite{stengel2013}
\begin{equation}\label{eq:long-range}
\Phi^{\mathcal{L}}(\mathbf{q})=\lim_{\mathbf{q}\to \mathbf{0}}4\pi\Omega\frac{\ket{\overline{\rho}(\mathbf{q})}\bra{\overline{\rho}(\mathbf{q})}}{\xi(\mathbf{q})},
\end{equation}
where $\ket{\overline{\rho}(\mathbf{q})}$ is the unscreened charge density response to a phonon perturbation $\mathbf{q}$ after removing the macroscopic potential, and $\xi(\mathbf{q})$ is the dielectric screening function.
They are both analytic in the long-wavelength limit ($\mathbf{q}\to\mathbf{0}$), which can be expanded as a Taylor series:
\begin{multline}\label{eq:charge_density}
\overline{\rho}_{\kappa\alpha}(\mathbf{q})=\frac{e}{\Omega}\Big(-i  q_\beta Z^\beta_{\kappa\alpha}-\frac{q_\beta q_\gamma}{2}Q^{\beta\gamma}_{\kappa\alpha}\\
+i\frac{q_\beta q_\gamma q_\delta}{6}O^{\beta\gamma\delta}_{\kappa\alpha}\Big)+\mathcal{O}(q^4)
\end{multline}
and
\begin{align}\label{eq:screening_func}
\!\!\!\!\!\xi(\mathbf{q}) =& \frac{|\mathbf{q}|^2}{\epsilon^{-1}({\mathbf{q}})} \nonumber \\
=& \boldsymbol{\epsilon}^{\infty}\cdot(\mathbf{q}\otimes\mathbf{q}) +\boldsymbol{\epsilon}^{(4)}\cdot(\mathbf{q}\otimes\mathbf{q}\otimes\mathbf{q}\otimes\mathbf{q})+ \mathcal{O}(q^6),
\end{align}
where $Z^\beta_{\kappa\alpha}$, $Q^{\beta\gamma}_{\kappa\alpha}$ and $O^{\beta\gamma\delta}_{\kappa\alpha}$ are the dynamical multipole tensors due to the displacement of the atom $\kappa$ along the Cartesian direction $\alpha$, corresponding to the dynamical Born effective charge, quadrupole and octupole tensors, respectively, $\epsilon^{-1}({\mathbf{q}})$ is the macroscopic inverse dielectric function, $\boldsymbol{\epsilon}^{\infty}$ is the second-order high-frequency electronic (clamped-ion) dielectric permittivity tensor, and $\boldsymbol{\epsilon}^{(4)}$ is the fourth-order dielectric dispersion tensor.
We note that the dielectric screening $\xi(\mathbf{q})$ is an even function of $\mathbf{q}$.

In practice, to numerically compute the multipole and dielectric tensors, it is advantageous to rewrite Eqs.~\eqref{eq:charge_density} and \eqref{eq:screening_func} into a cluster expansion form
with a least-squares fit to obtain the irreducible tensor components.
This rewriting using a cluster expansion model is detailed in Appendix~\ref{multipole_cluster_expansion} and is the method we used in this work.
Then, by injecting Eqs.~\eqref{eq:charge_density} and \eqref{eq:screening_func} into Eq.~\eqref{eq:long-range}, we have to $\mathcal{O}(q^2)$~\cite{stengel2013,royo2020}:
\begingroup
\allowdisplaybreaks
\begin{align}
\Phi^{\mathcal{L}}(\mathbf{q})&\simeq\Phi^{\rm{DD}}(\mathbf{q})+\Phi^{\rm{DQ}}(\mathbf{q})+\Phi^{\rm{QQ}}(\mathbf{q}) \nonumber \\
&\qquad\qquad\;\:\,+\Phi^{\rm{DO}}(\mathbf{q})+\Phi^{\rm{D}\epsilon\rm{D}}(\mathbf{q}), \label{eq:Phi_expand} \\
\Phi^{\rm{DD}}(\mathbf{q})&=\frac{4\pi e^2}{\Omega}\frac{\ket{Z(\mathbf{q})}\bra{Z(\mathbf{q})}}{\mathbf{q}\cdot\boldsymbol{\epsilon}^{\infty}\cdot\mathbf{q}}, \label{eq:DD} \\
\Phi^{\rm{DQ}}(\mathbf{q})&=-i\frac{4\pi e^2}{2\Omega}\frac{\ket{Z(\mathbf{q})}\bra{Q(\mathbf{q})}-\ket{Q(\mathbf{q})}\bra{Z(\mathbf{q})}}{\mathbf{q}\cdot\boldsymbol{\epsilon}^{\infty}\cdot\mathbf{q}}, \label{eq:DQ} \\
\Phi^{\rm{QQ}}(\mathbf{q})&=\frac{4\pi e^2}{4\Omega}\frac{\ket{Q(\mathbf{q})}\bra{Q(\mathbf{q})}}{\mathbf{q}\cdot\boldsymbol{\epsilon}^{\infty}\cdot\mathbf{q}},  \label{eq:QQ} \\
\Phi^{\rm{DO}}(\mathbf{q})&=-\frac{4\pi e^2}{6\Omega}\frac{\ket{Z(\mathbf{q})}\bra{O(\mathbf{q})}+\ket{O(\mathbf{q})}\bra{Z(\mathbf{q})}}{\mathbf{q}\cdot\boldsymbol{\epsilon}^{\infty}\cdot\mathbf{q}}, \label{eq:DO} \\
\Phi^{\rm{D}\epsilon\rm{D}}(\mathbf{q})&=\frac{-4\pi e^2}{\Omega}\frac{\ket{Z(\mathbf{q})}\boldsymbol{\epsilon}^{(4)}\cdot(\mathbf{q}\!\otimes\!\mathbf{q}\otimes\mathbf{q}\otimes\mathbf{q})\bra{Z(\mathbf{q})}}{(\mathbf{q}\cdot\boldsymbol{\epsilon}^{\infty}\cdot\mathbf{q})^2}, \label{eq:DeD}
\end{align}
\endgroup
where the superscripts D, Q, and O denote the dynamical dipoles, quadrupoles and octupoles, respectively, and we have introduced the notations:
\begin{align}
\braket{\kappa\alpha|Z(\mathbf{q})}&=q_\beta Z^\beta_{\kappa\alpha},\label{eq:Zq} \\
\braket{\kappa\alpha|Q(\mathbf{q})}&=q_\beta q_\gamma Q^{\beta\gamma}_{\kappa\alpha},\label{eq:Qq} \\
\braket{\kappa\alpha|O(\mathbf{q})}&=q_\beta q_\gamma q_\delta O^{\beta\gamma\delta}_{\kappa\alpha}.\label{eq:Oq}
\end{align}
Specifically, $\Phi^{\rm{D}\epsilon\rm{D}}(\mathbf{q})$ in Eq.~\eqref{eq:DeD} represents the dipole-dipole interaction mediated by dielectric spacial dispersion~\cite{stengel2013,royo2020}.
For practical calculations of the long-range dynamical matrix, the Ewald summation technique as described in Ref.~\cite{gonze1997} is adopted, and the multipolar electrostatic contributions to dynamical matrix can be rewritten into a summation over reciprocal lattice $\mathbf{G}$ vectors as
\begin{align}\label{eq:ewald_sum}
C^{\mathcal{L}}_{\kappa\alpha,\kappa'\beta}(\mathbf{q})=&\sum_{\mathbf{G}\neq\mathbf{-q}}\braket{\kappa\alpha|\Phi^{\mathcal{L}}(\mathbf{q+G})|\kappa'\beta}\times \nonumber \\
&\rm{e}^{i(\mathbf{q+G})\cdot(\boldsymbol{\tau}_\kappa-\boldsymbol{\tau}_{\kappa'})}\rm{e}^{-\frac{(\mathbf{q+G})\cdot\boldsymbol{\epsilon}^{\infty}\cdot(\mathbf{q+G})}{4\Lambda^2}},
\end{align}
where $\Lambda$ is the range separation parameter.
Therefore, the long-range parts of dynamical matrix can be written as~\cite{baroni2001,gonze1997}
\begin{equation}\label{eq:Phi_LR_ewald}
\Phi^{\mathcal{L}}_{\kappa\alpha,\kappa'\beta}(\mathbf{q})=C^{\mathcal{L}}_{\kappa\alpha,\kappa'\beta}(\mathbf{q})-\delta_{\kappa\kappa'}\sum_{\kappa''}C^{\mathcal{L}}_{\kappa\alpha,\kappa''\beta}(\mathbf{q=0}),
\end{equation}
and the second term is used to ensure the translational invariance of crystals.
In order to have a smooth and accurate interpolation of dynamical matrix and phonon dispersions, the resulting short-range IFC matrix $\Phi^{\mathcal{S}}(\mathbf{R}_{l})$ are required to be localized in real space.
This can be achieved by minimizing the summation of the absolute values of $\Phi^{\mathcal{S}}(\mathbf{R}_l)$ with respect to $\Lambda$ as~\cite{royo2021,ponce2023}
\begin{equation}\label{eq:sum_IFC_SR}
\Phi^{\mathcal{S}}(\mathbf{R}_l)=\overunderset{*}{\Lambda}{\arg\min}\big\Vert\Phi^{\mathcal{S}}(\mathbf{R}_l)\big\Vert_1,
\end{equation}
where $\Vert\cdots\Vert_1$ represents the $\ell_1$-norm (or Manhanttan distance) and the asterisk means the self-interaction terms are excluded.
We use the optimal $\Lambda$ that minimizes Eq.~\eqref{eq:sum_IFC_SR} to calculate the long-range part of dynamical matrix and interpolate phonon dispersions shown in Sec.~\ref{results}, whose value depends on the material of interest and the included multipolar terms.
However, in the case of elastic tensor calculations using short-circuit boundary conditions, $\Lambda$ needs to be small such that it only removes the macroscopic $\mathbf{G=0}$ component of the long-range Coulomb interaction.
The optimal $\Lambda$ obtained from Eq.~\eqref{eq:sum_IFC_SR} generally will be too large, which removes a certain portion of short-range components of Coulomb interactions as well.
Therefore, to obtain the short-circuit elastic tensor we minimize the sum of short-range IFCs in Eq.~\eqref{eq:sum_IFC_SR} with the only $\mathbf{G=0}$ taken into account to find the optimal value of $\Lambda$.
We find and show in Sec.~\ref{results} that the elastic tensor is not very sensitive to the choice of the range separation parameters.
The removal of such $\mathbf{G=0}$ component is also consistent with the DFPT formulation of elastic tensor, which excludes the macroscopic Hartree, local pseudopotential, and Ewald terms~\cite{stengel2013}.

Importantly, the multipole tensors of low-symmetry solids are not guaranteed to be symmetric when permuting the direction of phonon perturbation with the direction of the induced multipolar response.
For instance, the Born effective charge tensor in low-symmetry crystals is not always a symmetric matrix if their Wyckoff positions do not respect certain site symmetry constraints~\cite{cockayne2000,zhou2019-2}.
This will lead to the non-hermiticity of the long-range part of dynamical matrix.
However, we insist that the \emph{total} dynamical matrix should always be hermitian~\cite{powell1970,martin1971,scheringer1974} but might not be in practice due to numerical reasons.
To guarantee hermiticity, the following permutation symmetry of total IFCs must be satisfied: $\Phi^{0l}_{\kappa\alpha,\kappa'\beta}=\Phi^{l0}_{\kappa'\beta,\kappa\alpha}$, which should always hold as a result of the commutativity of the derivatives of a crystal potential.
Furthermore, the on-site (self-interaction) term $\Phi^{0}_{\kappa\alpha,\kappa\beta}$ that is determined from the translation invariance Eq.~\eqref{eq:ti_2nd} thus becomes~\cite{martin1971,scheringer1974}
\begin{equation}\label{eq:onsite_Herm_cond}
\frac{\Phi^{00}_{\kappa\alpha,\kappa\beta}+\Phi^{00}_{\kappa\beta,\kappa\alpha}}{2}=-\sum^{*}_{l\kappa'}\Phi^{l}_{\kappa\alpha,\kappa'\beta},
\end{equation}
where the asterisk indicates that the $\kappa'=\kappa$ term within the reference unit cell $l=0$ is excluded from the summation.
As an example of a case where hermiticity is broken due to numerical approximations, we have computed the lattice vibration properties of the low-symmetry rhombohedral BaTiO$_3$ with space group \textit{R3m} using the \textsc{Quantum ESPRESSO}~\cite{giannozzi2009,giannozzi2017} software
and find that the interpolated dynamical matrix is not hermitian when translational invariance is imposed by correcting only the on-site IFCs, even without the subtraction of the non-analytic long-range electrostatic part.
In this particular case, the observed non-hermitian behavior of the dynamical matrix is almost negligible and left for future study.

Before going further, we would like to stress the significance of the developed long-wavelength perturbation framework.
What we have achieved is an expression for the elastic constants, which only requires the knowledge of IFCs in real space, avoiding the finite difference of stress tensor, and IFCs can be obtained via DFPT~\cite{gonze1997,baroni2001} or small displacement method~\cite{Togo2023}, making it amenable to any atomistic code that provides forces.
This, by itself, has many advantages even if elastic constants are routinely computed by many \emph{ab initio} codes.
In particular, the Huang formula is independent on the PBCs, which are implicitly treated by constructing the optimal Wigner-Seitz supercell in lattice dynamical calculations~\cite{lin2022}, and this means Eq.~\eqref{eq:elastic_tensor} can be applied equivalently to any dimension of systems.
However, the benefit of our formulation is that it also gives access to other quantities that are otherwise almost impossible to compute, one such quantity being the bending rigidity tensor.
Furthermore, the extracted high-order multipole tensors from our generalized cluster expansion approach will enable the study of piezoelectricity~\cite{martin1972} and flexoelectricity~\cite{stengel2013}, and provide an accurate description of long-range electron-phonon coupling beyond Fr\"{o}hlich interactions.

\section{Extension to bending rigidities}\label{theory_bending_rgd}
We begin the study of flexural properties of 2D materials by briefly reviewing the classical Kirchhoff plate theory~\cite{reddy2006,ugural2017,mittelstedt2023}.
For a plate of thickness $h$, it is common to define an auxiliary midplane that bisects the plate into two equal halves (i.e. $-h/2 \le z \le h/2$), and we shall only study the case of such thin plate under small deflections.
The Kirchhoff plate theory relies on three assumptions known as \emph{Kirchhoff hypotheses}~\cite{reddy2006,ugural2017,mittelstedt2023}: (i) the deflection is small compared to the thickness of plate and the midplane remains unstrained; (ii) the plane cross section remains normal to the midplane and to the surfaces before and after bending deformations; and (iii) the plate behaves like linear-elastic materials and the Hooke's law is valid.
Importantly, (ii) implies that the thickness direction is not extensible and the strain components $\varepsilon_{zz}$, $\varepsilon_{xz}$ and $\varepsilon_{yz}$ are vanishing.
Consequently, the out-of-plane displacement $w$ becomes independent of the position $z$ as $w(x,y)$, while the in-plane displacement varies linearly in $z$ as $\mathbf{u}^{\parallel}(x,y)=-z\nabla w(x,y)$.
For the in-plane stress-strain fields, the constitutive relation described by the generalized Hooke's law and the corresponding equations of motion for longitudinal and transverse elastic waves in Sec.~\ref{strain_eq_motion} still hold for 2D materials, but with the additional flexural contributions to strain tensor.
According to the Kirchhoff plate theory, the flexural in-plane strain tensor is given by $\varepsilon_{\alpha\beta}=-z\kappa_{\alpha\beta}$, where $\kappa_{\alpha\beta}=\partial^2 w/(\partial \tau_\alpha \partial \tau_\beta)$ is the curvature tensor (which should not be confused with the one for atom sites in the unit cell).
We therefore focus here on the out-of-plane displacements $w$, bending waves $\omega^{}_{\mathrm{ZA}}(\mathbf{q})$, and their equations of motion.

\subsection{Equations of motion for bending waves}
In the case of bending, the stress-strain relationship of linear elasticity is replaced by the one between stress resultants (force per unit length) and curvature~\cite{ugural2017,mittelstedt2023}:
\begin{equation}\label{eq:moment_curvature_relation}
\mathcal{M}_{\alpha\beta}=-D_{\alpha\beta,\gamma\delta}\kappa_{\gamma\delta},
\end{equation}
where $D_{\alpha\beta,\gamma\delta}$ is the bending (flexural) rigidity tensor, and $\mathcal{M}_{\alpha\beta}$ is the bending ($\alpha=\beta$) or twisting ($\alpha\ne\beta$) moment.
By doing the tensor inverse of $D_{\alpha\beta,\gamma\delta}$, one can introduce the flexural equivalence of the compliance tensor $\mathcal{D}_{\alpha\beta,\gamma\delta}$, and the curvature-moment relation can be then written as
\begin{equation}\label{eq:curvature_moment-relation}
\kappa_{\alpha\beta}=-\mathcal{D}_{\alpha\beta,\gamma\delta}\mathcal{M}_{\gamma\delta}.
\end{equation}
For a Kirchhoff plate, the bending rigidity tensor by its definition can be calculated from the in-plane elastic stiffness tensor as
\begin{equation}\label{eq:bend_rgd_from_elastic_stiffness}
D_{\alpha\beta,\gamma\delta}=\int_{-\frac{h}{2}}^{\frac{h}{2}}z^2C_{\alpha\beta,\gamma\delta}\mathrm{d}z,
\end{equation}
and if the elastic stiffness tensor $C_{\alpha\beta,\gamma\delta}$ is homogeneous over the plate thickness $h$,  the above integral is then reduced to $D_{\alpha\beta,\gamma\delta}=h^3 C_{\alpha\beta,\gamma\delta} /12$.
As discussed in the literature~\cite{zhang2011,verma2016}, the validity of using such continuum approximation to calculate bending rigidities can be questionable for nanoscale materials, due to the ambiguous definition of the effective bending thickness which can be even direction-dependent in elastic anisotropic materials.
Therefore, we develop in Sec.~\ref{long_bending_wave} a microscopic theory of long-wavelength bending vibrations to predict the flexural properties of 2D materials.
We further note that the bending rigidity tensor defined in this work is the same as the one entering the expansion of the energy density in terms of curvatures, as explained in Eq.~(S41) of the SM~\cite{SM}, allowing for the connection between microscopic and macroscopic elasticity theory (see Chapter 7 of Ref.~\cite{mittelstedt2023} for further details and derivations).

The equations of motion for bending waves can be obtained via a similar divergence theorem for stress resultant tensor or based on the local equilibrium of moments loaded on an inﬁnitesimal plate element~\cite{ugural2017,mittelstedt2023}:
\begin{equation}\label{eq:resultants_eq_motion}
\rho\frac{\partial^2 w}{\partial t^2} = \frac{\partial^2\mathcal{M}_{\alpha\beta}}{\partial\tau_\alpha\partial\tau_\beta}.
\end{equation}
By substituting Eq.~\eqref{eq:moment_curvature_relation} into Eq.~\eqref{eq:resultants_eq_motion} with $\kappa_{\alpha\beta}=\partial^2 w/(\partial \tau_\alpha \partial \tau_\beta)$, we arrive at the following expression
\begin{equation}\label{eq:displace_field_motion}
 \rho^{\rm 2D}\frac{\partial^2 w}{\partial t^2} = -D_{\alpha\beta,\gamma\delta}\frac{\partial^4w}{\partial\tau_\alpha\partial\tau_\beta\partial\tau_\gamma\partial\tau_\delta},
\end{equation}
also known as the fourth-order displacement partial differential equation obtained first by Lagrange in 1811~\cite{ugural2017}, where $ \rho^{\rm 2D}=\rho h$ is the 2D mass density.
By applying a wave-like ansatz $w(\boldsymbol{\tau},t)=\overline{w}\exp(i\mathbf{q}\cdot\boldsymbol{\tau}-i\omega t)$ to Eq.~\eqref{eq:displace_field_motion}, we can then obtain
\begin{equation}\label{eq:eq_motion_bend_wave}
 \rho^{\rm 2D}\omega^2 = q_\alpha q_\beta q_\gamma q_\delta D_{\alpha\beta,\gamma\delta}.
\end{equation}
From such a macroscopic description of bending wave propagation, we see a quadratic dependence of frequencies on the wave vectors, consistent with the prediction from the microscopic lattice dynamical theory~\cite{lin2022,croy2020}.
As already pointed out from the assumptions made in the Kirchhoff plate theory, the deflection $w$ is independent of the plate thickness, all Cartesian components appearing in the equations of motion should be limited to the in-plane components.
In Sec.~\ref{long_bending_wave}, we will demonstrate that the dynamic equations obtained here are actually identical to the formulation from the lattice-dynamical approach of long-wavelength acoustic bending waves, where the exact formula of bending rigidity tensor is obtained.

Besides, it has been also shown that there can be a strong electromechanical coupling between the strain gradient (i.e. bending curvature) and macroscopic electric fields due to flexoelectric effects in piezoelectric 2D nanoplates~\cite{yan2012,zhang2014,yang2015}, which may further alter their mechanical properties and depends on the electrical boundary conditions.
The first-principles calculations of flexoelectric coefficients in 2D materials have only recently appeared~\cite{springolo2021,springolo2023} such that we restrict ourselves to the case of short-circuit boundary conditions, which corresponds to the removal of macroscopic long-range electrostatic interactions in the 2D formalism (see Sec.~\ref{2D_long_range} for details).
The effect of flexoelectricity on bending rigidities is left for future investigation.

\subsection{Long-wavelength flexural acoustic vibrations}\label{long_bending_wave}
To obtain the equations of motion for the bending vibrations in a system with a finite thickness, one needs to further expand the dynamical matrices $\Phi_{\kappa\alpha,\kappa'\beta}(\mathbf{q})$, mode eigenvectors $U_{\kappa\alpha}(\mathbf{q})$, and the corresponding frequencies $\omega(\mathbf{q})$ in terms of the momentum $\mathbf{q}\to\mathbf{0}$ up to the fourth order:
\begin{align}
\Phi_{\kappa\alpha,\kappa'\beta}(\mathbf{q}) \simeq& \Phi^{(0)}_{\kappa\alpha,\kappa'\beta}+i q_\gamma\Phi^{(1),\gamma}_{\kappa\alpha,\kappa'\beta}+\frac{q_\gamma q_\delta}{2}\Phi^{(2),\gamma\delta}_{\kappa\alpha,\kappa'\beta} \nonumber \\
+&\frac{i q_\gamma q_\delta q_\lambda}{6}\Phi^{(3),\gamma,\delta,\lambda}_{\kappa\alpha,\kappa'\beta} \!+\! \frac{q_\gamma q_\delta q_\lambda q_\mu}{24}\Phi^{(4),\gamma\delta\lambda\delta}_{\kappa\alpha,\kappa'\beta} \label{eq:dynmat_expand_4th} \\
U_{\kappa\alpha}(\mathbf{q}) \simeq& U^{(0)}_{\kappa\alpha}+ i  U^{(1)}_{\kappa\alpha}(\mathbf{q})+ U^{(2)}_{\kappa\alpha}(\mathbf{q})  \nonumber \\
&+i U^{(3)}_{\kappa\alpha}(\mathbf{q})+ U^{(4)}_{\kappa\alpha}(\mathbf{q}) \label{eq:eigenfunc_expand_4th} \\
\omega(\mathbf{q}) \simeq& \omega^{(1)}(\mathbf{q})+ \omega^{(2)}(\mathbf{q}),  \label{eq:freq_expand_2nd}
\end{align}
where the momentum derivatives of dynamical matrix up to the second order have been already given by Eqs.~\eqref{eq:D0} to \eqref{eq:D2}, and its third- and fourth-order derivatives are
\begin{align}
\Phi^{(3),\gamma,\delta,\lambda}_{\kappa\alpha,\kappa'\beta} &= \sum_{l}\Phi^{l}_{\kappa\alpha,\kappa'\beta}\tau^{l}_{\kappa\kappa'\gamma}\tau^{l}_{\kappa\kappa'\delta}\tau^{l}_{\kappa\kappa'\lambda}, \label{eq:D3} \\
\Phi^{(4),\gamma\delta\lambda\mu}_{\kappa\alpha,\kappa'\beta} &=  \sum_{l}\Phi^{l}_{\kappa\alpha,\kappa'\beta} \tau^{l}_{\kappa\kappa'\gamma}\tau^{l}_{\kappa\kappa'\delta}\tau^{l}_{\kappa\kappa'\lambda}\tau^{l}_{\kappa\kappa'\mu}.  \label{eq:D4}
\end{align}
In LD materials, if the rotational invariance and vanishing stress conditions are satisfied~\cite{croy2020,lin2022}, the phonon dispersion of ZA modes is always quadratic in the long-wavelength limit, leading  to the vanishing first-order term $\omega^{(1)}(\mathbf{q})$ in the frequency expansion.
Additionally, the displacement eigenvector $U_{\kappa\alpha}(\mathbf{q})$ of ZA modes becomes purely out-of-plane, fully decoupling from the in-plane vibrations.
Under these circumstances, the terms involving $\omega^{(1)}(\mathbf{q})$ have been dropped in writing the third
\begin{multline}\label{eq:eq_3rd}
0 = \frac{q_\gamma q_\delta q_\lambda}{6}\Phi^{(3),\gamma\delta\lambda}_{\kappa\alpha,\kappa'\beta}U^{(0)}_{\kappa'\beta} + \frac{q_\gamma q_\delta}{2}\Phi^{(2),\gamma\delta}_{\kappa\alpha,\kappa'\beta} U^{(1)}_{\kappa'\beta}(\mathbf{q}) \\
+ q_\gamma\Phi^{(1),\gamma}_{\kappa\alpha,\kappa'\beta} U^{(2)}_{\kappa'\beta}(\mathbf{q}) + \Phi^{(0)}_{\kappa\alpha,\kappa'\beta} U^{(3)}_{\kappa'\beta}(\mathbf{q}),
\end{multline}
and fourth-order long-wavelength equations:
\begin{multline}\label{eq:eq_4th}
M_\kappa[\omega^{(2)}(\mathbf{q})]^2 U^{(0)}_{\kappa\alpha} = \frac{q_\gamma q_\delta q_\lambda q_\mu}{24}\Phi^{(4),\gamma\delta\lambda\mu}_{\kappa\alpha,\kappa'\beta}U^{(0)}_{\kappa'\beta}  \\
-\frac{q_\gamma q_\delta q_\lambda}{6}\Phi^{(3),\gamma\delta\lambda}_{\kappa\alpha,\kappa'\beta} U^{(1)}_{\kappa'\beta}(\mathbf{q}) + \frac{q_\gamma q_\delta}{2}\Phi^{(2),\gamma\delta}_{\kappa\alpha,\kappa'\beta} U^{(2)}_{\kappa'\beta}(\mathbf{q})\\
-q_\gamma\Phi^{(1),\gamma}_{\kappa\alpha,\kappa'\beta} U^{(3)}_{\kappa'\beta}(\mathbf{q}) + \Phi^{(0)}_{\kappa\alpha,\kappa'\beta} U^{(4)}_{\kappa'\beta}(\mathbf{q}),
\end{multline}
which should be also solved iteratively to derive the dispersion relation $\omega^{(2)}(\mathbf{q})$ of bending modes near the Brillouin zone center.

As demonstrated in Appendix~\ref{solvability}, the solvability condition given by Eq.~\eqref{eq:solvability_4th_simple} for the fourth-order equation gives rise to the equations of motion for acoustic bending modes in LD materials as
\begin{equation}\label{eq:bend_eq_of_motion}
M[\omega^{(2)}(\mathbf{q})]^2 \overline{U}_{\alpha} = q_\gamma q_\delta q_\lambda q_\mu W_{\alpha\beta,\gamma\delta,\lambda\mu}\overline{U}_{\beta},
\end{equation}
where $W_{\alpha\beta,\gamma\delta,\lambda\mu}$ is the quantity related to the bending rigidity tensor of LD materials.
Using Eqs.~\eqref{eq:curvature_force_response} and \eqref{eq:curvature_force_response_CI}, the sixth-rank tensor $W_{\alpha\beta,\gamma\delta,\lambda\mu}$ can be further decomposed into the clamped-ion and lattice-mediated contributions to bending rigidities as
\begin{align}
W_{\alpha\beta,\gamma\delta,\lambda\mu} &= W^{\rm{CI}}_{\alpha\beta,\gamma\delta,\lambda\mu}+W^{\rm{LM}}_{\alpha\beta,\gamma\delta,\lambda\mu}, \label{eq:bend_rigidity_quantity} \\
W^{\rm{CI}}_{\alpha\beta,\gamma\delta,\lambda\mu} &= \sum_{\kappa}W^{\mathrm{CI},\kappa}_{\alpha\beta,\gamma\delta,\lambda\mu}=\frac{1}{24}\sum_{\kappa,\kappa'}\Phi^{(4),\gamma\delta\lambda\mu}_{\kappa\alpha,\kappa'\beta},  \label{eq:bending_rgd_CI} \\
W^{\rm{LM}}_{\alpha\beta,\gamma\delta,\lambda\mu}&=\sum_{\kappa}W^{\mathrm{LM},\kappa}_{\alpha\beta,\gamma\delta,\lambda\mu}, \label{eq:bending_rgd_LM}
\end{align}
where the lattice-mediated contribution $W^{\mathrm{LM},\kappa}_{\alpha\beta,\gamma\delta,\lambda\mu}$ is lengthy and given by Eqs.~\eqref{eq:curvature_force_response_LM} to \eqref{eq:curvature_force_response_LM3}.
For 2D materials, considering the out-of-plane direction set to $z$, the quadratic dispersion relation for ZA modes then reads
\begin{equation}\label{eq:dispersion_ZA}
[\omega^{(2)}_{\rm ZA}(\mathbf{q})]^2 = q_\alpha q_\beta q_\gamma q_\delta \frac{W_{zz,\alpha\beta,\gamma\delta}}{M},
\end{equation}
where we have used the fact that the bending acoustic modes have a pure out-of-plane polarization~\cite{lin2022,croy2020}, resulting in the displacement eigenvectors with $\overline{U}_{x,y}=0$ and $\overline{U}_{z}=1$ if properly normalized.

\subsection{Closed-form expression for bending rigidity tensor}\label{bend_rdg_formula}

The microscopic expression for the bending rigidity tensor of 2D materials can be obtained by connecting the equations of motion for bending waves derived from the Kirchhoff plate theory and the lattice-dynamical theory, i.e. Eqs.~\eqref{eq:eq_motion_bend_wave} and \eqref{eq:dispersion_ZA}.
By introducing the 2D mass density $\rho^{}_{2D}=M/A$ with the primitive unit cell area $A$, we can obtain
\begin{equation}\label{eq:ZA_dispersion}
[\omega^{(2)}_{\rm ZA}(\mathbf{q})]^2 = q_\alpha q_\beta q_\gamma q_\delta\frac{1}{\rho^{}_{2D}} \frac{W_{zz,\alpha\beta,\gamma\delta}}{A}.
\end{equation}
In addition, using Eq.~\eqref{eq:bend_rigidity_quantity} we define the unsymmetrized bending rigidity tensor as
\begin{equation}\label{eq:bend_rgd_unsym}
\tilde{D}_{\alpha\beta,\gamma\delta} = \frac{1}{A}\Big(W^{\rm{CI}}_{zz,\alpha\beta,\gamma\delta}+W^{\rm{LM}}_{zz,\alpha\beta,\gamma\delta}\Big),
\end{equation}
where $z$ is assumed to be the out-of-plane direction.
The different formulations of bending rigidity tensor again stem from its definition in reciprocal or real space, similar to the case of strain-gradient elasticity tensor~\cite{stengel2016,divincenzo1986}.
In general, the \emph{dynamic} bending rigidity tensor from the reciprocal-space formulation is totally symmetric with respect to any permutation of its four indices, $\alpha$, $\beta$, $\lambda$, and $\delta$.
However, this is only true within the clamped-ion approximation, since the pure electronic contribution $W^{\rm{CI}}_{zz,\alpha\beta,\gamma\delta}$ is inherently fully symmetric through its definition as a fourth-order derivative of the zone-center dynamical matrix.
When the lattice-mediated part $W^{\rm{LM}}_{zz,\alpha\beta,\gamma\delta}$ is included, the resulting total tensor is only symmetric upon exchanging indices within the two pairs $\alpha\beta$ and $\gamma\delta$, if $W^{\rm{LM}}_{zz,\alpha\beta,\gamma\delta}$ is properly symmetrized [see Eqs.~\eqref{eq:curvature_force_response_LM} to \eqref{eq:curvature_force_response_LM3} and the related discussion in Appendix~\ref{solvability}].
This can be further understood by noticing that $W^{\rm{LM}}_{zz,\alpha\beta,\gamma\delta}$ is a mixed derivative indicated by the internal strain effects.
On the other hand, using the real-space formulation, the \emph{static} bending rigidity tensor introduced in Eq.~\eqref{eq:bend_rgd_from_elastic_stiffness} must adhere to the same symmetry relations as the elastic stiffness tensor described in Eq.~\eqref{eq:elastic_sym_relation}.
Thus, by imposing the additional symmetry $\alpha\beta\leftrightarrow\gamma\delta$, the symmetrized bending rigidity tensor can be further introduced as
\begin{equation}\label{eq:bend_rgd_sym}
\!\!\!D_{\alpha\beta,\gamma\delta} = \frac{1}{A}\left[W^{\rm{CI}}_{zz,\alpha\beta,\gamma\delta}+\frac{W^{\rm{LM}}_{zz,\alpha\beta,\gamma\delta}+W^{\rm{LM}}_{zz,\gamma\delta,\alpha\beta}}{2}\right],
\end{equation}
which satisfies the symmetry relation of elastic stiffness tensor and is consistent with the fact that only the symmetric parts contribute to strain-energy density.
The applicability of a dynamic theory to study the static response is based on the establishment of mechanical equilibrium conditions between the applied external load and the internal strain response.
Although individual atomic components of the bending rigidity tensor are dynamic, their overall contribution becomes static, similar to the case of strain-gradient elasticity~\cite{stengel2016}.
The bending rigidity tensor defined in Eq.~\eqref{eq:bend_rgd_sym} is expected to be general and should also be applicable for one-dimensional (1D) materials, where the unit cell area $A$ is replaced by the unit cell length $L$ along the 1D axis.

Importantly, as a perturbative approach, our obtained close-form expression for the fourth-rank bending rigidity tensor in Eq.~\eqref{eq:bend_rgd_sym} adopts only the primitive unit cell of 2D materials, which avoids adding an explicit bending curvature to monolayer structures and allows the accurate calculations of bending rigidities in the perturbative limit.
These advantages make our approach more favorable to the supercell method~\cite{wei2013} and cyclic DFT calculations~\cite{banerjee2016,ghosh2019,kumar2020}, which both fit the total energy of bent systems to curvature radii, requiring careful tests of radius range and magnitude to ensure the accuracy and convergence of bending rigidity in the small curvature limit.
Although an implementation of DFPT into cyclic DFT would also allow for the calculation of bending rigidities, the recent DFPT development of cyclic DFT is focused only on phonon frequencies due to its complexity~\cite{Sharma2026}.
In addition, we note our bending rigidity formalism is only exact in the linear asymptotic limit where non-linear effects are neglected, and the bending ZA modes in 2D materials can only emerge when the rotational invariance and vanishing stress conditions (also known as the equilibrium conditions) introduced in Appendix~\ref{inv_conds} are both satisfied~\cite{lin2022,croy2020}.
This statement agrees with a recent study~\cite{aseginolaza2024} which shows that the rotational invariance protects the quadratic ZA phonons in the long-wavelength limit against the phonon-phonon interactions, prohibiting the divergence of bending rigidity as well as allowing the sound propagation in 2D system.
Besides, our numerical simulations reveal non-zero components of the elastic tensor involving the out-of-plane direction in 2D materials when the aforementioned invariance conditions are not met, further highlighting their importance.

\subsection{Separation of 2D long-range electrostatic interactions}\label{2D_long_range}

As already pointed out in Sec.~\ref{multipole} for bulk materials, the microscopic interatomic interactions in semiconductors and insulators are inherently long-ranged~\cite{Pick1970}.
The extension to 2D materials was recently investigated for the dynamical matrices~\cite{sohier2017,sohier2016,macheda2023} and electron-phonon coupling~\cite{sohier2016,deng2021,macheda2023,sio2022,ponce2023,ponce2023a}.
In particular, Royo and Stengel~\cite{royo2021} developed an elegant 2D framework based on a range separation of 2D long-range kernel using the image-charge construction and small-space representation.
In addition, for practical DFT calculations, 2D materials are simulated with a vacuum in its out-of-plane direction that is large enough to allow a sufficient decay of electronic wave function and avoid spurious interaction between periodic images.
However, due to the out-of-plane open-circuit boundary condition, the existence of surface charges will induce the long-range fields that can be addressed by a 2D Coulomb truncation scheme~\cite{sohier2017-2}.

In this work, we use Royo and Stengel's formulation of the 2D long-range Coulomb interactions~\cite{royo2021}:
\begin{multline}\label{eq:2D_long_range_dynmat}
\Phi^\mathcal{L}_{\kappa\alpha,\kappa'\beta}(\mathbf{q}) = \frac{2\pi f(\mathbf{q})}{Aq}\Bigg[ \frac{\left(\mathbf{q}\cdot\boldsymbol{\mathcal{Z}}\right)^*_{\kappa\alpha}\left(\mathbf{q}\cdot\boldsymbol{\mathcal{Z}}\right)_{\kappa'\beta}}{1+\frac{2\pi f(\mathbf{q})}{q}\mathbf{q}\cdot\boldsymbol{\alpha}\cdot\mathbf{q}}\\
- \frac{|\mathbf{q}|^2\left(\mathcal{Z}^z_{\kappa\alpha}\right)^*\mathcal{Z}^z_{\kappa'\beta}}{1-2\pi |\mathbf{q}|f(\mathbf{q})\alpha_{zz}}  \Bigg]\mathrm{e}^{-i\mathbf{q}\cdot(\boldsymbol{\tau}_{\kappa'}-\boldsymbol{\tau}_\kappa)},
\end{multline}
where $f(\mathbf{q})=1-\tanh(|\mathbf{q}|L/2)$ is the range separation function with $L$ denoting the characteristic length scale of the range separation, and $\boldsymbol{\mathcal{Z}}$ are the generalized effective charge tensor that can be multipolarly expanded as
\begin{align}
\mathcal{Z}_{\kappa\alpha}^\beta(\mathbf{q}) &\simeq \hat{Z}^\beta_{\kappa\alpha} - \frac{iq_\gamma}{2}\left(\hat{Q}^{\beta\gamma}_{\kappa\alpha} - \delta_{\beta\gamma} \hat{Q}^{zz}_{\kappa\alpha}\right) \\
\mathcal{Z}_{\kappa\alpha}^{z}(\mathbf{q}) &\simeq \hat{Z}^z_{\kappa\alpha} - i q_\beta \hat{Q}^{\beta z}_{\kappa\alpha},
\end{align}
corresponding to the in-plane and out-of-plane polarization response, respectively.
It should be noted that the dynamical Born effective charge $\hat{\mathbf{Z}}_\kappa$ and quadrupole tensors $\hat{\mathbf{Q}}_\kappa$ entering the above two equations are calculated at the mixed electrical boundary conditions~\cite{royo2021}, which are different from those defined in Eq.~\eqref{eq:charge_density}, as indicated by an additional hat symbol.
Specifically, the short-circuit boundary condition is set for the in-plane directions, while the out-of-plane direction is set to be open-circuit.
Furthermore, we have truncated the expansion of short-range charge density response and polarizability function to the terms of quadrupole and macroscopic polarizability, respectively.
Given the ion-clamped dielectric permittivity tensor $\boldsymbol{\epsilon}^\infty$ of the 2D system, the corresponding macroscopic polarizability tensor can be calculated as
\begin{equation}
\boldsymbol{\alpha} = \frac{t}{4\pi}(\boldsymbol{\epsilon}^\infty-\mathbb{I}),
\end{equation}
where $t$ is the vacuum thickness and $\mathbb{I}$ is an identity tensor.
One can find that $\boldsymbol{\epsilon}^\infty$ is not a well-defined quantity in 2D materials, since it depends on the vacuum thickness $t$, while $\boldsymbol{\alpha}$ is thickness-independent.
Also, the definition of $\alpha_{zz}$ here differs from the one $\alpha_{zz} = \frac{t}{4\pi}(1-\epsilon^{\infty,-1}_{zz})$ that can be found in the literature~\cite{royo2021,tian2019}, which is due to the fact that we impose the open-circuit boundary condition in the out-of-plane direction through the 2D Coulomb truncation technique~\cite{sohier2017-2}, and this is equivalent to multiplying $\alpha_{zz}$ by a factor of $\epsilon^{\infty}_{zz}$.
As a result, one can readily use the short-circuit multipole tensors in Eq.~\eqref{eq:2D_long_range_dynmat} that are the standard outputs of most DFT codes.
For the practical calculation of the long-range part of dynamical matrix in 2D crystals, one can still apply the Ewald summation technique described in Sec.~\ref{multipole}, and the range-separation function $f(\mathbf{q})\simeq1-|\mathbf{q}|L/2$ varies linearly for small momenta, unlike a quadratic behavior of the three-dimensional (3D) counterpart.
We find that such a difference in the long-wavelength limit has a significant effect on choosing a suitable range-separation parameter $L$ in 2D materials for the accurate calculation of elastic tensors.
In contrast to bulk solids, the short-circuit elastic constants of 2D materials become very sensitive to $L$.
In order to address this obstacle, we utilize the same strategy as proposed for the bulk case to determine the optimal $L$, which is achieved through a minimization of real-space IFCs in Eq.~\eqref{eq:sum_IFC_SR} with only the macroscopic $\mathbf{G=0}$ component removed.
Rigorously, $\mathbf{G=0}$ is the only long-range component that induces the macroscopic electric field effects, and any $\mathbf{G\ne0}$ corresponds to the short-range local field effects~\cite{venkataraman1975}.
Lastly, it should be emphasized again that the implementation of Ewald summation with only $\mathbf{G=0}$ would lead to the real-space IFCs having some imaginary parts as a signature of the breakdown of the Fourier transform of the resulting short-range dynamical matrix.
Such an issue is understood in the sense that the non-analytic correction to dynamical matrix is not only limited to the vicinity of the Brillouin zone center if only $\mathbf{G=0}$ is included in the Ewald summation; the non-analytic correction on the zone boundary also becomes unsymmetric with an even $\Gamma$-centered $\mathbf{q}$ grid, which consequently results in the presence of imaginary parts in the real-space IFCs.
Therefore, the calculation of short-circuit elastic constants in 2D materials are performed with the real-space IFCs where the long-range effects are removed from a regular Ewald summation over a $\mathbf{G}$-grid, but with the range-separation parameter $L$ obtained with only $\mathbf{G=0}$ removed.

\section{Results and discussion}\label{results}

To validate our new method for computing elastic constants in Sec.~\ref{atomic_theory}, we benchmark eight representative solids, which have been widely explored by both theoretical calculations and experimental measurements: silicon, NaCl, GaAs, and rhombohedral BaTiO$_3$ are chosen as the bulk materials, while monolayer graphene, \emph{h}-BN, MoS$_2$, and InSe are selected for 2D materials.
We also compute and validate the bending rigidities of graphene, \emph{h}-BN, MoS$_2$, and InSe.
Specifically, in the case of bulk system, silicon is chosen as an infrared-inactive material; NaCl is selected as an infrared-active solid without quadrupoles, while GaAs is infrared-active with also the contribution from quadrupoles.
Finally, BaTiO$_3$ is chosen as a representative example of a low-symmetry system.
Moreover, graphene is chosen as a prototype of the 2D system, and \emph{h}-BN is an example for the infrared-active 2D materials; MoS$_2$ represents the well-known 2D family of transition metal dichalcogenides, while InSe is selected for another 2D class of III-VI metal chalcogenides.
The calculation details of our first-principles simulations are summarized in Appendix~\ref{comp_details}, except for \emph{h}-BN, MoS$_2$ and InSe where we have used the computational parameters from Ref.~\cite{ponce2023}.

\begin{figure}[t!]
\centering
\includegraphics[width=0.85\linewidth]{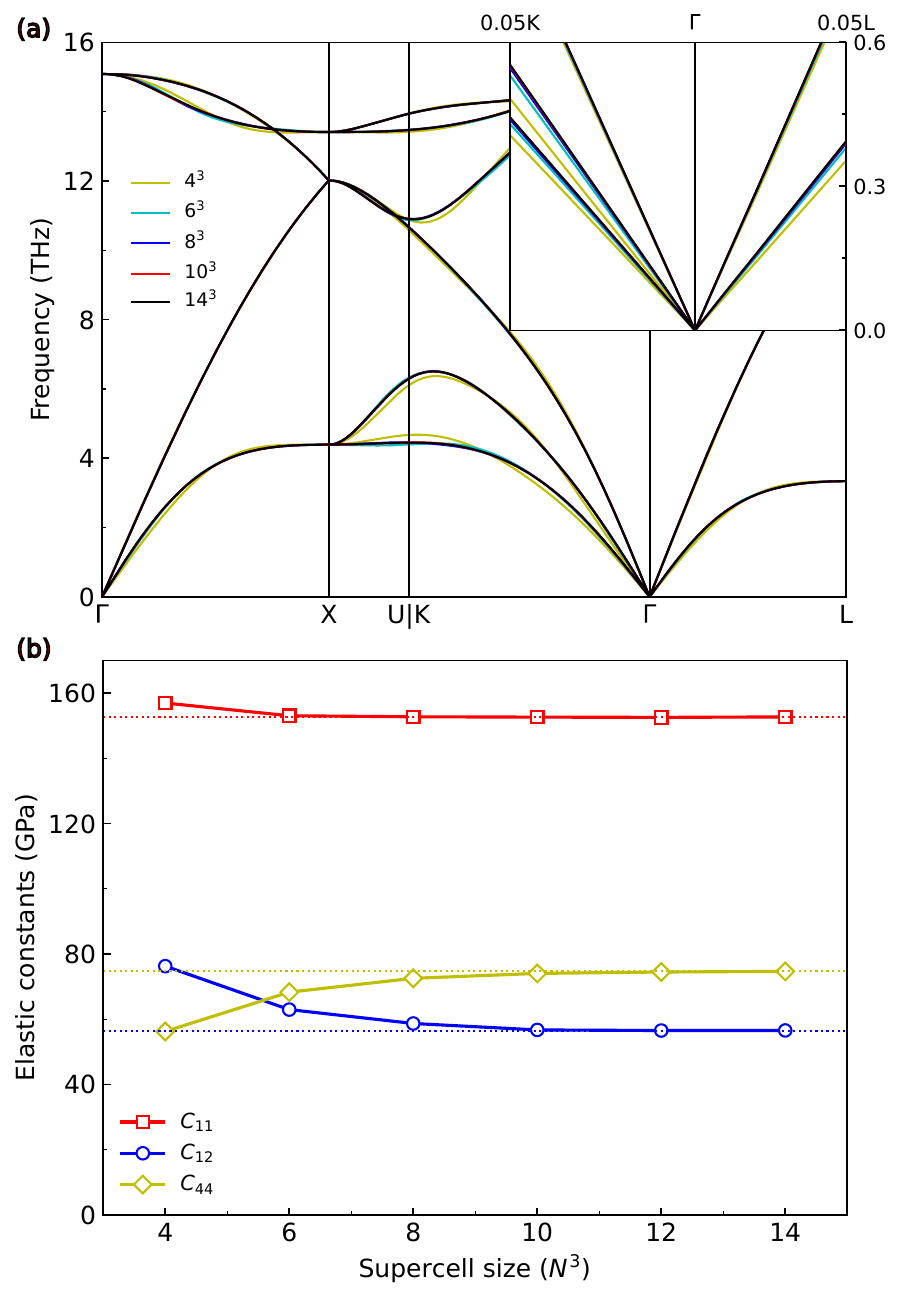}
\caption{(a) Phonon dispersion and (b) independent components of the elastic tensor for silicon as a function of supercell size.
The inset of panel (a) shows the enlarged acoustic branches around the $\Gamma$ point along the K and L directions.
The dotted lines in the panel (b) are the reference values of elastic constants from the finite-difference calculations using the $\textsc{thermo\_pw}$ code~\cite{dal2016}.
\label{fig:Si}
}
\end{figure}

\begin{table}[t]
\centering
\caption{The dynamical Born effective charge $Z$ [$e$], quadrupole $Q$ [$e~\mathrm{bohr}$], octupole $O$ [$e~\mathrm{bohr}^2$], clamped-ion dielectric permittivity $\boldsymbol{\epsilon}^{\infty}$, and dielectric dispersion tensors $\boldsymbol{\epsilon}^{(4)}$ [bohr$^2$] tensors of Si, NaCl and GaAs.
The values in the column \textit{this work} are obtained via a least-squares fit to the Taylor expansion of unscreened charge density response in Eq.~\eqref{eq:charge_density} and dielectric screening function in Eq.~\eqref{eq:screening_func}, compared with those calculated from DFPT.
Only symmetry-independent components are shown.
The DFPT quadrupole values for silicon and GaAs are taken from Ref.~\cite{brunin2020}.
}
\label{table:multipoles}
\begin{ruledtabular}
\begin{tabular}{c l c c}
\multicolumn{2}{c}{Elements} & This work & DFPT \\
\hline
\multirow{6}{*}{Si} & $Q_{x}^{yz}$ & 13.762  & 13.67 \\
& $O_{x}^{xxx}$ & -338.035 & -- \\
& $O_{x}^{xyy}$ & -123.629 & -- \\
& $\epsilon^\infty_{xx}$ & 13.087  & 13.114 \\
& $\epsilon^{(4)}_{xxxx}$ & -172.780 & -- \\
& $\epsilon^{(4)}_{xxyy}$ & -90.933 & -- \\
\hline
\multirow{3}{*}{Na} & $Z_{x}^{x}$ & 1.107 & 1.107 \\
& $O_{x}^{xxx}$ & -16.069 & -- \\
& $O_{x}^{xyy}$ & -14.562 & -- \\
\multirow{3}{*}{Cl} & $Z_{x}^{x}$ & -1.107 & -1.107 \\
& $O_{x}^{xxx}$ & -74.762 & -- \\
& $O_{x}^{xyy}$ & -28.066 & -- \\
\multirow{3}{*}{NaCl} & $\epsilon^\infty_{xx}$ & 2.484  & 2.485 \\
& $\epsilon^{(4)}_{xxxx}$ & -2.353 & -- \\
& $\epsilon^{(4)}_{xxyy}$ & -1.184 & -- \\
\hline
\multirow{4}{*}{Ga} & $Z_{x}^{x}$ & 2.112 & 2.115 \\
& $Q_{x}^{yz}$ & 17.469 & 16.54 \\
& $O_{x}^{xxx}$ & -356.233 & -- \\
& $O_{x}^{xyy}$ & -86.146 & -- \\
\multirow{4}{*}{As} & $Z_{x}^{x}$ & -2.112 & -2.115 \\
& $Q_{x}^{yz}$ & -8.880  & -8.57 \\
& $O_{x}^{xxx}$ & -726.922 & -- \\
& $O_{x}^{xyy}$ & -305.694 & -- \\
\multirow{3}{*}{GaAs} & $\epsilon^\infty_{xx}$ & 14.274  & 14.282 \\
& $\epsilon^{(4)}_{xxxx}$ & -241.495 & -- \\
& $\epsilon^{(4)}_{xxyy}$ & -202.947 & -- \\
\end{tabular}
\end{ruledtabular}
\end{table}

\begin{table*}[t]
\centering
\caption{Theoretical and experimental elastic constants ($C_{ij}$), bulk ($K$), shear ($G$), and Young's ($E$) moduli [GPa], Poisson's ratio ($\nu$), and longitudinal ($v_{\rm l}$) and transverse ($v_{\rm t}$) sound velocities [m/s] of silicon, NaCl and GaAs.
For \textit{this work}, a $14\times14\times14$ supercell is adopted to obtain the elastic tensor of silicon and GaAs, while a $12\times12\times12$ supercell is used for NaCl.
The results of $\textsc{thermo\_pw}$ are from the finite-difference of strain-stress relation.
The superscript ``CI'' denotes the clamped-ion approximation, and the relaxation of internal atomic positions is only allowed in silicon and GaAs for their $C_{44}$ component.
To obtain the elastic moduli, the polycrystalline average based on the Voigt-Reuss-Hill scheme is utilized.
The experiment data for Si and NaCl were measured at 298 and 300 K, respectively, while those for GaAs were extrapolated to zero temperature.
The experimental Young's modulus, Poisson's ratio, and speed of sound for GaAs are single-crystal results, along the [100] crystallographic direction.
}
\label{table:elastic_prop}
\begin{ruledtabular}
\begin{tabular}{cccccccccccc}
\multicolumn{2}{c}{Compounds} & $C_{11}$ & $C_{12}$ & $C_{44}$ & $C_{44}^{\rm CI}$ & $K$ & $G$ & $E$ & $\nu$ & $v_{\rm l}$ & $v_{\rm t}$ \\
\hline
\multirow{3}{*}{Si} & this work & 152.70 & 56.56 & 74.66 & 99.64 & 88.60 & 62.58 & 151.97 & 0.21 & 8685 & 5238 \\
& $\textsc{thermo\_pw}$ & 152.55 & 56.26 & 74.75 & 99.83 & 88.36 & 61.21 & 149.20 & 0.21 & 8681 & 5241 \\
& experiment~\cite{hall1967,hopcroft2010} & 165.64 & 63.94 & 79.51 & -- & 97.8 & 65 & 160 & 0.22 & 8434 & 5843 \\
\hline
\multirow{3}{*}{NaCl} & this work & 48.08 & 11.69 & 12.77 & 12.77 & 23.82 & 14.72 & 36.61 & 0.24 & 4550 & 2648 \\
& $\textsc{thermo\_pw}$ & 47.29 & 11.87 & 12.34 & 12.34 & 23.67 & 14.27 & 35.64 & 0.25 & 4511 & 2607 \\
& experiment~\cite{fukui2020} & 48.50 & 10.42 & 12.50 & -- & 23.11 & 14.81 & 36.61 & 0.24 & 4456 & 2620 \\
\hline
\multirow{3}{*}{GaAs} & this work & 109.38 & 49.88 & 54.04 & 74.62 & 69.71 & 42.53 & 106.04 & 0.25 & 4886 & 2834 \\
& $\textsc{thermo\_pw}$ & 109.30 & 49.13 & 54.64 & 74.78 & 69.19 & 43.00 & 106.83 & 0.24 & 4888 & 2850 \\
& experiment~\cite{blakemore1982} & 112.6 & 57.1 & 60.0 & -- & 78.9 & 32.8 & 86.3 & 0.32 & 4784 & 3350 \\
\end{tabular}
\end{ruledtabular}
\end{table*}

\begin{figure}[t!]
\centering
\includegraphics[width=0.85\linewidth]{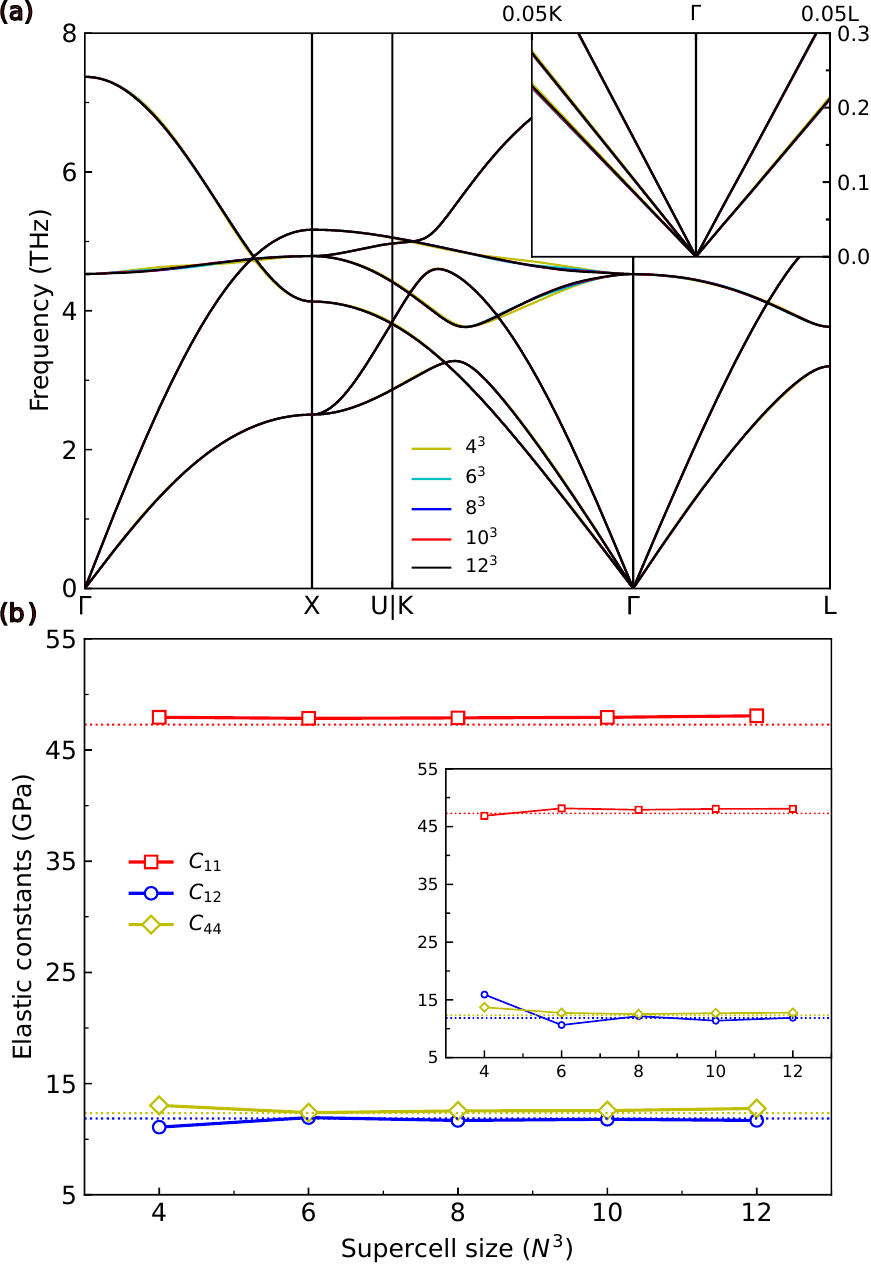}
\caption{(a) Phonon dispersion with enlarged inset around the $\Gamma$ point and (b) independent components of elastic tensor for NaCl as a function of supercell size where the standard (DD) long-range has been removed.
In the panel (b), the elastic constants of NaCl obtained without the long-range removal are shown in the inset, and the dotted lines are the reference values from the $\textsc{thermo\_pw}$ finite-difference calculations.
\label{fig:NaCl}
}
\end{figure}

\subsection{Dipole-free materials}

Our first example is silicon, which is a simple covalent and infrared-inactive solid.
The optimized lattice constant of its diamond-like structure from DFT is 5.469~\r{A}, in close agreement with the experimental value 5.431~\r{A}~\cite{okada1984}.
Fig.~\ref{fig:Si} (a) shows the phonon dispersion of silicon with increasing supercell size of IFCs from $4\times4\times4$ to $14\times14\times14$.
Although there is no LO-TO splitting due to its infrared-inactive nature and vanishing Born effective charges, the phonon dispersion only converges from a supercell size of $8\times8\times8$.
As demonstrated in the inset of Fig.~\ref{fig:Si} (a), this situation is more severe for long-wavelength acoustic branches around the Brillouin zone center, which is closely linked to the macroscopic elastic properties.
The lowest order in the multipole expansion is the quadrupole term, and the QQ term is the only long-range contribution to the dynamical matrix up to the second order in $\mathbf{q}$ in the expansion of Eq.~\eqref{eq:long-range}.
The dynamical quadrupole tensor of silicon can be expressed as $Q_{\kappa\alpha}^{\beta\gamma}=(-1)^{\kappa+1}Q|\varepsilon_{\alpha\beta\gamma}|$ with $\varepsilon_{\alpha\beta\gamma}$ being the Levi-Civita symbol, so it only has one independent component when $\alpha\neq\beta\neq\gamma$.
Our fitting procedure based on the Taylor expansion of unscreened charge density response in Eq.~\eqref{eq:charge_density} gives a value of 13.76 $e~\rm{bohr}$ for $|Q_{x}^{yz}|$ which agrees well with the previous calculations either from DFPT (13.67~$e~\rm{bohr}$~\cite{brunin2020}) or a fit to the analytic long-range electron-phonon matrix elements (13.66~$e~\rm{bohr}$~\cite{Ponce2021}).
However, our further interpolation of dynamical matrix with the QQ term taken into account shows an almost unchanged phonon dispersion for silicon,  except for the case using a $4\times4\times4$ supercell which is too small to ensure the good decay of IFCs in real space.
Therefore, the inclusion of quadrupole term does not help to improve the convergence rate of lattice-dynamical properties for silicon, and only the phonon dispersion based on the standard interpolation scheme is illustrated in Fig.~\ref{fig:Si} (a).
We report the dynamical multipole tensors and dielectric properties of silicon in Table~\ref{table:multipoles}, although the octupole tensor is not used here due to the vanishing DO interactions in silicon.

Silicon has the highest cubic symmetry and there is only three independent components ($C_{11}$, $C_{12}$ and $C_{44}$) for its elastic tensor (see Sec.~II of the SM~\cite{SM} for a brief introduction of Voigt notation).
The calculated elastic stiffness tensor based on the real space IFCs and long-wavelength expansion as a function of supercell size is shown in Fig.~\ref{fig:Si} (b).
Similar to the results of phonon dispersion, the elastic constants of silicon also begin to converge after a supercell of $8\times8\times8$ with the relative error of $4.3\%$ and $2.9\%$ to the reference values from finite-difference calculations for $C_{12}$ and $C_{44}$, respectively, while the $C_{11}$ component has already converged after a $6\times6\times6$ phonon grid.
The results displayed here are calculated based on the total IFCs as silicon is infrared-inactive, without the removal of any high-order multipolar interaction.
A further subtraction of the QQ term in the dynamical matrix of silicon does not modify its elastic tensor, which implies that the quadrupole will not couple with the mechanical field therein; this observation is in agreement with the fact that silicon is not a piezoelectric crystal because of the cubic symmetry and the existence of spatial inversion.
Using the obtained quadrupole values, the calculated piezoelectric tensor via Martin's formula~\cite{martin1972} is indeed zero.
In addition, the detailed comparison of the elastic constants, moduli and sound velocities of silicon calculated from our method based on a supercell of $14\times14\times14$ and the finite difference implemented in the \textsc{thermo\_pw} code~\cite{dal2016} as well as the experimental values~\cite{hall1967,hopcroft2010} are listed in Table~\ref{table:elastic_prop}.
The Voigt-Reuss-Hill (VRH) scheme of polycrystalline average is adopted to calculate the bulk ($K$) and shear ($G$) moduli, from which the Young's modulus ($E$) and Poisson's ratio ($\nu$) are further obtained under the isotropic approximation (see Sec.~II~A of the SM~\cite{SM} for more details).
In solids, sound waves can be generated by either compressional (volumetric) or shear deformation, and two such deformation modes correspond to the pressure (longitudinal) and shear (transverse) waves, respectively.
As a consequence,  the speed of sound in the same solid material can be different, depending on the specific deformation mode.
With the knowledge of elastic moduli, we also calculate the longitudinal ($v_{\rm l}$) and transverse velocities ($v_{\rm t}$), where the detailed formula can be found in Sec.~III~A of the SM~\cite{SM}.
Overall, our results for the mechanical and elastic properties of silicon are in good agreement with both the finite-difference calculations and experimental measurements.

\subsection{Quadrupole-free materials}

We then focus on the phonon spectra and elastic properties of NaCl, as depicted in Fig.~\ref{fig:NaCl}, and our DFT calculation yields a lattice constant of 5.698~\r{A},  close to the experimentally measured value of 5.628~\r{A}~\cite{boswell1951}.
Unlike silicon, NaCl is an infrared-active solid where the non-zero Born effective charges give rise to the non-analytic LO-TO splitting at the Brillouin zone center; however, it has a vanishing dynamical quadrupole tensor due to spacial inversion symmetry.
The calculated dipole, octupole, and dielectric tensors of NaCl can be found in Table~\ref{table:multipoles}.
As can be seen from Fig.~\ref{fig:NaCl}(a), where the solid and dotted lines represent the phonon dispersions using the standard DD and the advanced multipolar interpolation, Eq.~\eqref{eq:Phi_expand}, the phonon spectra have already converged on a relatively small $4\times4\times4$ supercell, except for the dispersions of TO modes.
While this convergence of TO branches could be solved using a larger supercell, the inclusion of high-order multipolar interactions (DO+D$\epsilon$D) in the Fourier interpolation can already capture its correct dispersion based on a $4\times4\times4$ supercell, see Fig.~S1 of the SM~\cite{SM}.

Our calculated elastic constants are validated against those obtained by finite-difference using \textsc{thermo\_pw} in Fig.~\ref{fig:NaCl}(b).
If the long-range interactions are not subtracted, we see in the inset of Fig.~\ref{fig:NaCl}(b) that the elastic constants oscillates.
To guarantee convergence, the IFCs must decay faster than $1/d^5$ with increasing interatomic distance $d$~\cite{royo2020}.
A subtraction of long-range multipolar interactions up to $\mathcal{O}(q^2)$ is then required, which corresponds to the short-range IFCs decaying at least as $1/d^6$.
In practice, for the case of NaCl, we find that removing DD only is sufficient to reach convergence for a $6\times6\times6$ supercell and that further removing the higher order DO and D$\varepsilon$D terms only have a negligible impact.
This observation implies that the long-range multipolar interactions in NaCl are weak due to inversion symmetry, which eliminates the odd quadrupole tensor.
Overall, our results for the elastic constants of NaCl are in excellent agreement with the finite-difference calculations and the experiment measurements, as shown in Table~\ref{table:elastic_prop}.

\subsection{High-symmetry materials}

We now study the elastic tensor of GaAs, which is also an infrared-active material.
Although it has a similar cubic structure as silicon and NaCl, the lack of inversion symmetry in GaAs makes it piezoelectric  with a larger electromechanical coupling effect~\cite{fricke1991,soderkvist1994}.
In particular, the calculated dielectric function is very sensitive to the change of lattice constant, and our optimized lattice constant from the PBEsol functional yields a value of 5.661~\r{A}, in excellent agreement with the measured 5.653~\r{A}~\cite{blakemore1982} in experiments (0.14\% error).
We have compared the phonon dispersion of GaAs obtained with the standard dipole and advanced multipole interpolations, where the additional DQ and QQ interactions are considered, see Fig.~S1 of the SM~\cite{SM}.
It can be noticed that the inclusion of such multipolar interactions have improved the quality of the interpolated optical branches, especially the two TO modes near the Brillouin zone center.
When it comes to acoustic phonons, we find that the long-range multipole interactions (i.e. DQ and QQ) have almost no influence on their dispersions, and there is only a slight change in the slope in the long-wavelength limit (see the inset of those two plots).
For both cases, the acoustic branches reach convergence after a $6\times6\times6$ supercell.
In addition, the further introduction of DO and D$\epsilon$D interactions plays a negligible role in improving the convergence efficiency of the phonon dispersion of GaAs, although they are at the same $\mathcal{O}(q^2)$ order  as the QQ term.
Last but not least, the calculated dynamical quadrupole tensor of GaAs is $Q_{\kappa\alpha}^{\beta\gamma}=Q_{\kappa}|\varepsilon_{\alpha\beta\gamma}|$ with values of 17.47 and -8.88 $e~\rm{bohr}$ for Ga and As, respectively, in line with the previous DFPT prediction, that is 16.54 and -8.57 $e~\rm{bohr}$~\cite{brunin2020}, respectively.
The full multipolar and dielectric properties of GaAs are summarized in Table~\ref{table:multipoles}.

We proceed to analyze the elasticity of GaAs and the influence of long-range multipolar interactions on its elastic tensor, also shown in Fig.~S1 of the SM~\cite{SM}.
The elastic constants are calculated based on the short-range IFCs after removing the DD, DQ and QQ terms, since the IFCs must decay faster than $1/d^5$ in order to guarantee the convergence with respect to supercell size.
Importantly, this treatment can lead to the appropriate short-circuit electric boundary conditions that we are interested in,  ensuring the correct values of elastic constants, i.e. without the macroscopic electric field effect.
Indeed, the $C_{11}$ component converges to the correct value with a supercell of $8\times8\times8$, while $C_{12}$ and $C_{44}$ need a larger $10\times10\times10$ supercell.
All of these three independent components of the elastic tensor are in good agreement with the reference values obtained from the finite-difference calculations of strain-stress relation (the dotted lines in the plot), and the detailed comparison among them and the measured experiment data are listed in Table~\ref{table:elastic_prop}.
Crucially, the $C_{12}$ and $C_{44}$ coefficient converge to incorrect values if dipoles only are subtracted, and deviate from the finite-difference reference by 11.38\% and 5.13\%, respectively.
We insist that this is a qualitatively different result.
While for our first two cases, the addition of higher multipoles speeds up the convergence, here quadrupoles are required to obtain the correct result.
Indeed, in general, convergence may never be reached, even for infinite supercell size, if not all non-analyticies have been removed~\cite{brunin2020,royo2020}.
However, for the case of GaAs, we find that instead the DO and D$\epsilon$D removal in IFCs have no influence on the calculated elastic tensor of GaAs, which is consistent with the similar observation in its phonon dispersion.

\begin{figure}[t!]
\centering
\includegraphics[width=0.99\linewidth]{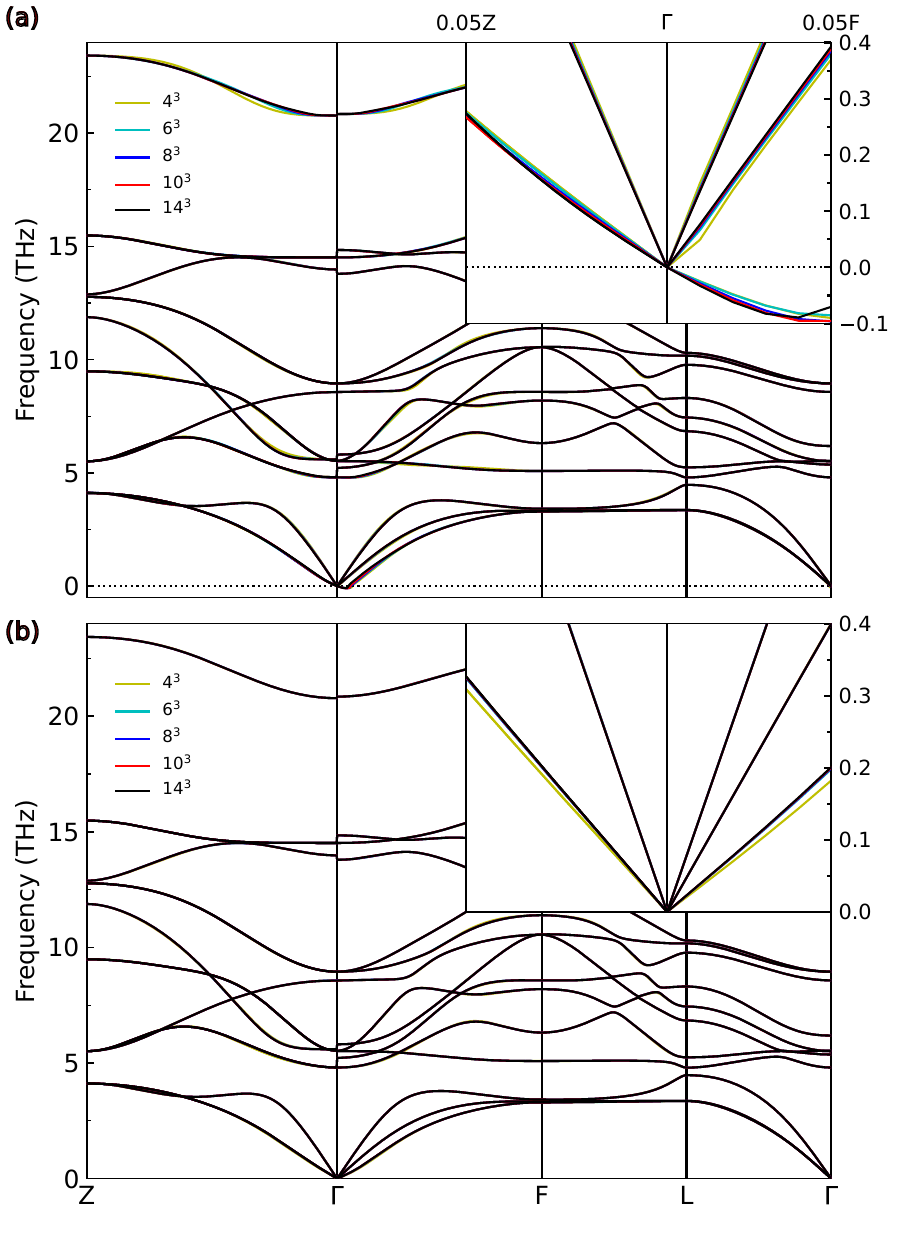}
\caption{Phonon dispersions of BaTiO$_3$ as a function of supercell size using (a) the standard (DD) interpolation and the multipolar interpolation with (b) DD+DQ+QQ+DO+D$\epsilon$D where the inset shows an enlarged version of the acoustic branches.}
\label{fig:BaTiO3_phonon}
\end{figure}

\begin{figure*}[ht]
\centering
\includegraphics[width=0.99\linewidth]{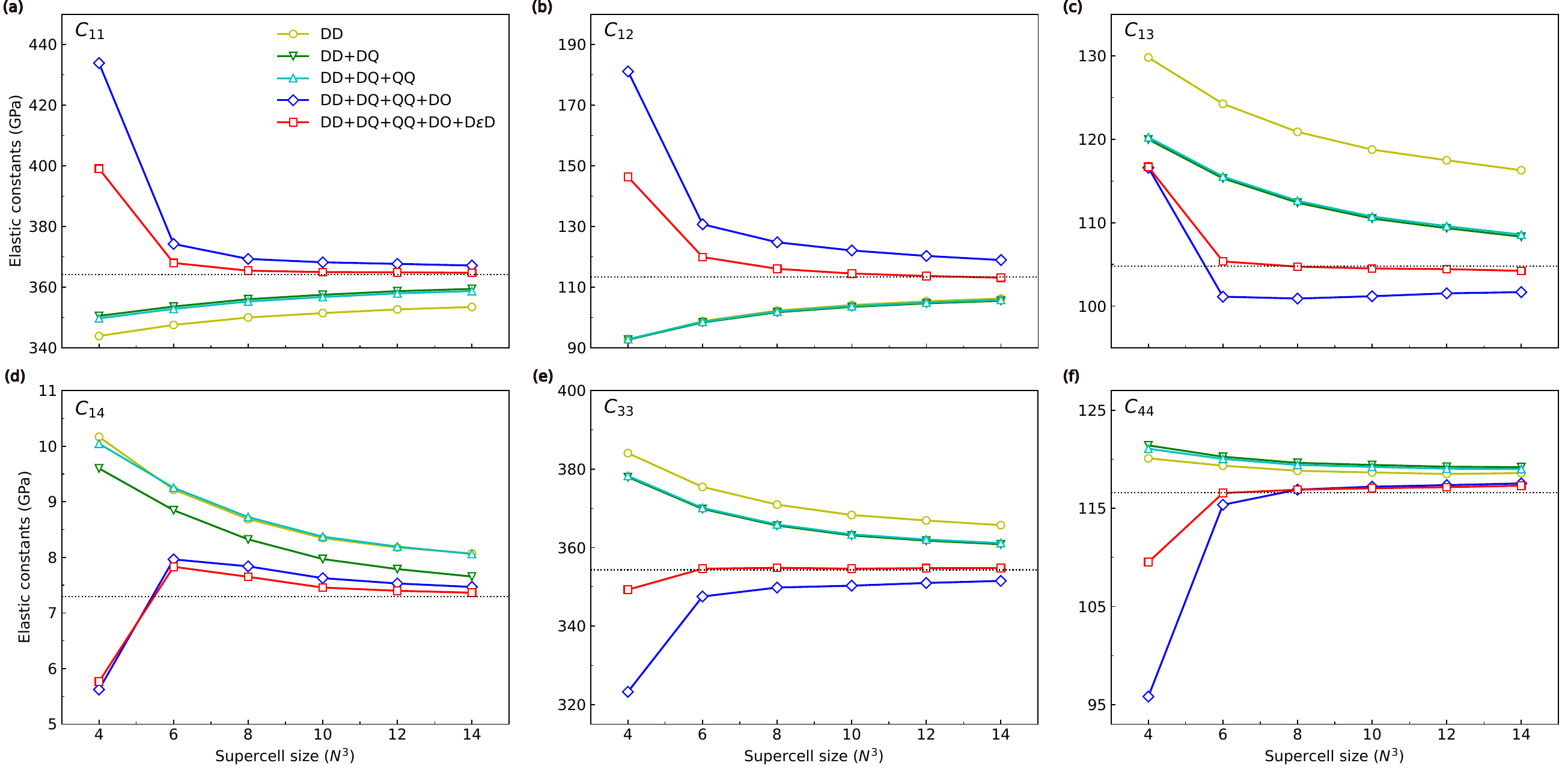}
\caption{(a-f) Independent components of the clamped-ion elastic tensor of BaTiO$_3$ as a function of supercell size.
The elastic constants are calculated based on the short-range IFCs in real space by increasing the level of treatment for the long-range multipolar interactions.
The dotted horizontal line in each subplot is the corresponding reference value from the finite-difference calculations using the \textsc{thermo\_pw} code.
}
\label{fig:BaTiO3_elastic}
\end{figure*}

\subsection{Low-symmetry materials}

Now we apply our approach to study the elasticity of the ferroelectric rhombohedral barium titanate (BaTiO$_3$) and the related lattice-dynamical properties.
Upon cooling, BaTiO$_3$ displays three consecutive ferroelectric phase transitions: it has the perovskite cubic structure in the high-temperature paraelectric phase, and first undergoes a cubic-to-tetragonal transition around 393 K; the tetragonal ferroelectric phase is then stable until 278 K, where it further transforms into the orthorhombic phase, and below 183 K,  BaTiO$_3$ is finally stablized in the rhombohedral phase~\cite{kay1948,kwei1993}.
In this work, we focus on the low-temperature rhombohedral phase of BaTiO$_3$, with the space group of \textit{R3m}.
Our optimized lattice constant and angle of rhombohedral cell at the LDA level are 3.948 \r{A} and 89.935$^{\circ}$, respectively, which are close to the measured lattice parameters, 4.004~\r{A} and 89.839$^{\circ}$~\cite{kwei1993}.
The phonon dispersions of BaTiO$_3$ are shown in Fig.~\ref{fig:BaTiO3_phonon} as a function of the supercell size with increasing level of treatment for the long-range multipolar interactions.
As can be seen in Fig.~\ref{fig:BaTiO3_phonon}(a), the lowest acoustic branch exhibits an imaginary frequency (about -0.1 THz) in the vicinity of the $\Gamma$ point if the standard interpolation is used with only the DD interaction removed.
This issue has been attributed to the strong effect of high-order multipole interactions in BaTiO$_3$~\cite{royo2020}, such as quadrupole- and octupole-related terms in the expansion of macroscopic electric field effects.
Hence, we have performed a Taylor expansion fit to the unscreened charge density response and the dielectric screening function due to a phonon perturbation up to $\mathcal{O}(q^2)$ accuracy, and the obtained dynamical multipole tensors and dielectric tensors are summarized in Table~\ref{table:multipole_BTO}.
Due to the relatively low symmetry of the rhombohedral BaTiO$_3$, the number of independent components for quadrupole are 4, 4, and 10, respectively, for Ba,  Ti and O atoms, while their corresponding octupole tensors have 6, 6, and 16 symmetry-distinct components.
It should be reiterated that the dynamical matrix of BaTiO$_3$ has been found to be non-hermitian because of the unsymmetric multipole tensor regarding to the permutation involving the direction of phonon perturbation, which is a known issue of low-symmetry infrared-active solids~\cite{cockayne2000,zhou2019-2,powell1970,martin1971,scheringer1974}.
For instance, the second-rank Born effective charge tensor of the O atom is not a symmetric matrix, i.e. $Z_{y}^{z}\neq Z_{z}^{y}$.

\begin{table*}[th!]
\centering
\caption{The calculated elastic constants ($C_{ij}$), bulk ($K$), shear ($G$) and Young's ($E$) moduli [GPa], Poisson's ratio ($\nu$) and longitudinal ($v_{\rm l}$) and transverse ($v_{\rm t}$) sound velocities [m/s] of BaTiO$_3$ from the IFCs of a $14\times14\times14$ supercell (\textit{this work}) and the finite difference of strain-stress relation in $\textsc{thermo\_pw}$.
Both the clamped-ion and relaxed-ion results are shown.
The elastic moduli are estimated using the Voigt-Reuss-Hill scheme of polycrystalline average.
}
\label{table:elastic_prop_BTO}
\begin{ruledtabular}
\begin{tabular}{cccccccccccccc}
\multicolumn{2}{c}{Methods} & $C_{11}$ & $C_{12}$ & $C_{13}$ & $C_{14}$ & $C_{33}$ & $C_{44}$ & $K$ & $G$ & $E$ & $\nu$ & $v_{\rm l}$ & $v_{\rm t}$ \\
\hline
\multirow{2}{*}{Clamped} & this work & 364.72 & 113.12 & 104.23 & 7.37 & 354.78 & 117.31 & 191.87 & 122.64 & 303.31 & 0.24 & 7518 & 4416 \\
& $\textsc{thermo\_pw}$ & 364.14 & 113.29  & 104.77 & 7.28 & 354.32 & 116.62 & 192.03 & 122.38 & 302.82 & 0.24 & 7511 & 4406 \\
\multirow{2}{*}{Relaxed} & this work & 300.63 & 93.72 & 56.44 & 44.14 & 293.61 & 54.86 & 145.05 & 71.16 & 183.49 & 0.29 & 6177 & 3364 \\
& $\textsc{thermo\_pw}$ & 299.99 & 93.07 & 56.77 & 43.39 & 292.80 & 54.19 & 144.85 & 71.03 & 181.90 & 0.28 & 6172 & 3361 \\
\end{tabular}
\end{ruledtabular}
\end{table*}

To illustrate the subtlety of high-order multipoles on the lattice dynamics of BaTiO$_3$, we start by taking into account the long-range interactions related up to dynamical quadrupoles order, see Fig.~S2 of the SM~\cite{SM}, which produces a positive definite phonon dispersion.
However, the convergence of the phonon dispersion with respect to supercell sizes is still poor, especially the low-energy acoustic modes near the zone center where there is a change in their slopes even using a large $14\times14\times14$ supercell.
This means that BaTiO$_3$ has a pronounced long-range Coulomb interactions which makes the general DD interpolation fail, since the  removal of DD term can only ensure IFCs decay faster than $1/d^3$~\cite{gonze1997,gonze1994}, which is not enough to guarantee the locality of IFCs in real space.
Although the additional elimination of DQ and QQ terms in the dynamical matrix makes real-space IFCs decay faster than $1/d^4$ and improve the situation, one needs to remove DO and D$\epsilon$D as well to have fast convergence ($6\times6\times6$), as shown in Fig.~\ref{fig:BaTiO3_phonon}(b) for both the low-energy acoustic and high-energy optical modes.

Last but not least, we compute the six independent components of the clamped-ion elastic constants of rhombohedral BaTiO$_3$ and present their convergence rate and values in Fig.~\ref{fig:BaTiO3_elastic} with increasing level of treatment for the long-range multipolar interactions.
We can see that perfect agreement and fast convergence ($10\times10\times10$) when subtracting all the long-range between our result and the reference finite-difference results using \textsc{thermo\_pw}.
We also find that the QQ term has little impact in that material, which is in sharp contrast to the results in GaAs where the long-range QQ term plays a vital role in the determination of the correct short-circuit elastic tensor.
The lattice-mediated part, Eq.~\eqref{eq:curly}, can also be computed with both methods and is reported in Table~\ref{table:elastic_prop_BTO} where some difference can be observed.
Indeed, when estimating the contribution of lattice relaxation to the elastic tensor, the structure optimization under a small strain perturbation in DFT simulations does not always reach the global minimum of the Born-Oppenheimer potential energy surface, which gives rise to a difference in the calculated elastic tensor within the finite-difference implementation.
Such an issue is a limitation of studying elastic properties with finite difference of strain-stress or strain-energy relation~\cite{sluiter1998},
while our approach is able to include the lattice-mediated contributions exactly by using the analytic expression in Eq.~\eqref{eq:curly}.
Nonetheless, the converged results from the $14\times14\times14$ supercell are in relative agreement with the finite-difference elastic constants, where the largest absolute difference is 0.69 GPa and the largest relative difference is 1.24\%, as listed in Table~\ref{table:elastic_prop_BTO}.
The effective elastic moduli and sound velocities of BaTiO$_3$ are also given in Table.~\ref{table:elastic_prop_BTO}, and to our best knowledge the experimental measurements of its elastic properties are still unavailable.

\subsection{Elastic tensor and bending rigidity in semi-metallic graphene}

\begin{figure}[t!]
\centering
\includegraphics[width=0.99\linewidth]{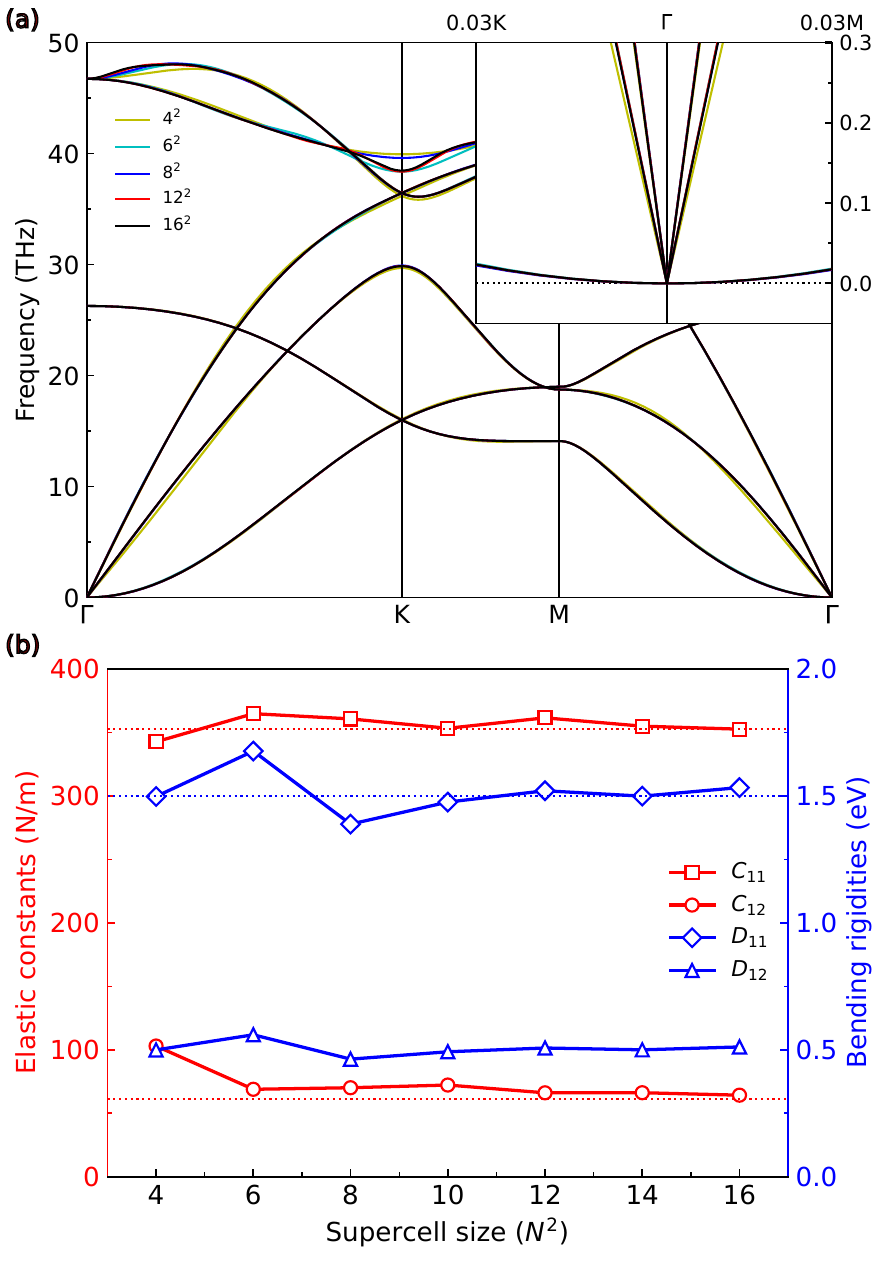}
\caption{(a) Phonon dispersion and (b) independent components of the elastic and bending rigidity tensors in graphene as a function of supercell size.
The inset of panel (a) shows the enlarged acoustic branches.
The dotted lines in the panel (b) represent the theoretical elastic constants $C_{11}$/$C_{12}$ and bending rigidity $D_{11}$ taken from Ref.~\cite{andrew2012} and Ref.~\cite{kumar2020}, respectively.
\label{fig:Gr}
}
\end{figure}

Building on the success of our approach for predicting the elastic tensor of bulk solids, we extend it for 2D solids and also compute their bending rigidities.
We start with graphene as the prototype 2D material, whose lattice constant is 2.466~\r{A} from our DFT simulation.
The calculated phonon dispersion and independent components ($C_{11}$ and $C_{12}$) of the elastic tensor as a function of the supercell size are shown in Fig.~\ref{fig:Gr}.
Since graphene is a semimetal with a Fermi surface formed by six Dirac points, the long-range effect has been suppressed by the metallic-like screening, and the separate treatment for multipolar interactions is thus not needed.
The phonon dispersions and elastic constants of graphene both converge from a 6$\times$6$\times$1 supercell.
However, the uppermost two optical branches are still slightly changing for a larger supercell size due to the well-known Kohn anomalies observed at $\Gamma-E_{2g}$ and $\mathrm{K}-A'_1$ modes~\cite{kohn1959,piscanec2004}.
For our largest 16$\times$16$\times$1 supercell, we find that $C_{11}$ = 352.42~N/m and $C_{12}$ = 64.15~N/m, in good agreement with finite-difference calculations from Ref.~\cite{andrew2012} and with experiment, see Table~\ref{table:2D_elastic_prop}.
We also note that lattice relaxation induced by the internal strain field only has a small effect and the clamped-ion results for $C_{11}$ and $C_{12}$ are found to be 360.19 and 56.38~N/m, respectively.

We then proceed to investigate the bending rigidities of graphene, which are also calculated based on the real-space IFCs but using the higher-order perturbation expansion presented in Sec.~\ref{long_bending_wave}.
As shown in Fig.~\ref{fig:Gr}(b), using the same Voigt notation, the fourth-rank bending rigidity tensor of graphene only has two independent components, $D_{11}$ and $D_{12}$.
Unlike the elastic tensor entering the second-order long-wavelength equation of acoustic vibrations, the bending rigidity tensor describes the fourth-order equations of motion for flexural modes, which is more sensitive to the accuracy of IFCs.
Furthermore, the bending rigidities of 2D materials can be inferred from the quadratic coefficients of the long-wavelength dispersion of ZA modes, which have a parabolic behavior near the zone center.
As can be seen from the inset of Fig.~\ref{fig:Gr}(a), the parabolic shape of ZA phonon remains unchanged when varying the supercell size from 4$\times$4$\times$1 to 16$\times$16$\times$1, which is consistent with the observation of a nearly constant bending rigidities of graphene throughout the entire region of supercell size.
The obtained $D_{11}$ and $D_{12}$ based on a 16$\times$16$\times$1 supercell are 1.53 and 0.51~eV,  respectively.
The principal bending rigidity of graphene $D^\mathrm{P}=D_{11}=D_{22}$ was computed previously using the Helfrich Hamiltonian model for membranes~\cite{helfrich1973,lipowsky1991}, which gave a value of 1.44~eV~\cite{wei2013}, in good agreement with our result.
In addition, using the Helfrich Hamiltonian method within the cyclic DFT formalism~\cite{banerjee2016,ghosh2019}, the authors of Ref.~\cite{kumar2020} obtained a principal bending rigidity of  1.50~eV.
Besides, experimental measurement yielded a bending stiffness of monolayer graphene between 1.2 and 1.7~eV~\cite{han2020}, further corroborating the accuracy and reliability of our long-wavelength perturbation method.

The tensorial nature of our bending rigidity formalism also allows for a direct evaluation of the Gaussian bending rigidity or Gaussian modulus $D^\mathrm{G}$.
This quantity describes non-pure bending such as twisting which is dominated by the Gaussian curvature defined as the product of two principal curvatures.
In the case of an isotropic system, $D^\mathrm{G}=-2D_{66}$ as derived in Sec.~III~B of the SM~\cite{SM} is used to calculate the Gaussian modulus.
We obtain a Gaussian modulus of -1.02~eV, in reasonable agreement with the value of -0.70~eV~\cite{koskinen2010} from the Helfrich Hamiltonian method within the density functional tight-binding calculation, while another DFT calculation using the same Helfrich Hamiltonian method gave a value of \mbox{-1.52}~eV~\cite{wei2013}.
We also notice that a direct measurement of Gaussian modulus is challenging and was only available a few years ago, where atomic force microscopy was used to extract the bending rigidities from a quadratic relationship of adhesion energy between the rolled and unrolled monolayer graphene~\cite{ashino2021}, which gave -1.82 and -1.93~eV for a roll-up nanotube and a bent nanoribbon, respectively.
This experimental result is rather different from our computed value.
Interestingly, we find that there is no lattice relaxation effect on the bending rigidities of graphene, which results in an additional symmetry relation $D_{12}=D_{66}$.
Since $D_{11}-D_{12}=2D_{66}$ holds in hexagonal lattices~\cite{mazdziarz2019}, this means $D_{11}=3D_{66}$ and hence $D^\mathrm{G}=-2D^\mathrm{P}/3$ in the case of graphene.
In our computational case, such relationship is verified exactly but this is not the situation for the current experimental values, therefore awaiting further experimental verification in the future.
One possibility for the observed discrepancy in the Gaussian modulus of graphene compared to those from the Helfrich Hamiltonian method is the existence of local geometric distortions due to the use of a relatively large curvature in their calculations, i.e. not in the linear response region.
In addition, the calculations using fullerene-like structures require an additional energy correction due to the existence of 12 pentagon rings, which causes a poor fitting performance for the Gaussian modulus~\cite{wei2013}.
Instead, our perturbative approach uses only a single unit cell and is exact in the infinitely small curvature limit.
Moreover, the probe-induced topological defects and their attractive interactions could explain the more negative Gaussian modulus obtained in experiments~\cite{ashino2021}.
Importantly, the obtained complete bending rigidity tensor of a general monolayer can enable the exploration of bending anisotropy as well as building authentic microscopic to macroscopic models to understand the bending of nanoplates and biological membranes.

\begin{table*}[th!]
\centering
\caption{Theoretical and experimental elastic constants ($C_{ij}$), layer ($K$), shear ($G$) and Young's ($E$) moduli [N/m], Poisson's ratio ($\nu$) and bending rigidities ($D_{ij}$) [eV] of 2D graphene (C), \emph{h}-BN, MoS$_2$, and InSe.
We used a $16\times16\times1$ supercell to obtain the elastic and bending rigidity tensors of graphene and \emph{h}-BN,  a $12\times12\times1$ supercell is used for MoS$_2$, and a $10\times10\times1$ supercell is used for InSe.
To obtain the elastic moduli, the 2D polycrystalline average based on the Voigt-Reuss-Hill scheme is utilized.
For isotropic 2D materials, the principal bending rigidity is $D^\mathrm{P}=D_{11}=D_{22}$, and the Gaussian bending rigidity is calculated as $D^\mathrm{G}=-2D_{66}$.
We note that the layer modulus is the 2D equivalent of the bulk modulus and the experiment data are all assumed to be obtained at room temperature including those specifically reported as such.
}
\label{table:2D_elastic_prop}
\begin{ruledtabular}
\begin{tabular}{ccccccccccc}
\multicolumn{2}{c}{Compounds} & $C_{11}$ & $C_{12}$ & $D_{11}$ & $D_{12}$ & $D^\mathrm{G}$ & $K$ & $G$ & $E$ & $\nu$ \\
\hline 
\multirow{5}{*}{C} & this work & 352.42 & 64.15 & 1.53 & 0.51 & -1.02 & 208.28 & 144.13 & 340.74 & 0.18 \\
& theory  & 352.7~\cite{andrew2012} & 60.9~\cite{andrew2012} & 1.44~\cite{wei2013} &  & -1.52~\cite{wei2013}  & 206.8~\cite{andrew2012} & 145.9~\cite{andrew2012} & 342.2~\cite{andrew2012} & 0.17~\cite{andrew2012} \\
&         &                         &                        &   1.50~\cite{kumar2020} &     & -0.70~\cite{koskinen2010}   &  &   &  &  \\
& experiment & 372$\pm$5~\cite{sun2021} & 47$\pm$12~\cite{sun2021}  & 1.2--1.7~\cite{han2020} &  & -1.82~\cite{ashino2021} &  & 116$\pm$12~\cite{liu2012} & 340$\pm$50~\cite{lee2008}  & 0.19~\cite{politano2015} \\
&         &                         &                        &    &     & -1.93~\cite{ashino2021}   &  &   & 342~\cite{politano2015} &  \\
\hline
\multirow{5}{*}{\emph{h}-BN} & this work & 295.36 & 65.69 & 1.06 & 0.28 & -0.79 & 180.53 & 114.83 & 280.76 & 0.22 \\
& theory  & 293.2~\cite{peng2012} & 66.1~\cite{peng2012} & 1.02~\cite{shirazian2022} &  &  & 179.6~\cite{peng2012} & 113.6~\cite{peng2012} & 278.3~\cite{peng2012} & 0.22~\cite{peng2012} \\
&         &                         &                        & 0.95~\cite{wu2013}   &    &   &  &   &  &  \\
& experiment  &  &   &  &  &  &  &  & 281$\pm$10~\cite{falin2023}  &  \\
&         &                         &                        &    &    &   &  &   & 289$\pm$24~\cite{falin2017} &  \\
\hline
\multirow{5}{*}{MoS$_2$} & this work & 131.88 & 32.28 & 9.06 & 4.35 & -4.71 & 82.08 & 49.80 & 123.98 & 0.24 \\
& theory  & 132.3~\cite{singh2018} & 32.8~\cite{singh2018} & 9.10~\cite{lai2016} &  &  & 82.5~\cite{singh2018} & 49.5~\cite{singh2018} & 124.1~\cite{singh2018} & 0.25~\cite{singh2018} \\
&         &                         &                        & 9.01~\cite{kumar2020}   &    &   &  &   &  &  \\
& experiment  & &   & 10.2$\pm$0.6~\cite{zhao2015} &  &  &  &  & 160$\pm$40~\cite{lloyd2017} &  \\
&         &                         &                      &  10.5$\pm$3.8~\cite{yu2021}  &    &   &  &   & 180$\pm$60~\cite{bertolazzi2011} &  \\
\hline
\multirow{2}{*}{InSe} & this work & 49.27 & 13.38 & 22.60 & 8.34 & -14.25 & 31.33 & 17.95 & 45.64 & 0.27 \\
& theory & 49.38~\cite{chen2021} & 13.72~\cite{chen2021} & 15.3, 17.5~\cite{kumar2020} &  &  & 31.55~\cite{chen2021} & 17.83~\cite{chen2021} & 45.57~\cite{chen2021} & 0.28~\cite{chen2021} \\
\end{tabular}
\end{ruledtabular}
\end{table*}

\subsection{Elastic tensor and bending rigidity in semiconducting monolayers}\label{hbn}

To validate the generality of our bending rigidity approach, we also tested semiconducting monolayers, where long-range fields couple with lattice vibrations.
We start with \emph{h}-BN but remark that the calculation of quadrupole tensor in DFPT is not implemented in \textsc{Quantum ESPRESSO}~\cite{giannozzi2009,giannozzi2017} and the recent generalization of charge density response to 2D systems is limited to in-plane atomic displacements~\cite{macheda2023}.
Therefore, we use the quadrupole tensors of \emph{h}-BN, MoS$_2$ and InSe computed by some of us~\cite{ponce2023} using the DFPT implementation in the \textsc{abinit} package~\cite{royo2019,royo2021,gonze2016,gonze2020} and
reported in Table~S1 of the SM~\cite{SM}.
To guarantee result consistency, we therefore use the same computational parameters as Ref.~\cite{ponce2023}.
As a side remark, we find that the use of 2D Fourier interpolation with the long-range presented in Sec.~\ref{2D_long_range} combined with the rotational invariance and equilibrium conditions~\cite{lin2022}
gives small imaginary or linear ZA modes in the long-wavelength limit, due to out-of-plane dynamical dipoles, i.e. the second term in square parentheses in Eq.~\eqref{eq:2D_long_range_dynmat}.
We find that these out-of-plane dipoles have negligible contributions to both the phonon dispersions and elastic properties in the studied materials and leave this limitation for further studies.
We give additional details on the rotational invariance and equilibrium conditions in Appendix~\ref{inv_conds}.

As shown in Fig.~\ref{fig:bn}(a), the phonon dispersion  of \emph{h}-BN converges with a 4$\times$4$\times$1 supercell when the multipolar interactions up to dynamical quadrupoles are taken into account.
In contrast, if the standard Fourier interpolation with only DD interactions is adopted, we find that there is a slight modification in the slope of the long-wavelength TA phonon modes (not shown).
The short-circuit elastic constants are more sensitive and converge with a 12$\times$12$\times$1 supercell, but only the one with DD+DQ+QQ recovers previous DFT calculations~\cite{peng2012}.
Based on a 16$\times$16$\times$1 supercell, the two symmetry-independent components $C_{11}$ and $C_{12}$ calculated by removing only the DD interactions are 298.37 and 76.52~N/m, respectively, while the respective reference data are 293.2 and 66.1~N/m~\cite{peng2012}.
The deviation in $C_{12}$ is large with a relative error of 15.77\%, which implies the important role of dynamical quadrupole interactions in the electromechanical coupling in 2D \emph{h}-BN.
Indeed with DD+DQ+QQ removed, we obtain a value of 295.36 and 65.69~N/m for the $C_{11}$ and $C_{12}$ components of \emph{h}-BN.
The detailed comparison of our calculated elastic constants and moduli for \emph{h}-BN  with those from previous simulations as well as experimental measurements can be found in Table~\ref{table:2D_elastic_prop}.
Importantly, the convergence rate of elastic constants in 2D materials is different from their bulk counterpart due to $1/|\mathbf{q}|$ decay for the Coulomb kernel instead of $1/|\mathbf{q}|^2$ for the 3D case.
As a result, the lowest-order DD interactions is first-order such that the LO-TO splitting is vanishing at the zone-center of 2D materials and the DQ and QQ interactions are thus second- and third-order, respectively.
Meanwhile, we are left with a dipole-dipole interaction mediated by the macroscopic polarizability (similar to the D$\epsilon$D term for bulk systems), which is also second-order and can be included directly into the DD term if one does not truncate the denominator in Eq.~\eqref{eq:2D_long_range_dynmat}.
As the 2D elastic tensor is still determined from the second-order long-wavelength equation, the removal of multipolar interactions up to the second order is necessary to converge it with respect to the supercell size.
However, our simulation reveals that the third-order QQ term still has a non-trivial influence on the calculated short-circuit elastic tensor in \emph{h}-BN.
This further suggests the higher-order QQ interactions are important to keep the obtained IFCs short-ranged with a sufficient decay in real space, which is not a direct effect on the 2D elastic tensor.
We believe that only the DD and DQ interactions are necessary to converge the elastic constants of 2D crystals, but the further separation of an additional QQ term is required to guarantee the IFCs decay in the desired manner; in other words, the range separation using the Ewald summation technique in reciprocal space may be not exact.
As the elastic tensor is determined from the second-order long-wavelength equation, the removal of multipolar interactions up to at least second order is necessary to converge it with respect to the supercell size.

\begin{figure}[t]
\centering
\includegraphics[width=0.99\linewidth]{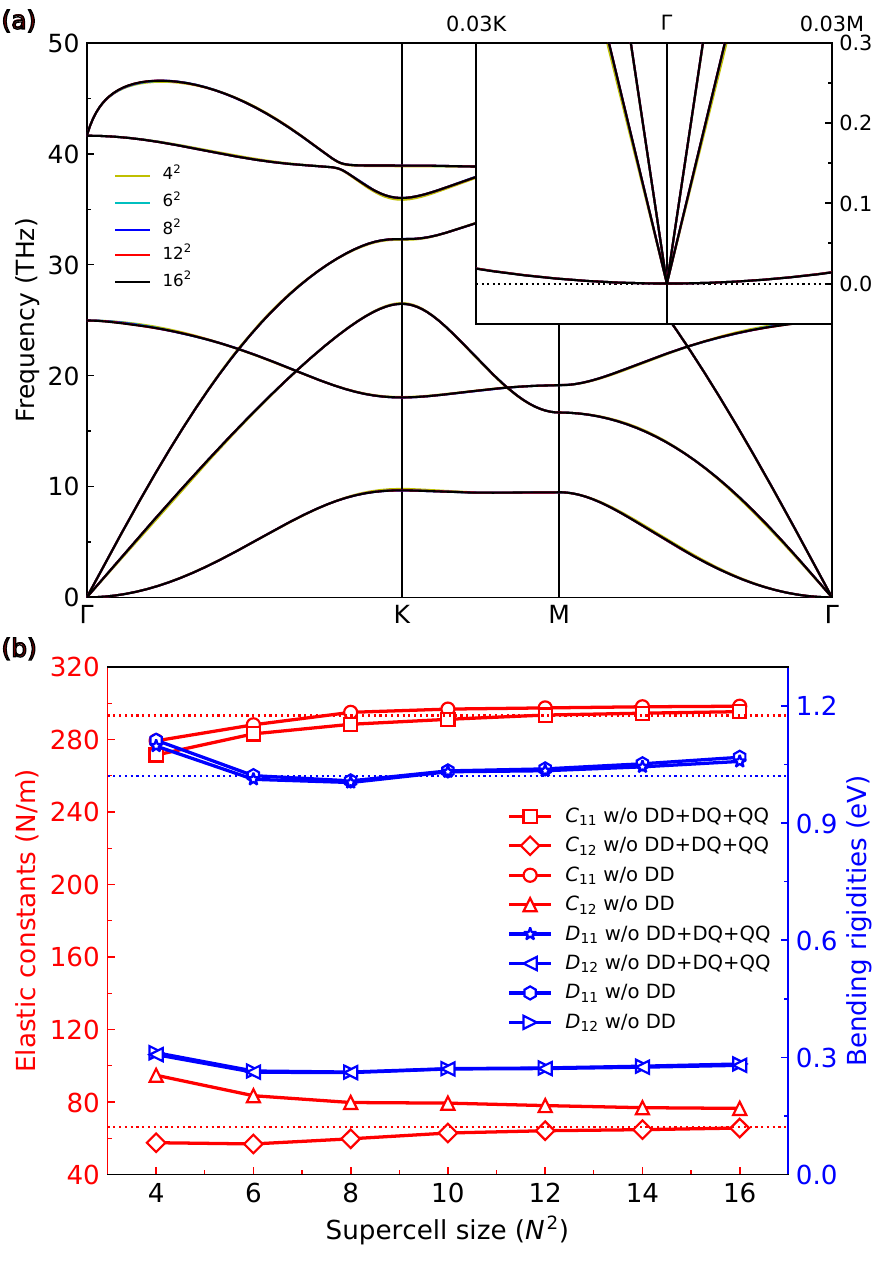}
\caption{(a) Phonon dispersion and (b) independent components of the elastic and bending rigidity tensors in \emph{h}-BN as a function of supercell size.
The multipolar interactions up to the quadrupole terms are considered for the long-range effect in the phonon dispersion of \emph{h}-BN,  and the inset of panel (a) shows the enlarged acoustic branches around the $\Gamma$ point along the K and M directions.
In the panel (b), the short-circuit elastic constants and bending rigidities of \emph{h}-BN are calculated at the levels of removing DD and DD+DQ+QQ interactions; the dotted lines represent the theoretical elastic constants $C_{11}$/$C_{12}$ and bending rigidity $D_{11}$ taken from Ref.~\cite{peng2012} and Ref.~\cite{shirazian2022}, respectively.
\label{fig:bn}
}
\end{figure}

\begin{figure}[ht!]
\centering
\includegraphics[width=0.99\linewidth]{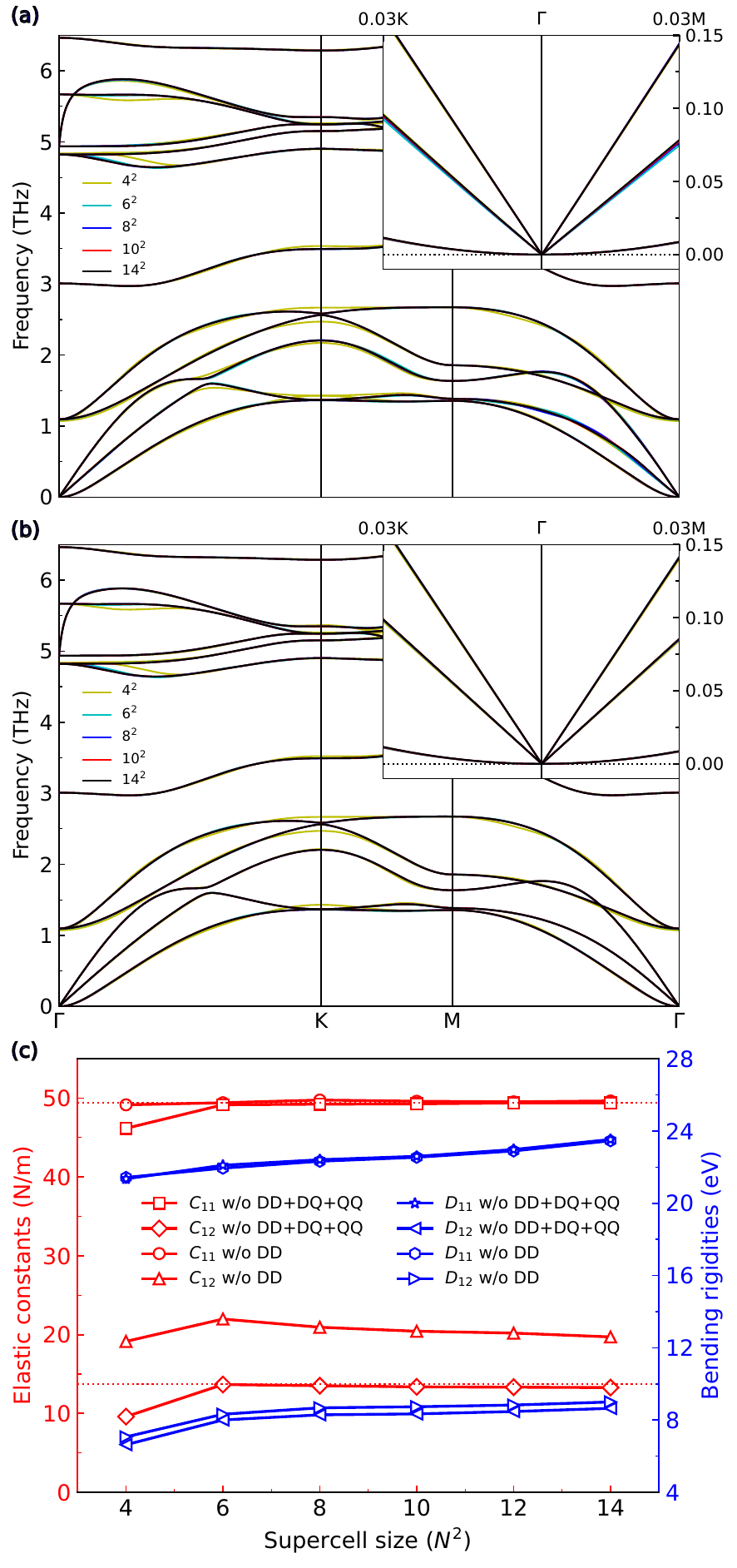}
\caption{Phonon dispersions and independent components of the elastic and bending rigidity tensors in InSe as a function of supercell size.
The phonon spectra in the panels (a) and (b) are obtained based on a Fourier interpolation with the DD and DD+DQ+QQ interactions taken into account, respectively, and their insets show the enlarged acoustic branches around the $\Gamma$ point along the K and M directions.
In the panel (c), the short-circuit elastic constants and bending rigidities of InSe are also calculated at the levels of removing DD and DD+DQ+QQ interactions; the dotted lines represent the theoretical elastic constants $C_{11}$ and $C_{12}$ taken from Ref.~\cite{chen2021}.
\label{fig:inse}
}
\end{figure}

When it comes to the bending rigidities of \emph{h}-BN shown in Fig.~\ref{fig:bn}(b), the convergence is faster than for elastic constants and reached for a 6$\times$6$\times$1 supercell, which is connected to  fast convergence of the long-wavelength dispersion of ZA modes as shown in the inset of Fig.~\ref{fig:bn}(a).
However, we notice a small drift at higher supercell sizes for the bending rigidities because in monolayers the necessary condition of an $n$-order moment of IFCs to converge is that the IFCs themselves decay faster than $1/d^{n+2}$, as the integral in Eq.~\eqref{eq:bending_rgd_CI} is performed over the in-plane area of a supercell.
This implies that the decay of short-range IFCs in real space must be faster than $1/d^6$ in order to converge the fourth-order moment of IFCs in 2D materials, which is equivalent to the requirement that all the multipolar interactions at $\mathcal{O}(q^4)$ should be be removed; these include the first-order DD, the second-order DQ, the third-order QQ and DO, and the fourth-order QO and DH interactions, where H denotes the hexadecapoles.
We leave the calculations of multipoles moments higher than quadrupoles in 2D for further studies but notice that our results range within 1.01--1.06~eV, which are still in good agreement with the existing theoretical results, 1.02~\cite{shirazian2022} and 0.95~eV~\cite{wu2013}.
Also, interestingly, the effects of DQ and QQ interactions on bending rigidities are negligible, as shown in Fig.~\ref{fig:bn}(b).
Besides, we predict for the first time $D_{12}=0.28$~eV and a Gaussian modulus of -0.79 eV~in \emph{h}-BN.
Lastly, unlike graphene, the lattice relaxation does contribute to the bending rigidities of \emph{h}-BN, that is, -0.11~eV for both $D_{11}$ and $D_{12}$, which is negative and thus lowers the energy of the system upon bending.

We continue to study the case of 2D MoS$_2$ monolayer, whose detailed results are shown in Fig.~S4 of the SM~\cite{SM}.
We find that convergence for that material is easier because the macroscopic long-range effects are weak, which we attribute to the relatively large in-plane dielectric permittivity but accompanied with small effective charges.
Based on the IFCs of a 12$\times$12$\times$1 supercell, we obtain the $C_{11}$ and $C_{12}$ of monolayer MoS$_2$ as 131.88 and 32.28~N/m, respectively, while the previous DFT values are 132.3 and 32.8~N/m~\cite{singh2018}.
The detailed mechanical and elastic properties of 2D MoS$_2$ are summarized in Table~\ref{table:2D_elastic_prop}, in excellent agreement with experiment.
In addition, the converged principal bending rigidity $D_{11}$ based on a 12$\times$12$\times$1 supercell is 9.06~eV, consistent with a value of 9.01~\cite{kumar2020} and 9.10~eV~\cite{lai2016}, which both relies on the Helfrich Hamiltonian method.
Notably, we find that the lattice-mediated contribution due to relaxation effects significantly impacts the bending rigidities of MoS$_2$, which is a crucial difference from the behavior observed in graphene and \emph{h}-BN.
Without the lattice-mediated contribution, both $D_{11}$ and $D_{12}$ become negative (i.e. -1.53 and -0.51~eV),  which indicates that such a material is not stable under a bending perturbation due to the failure of the clamped-ion approximation.
It further suggests that the motions of bending waves in MoS$_2$ can be only stabilized by the renormalization effect of the lattice, a phenomenon that has not been discussed before.
Finally, we here report for the first time the values of $D_{12}$ and Gaussian modulus $D^\mathrm{G}$ of MoS$_2$ are 4.35 and -4.71~eV, respectively, awaiting the future experimental verifications.

Our last example for demonstrating the accuracy of the proposed method is monolayer InSe, which serves as a prototype for the 2D family of III–VI metal chalcogenides.
The calculated phonon dispersions with the increasing supercell size are shown in Figs.~\ref{fig:inse}(a) and (b), where the long-range Coulomb electrostatics is treated at the levels of DD and DD+DQ+QQ interactions, respectively.
In the case of DD interaction only, the slope of the linear TA modes around the $\Gamma$ point changes slowly as the supercell size increases.
As can be seen from Fig.~\ref{fig:inse}(b) and the inset therein, when the DD+DQ+QQ interpolation scheme is used, the resulting phonon spectra can be readily converged using only a small 6$\times$6$\times$1 supercell.
In particular, the long-wavelength behaviors of all three acoustic modes remain constant with the increasing supercell size from 4$\times$4$\times$1 to 14$\times$14$\times$1.
For the short-circuit elastic constants, as illustrated in Fig.~\ref{fig:inse}(c), the values of $C_{11}$ and $C_{12}$ are 49.60 and 20.43~N/m, respectively.
While $C_{11}$ agrees well with the reference result of 49.38~N/m from the previous DFT calculations~\cite{chen2021}, $C_{12}$ shows a large error around 48.91\% with respect to the reference value of 13.72~N/m.
Such an inconsistency can be remedied by further taking into account the contributions from DQ and QQ interactions, as shown by the red dotted lines in Fig.~\ref{fig:inse}(c).
When the multipolar interactions up to the quadrupole terms are considered, values of 49.27 and 13.38~N/m are obtained, respectively, in excellent agreement with the previous reported values.
Also, if the clamped-ion approximation is adopted, the respective $C_{11}$ and $C_{12}$ are 73.38 and 23.39~N/m.
The detailed results of elastic constants and moduli for 2D InSe are summarized in Table~\ref{table:2D_elastic_prop}.

Similarly to the cases of \emph{h}-BN and MoS$_2$, the bending rigidities of InSe in Fig.~\ref{fig:inse}(c) suffer from the issue of divergence when using a large supercell, due to the existence of high-order mutipolar interactions beyond the quadrupole terms.
Although the removal of DQ and QQ interactions only results in a small positive shift in the $D_{12}$ component compared to the case with only DD interactions subtracted, the bending rigidities are still diverging, especially from the supercell size of 12$\times$12$\times$1.
We note that the impact of DD term itself is also relatively small, see Fig.~S5 of the SM~\cite{SM}, which suggests that the octupole tensor will be important in this material.
However, owing to the technical difficulty of computing higher-order multipolar interactions, we will leave this point for the future investigation.
Using a moderate 10$\times$10 supercell, our approximated principal bending rigidity $D_{11}$, Gaussian modulus $D^\mathrm{G}$, and the $D_{12}$ component of InSe monolayer are 22.60, -14.25, and 8.34~eV, respectively.
By comparison, the principal bending rigidities of InSe from the Helfrich Hamiltonian method within the cyclic DFT calculations are 15.3 and 17.5~eV, respectively, using the roll-up nanotubes along the zigzag and armchair directions~\cite{kumar2020}.
Such discrepancy could be attributed to many factors, including unconverged DFT parameters and numerical instability in extracting bending rigidities from the energy-curvature relation.
The latter is a drawback of the Helfrich Hamiltonian method, which can be corroborated by the observed symmetry breaking in the calculated bending rigidities along the two principal axes.
Furthermore, our simulation also reveals a significant renormalization effect from the lattice relaxation on bending rigidities of InSe, where the clamped-ion $D_{11}$ and $D_{12}$ are found to be 0.32 and 0.11~eV.
Interestingly, for 2D materials containing only a single layer such as \emph{h}-BN the lattice relaxation induced by a bending deformation tends to lower the bending rigidity, while it can significantly stiffen the bending modulus of multilayer 2D materials.
This finding will provide practical guidelines for designing future flexible electronics.

\section{Conclusions}\label{conclusions}

In this work, we propose a novel method to simultaneously compute the elastic constants and bending rigidities of bulk and LD materials.
Compared to existing methods, the approach is based on the long-wavelength perturbation theory and uses only the real-space IFCs, dynamical charges and dielectric tensors as input.
This provides the direct benefit to work for any dimensions unlike the finite-difference method based on the strain-stress and strain-energy relations where the elastic constants involving the out-of-plane Cartesian component should be removed in the constitutive equations for 2D materials.
Second, the electromechanical coupling is treated on the same footing through the calculated multipole and dielectric tensors.
One can readily obtain the open-circuit elastic constants via the piezoelectric tensor calculated directly from the quadrupole tensor, with the lattice-mediated contribution fully included.
Third, the bending rigidities of 2D materials can be easily obtained based on a higher-order long-wavelength perturbation expansion in the same manner.
Notably, the calculated bending rigidity is inherently tensorial, which further allows us to explore the elastic anisotropy of 2D plates, and currently this is the only way to obtain the complete fourth-rank bending rigidity tensor, to the best of our knowledge.
Fourth, temperature-dependent elastic constants and bending rigidities beyond the quasiharmonic approximation can be readily achieved via the self-consistent harmonic approximation~\cite{monacelli2021} by taking into account the full quantum anharmonicity, as our approach takes the real-space IFCs as the input.
Last, if phonon calculations have been already performed, the calculations of elastic constants and bending rigidities with the present approach can be performed at virtually no additional computational cost.
This enables the automated high-throughput calculations of elastic properties directly through existing phonon workflows that have been already implemented in
most of high-throughput workflow managers~\cite{huber2020,Bastonero2024,petretto2018}.

In summary, we present an efficient and accurate approach for calculating both the elastic and bending rigidity tensors of crystalline solids, which is rigorously formulated based on the secular equation of lattice dynamics and the long-wavelength perturbation theory.
In particular, we have provided a practical recipe in the framework of modern electronic structure calculations for implementing the Huang formula of elastic stiffness tensor that dates back to the atomic theory of elasticity introduced in 1950.
Besides, we extend the Huang's theory to describe also the equations of motion of bending wave in 2D materials and obtain the corresponding bending rigidity tensor, through the fourth-order long-wavelength expansion.
Significantly, in semiconductors and insulators, the induced macroscopic electric fields couple with the lattice vibrations in the long-wavelength limit, which must be treated separately in order to recover the elastic constants and bending rigidities at the correct electrical boundary conditions.
To obtain the short-circuit one, it is found that the multipolar interactions up to the octupole terms are necessary to remove the macroscopic electric field effects in elastic constants at $\mathcal{O}(q^2)$, while the higher-order terms such as the dipole-hexadecapole and quadrupole-octupole interactions are required to converge the bending rigidity tensor.
For this purpose, we have further developed a reliable cluster expansion model to efficiently extract the high-order multipole and dielectric tensors.
We then validate the effectiveness and accuracy of our methodology by demonstrating the example calculations on four bulk solids: Si, NaCl, GaAs and BaTiO$_3$, and four 2D materials: graphene, \emph{h}-BN, MoS$_2$ and InSe, where an overall good agreement has been achieved among other theoretical methods as well as the experimental measurements.
Interestingly, we discover that the lattice-mediated contributions due to the relaxation of internal atomic positions under bending deformation play a crucial role in stabilizing bent 2D materials and flexural modes, which has not yet been reported.
While the lattice relaxation effect tends to lower the bending rigidities of the single-layer 2D materials, it can stiffen the bending modulus of multilayer 2D materials.
Our findings will deepen the understanding of bending deformation of nanoplates and membranes, complementing the microscopic theory of flexural nanomechanics.

Last but not least, the computational framework presented here serves as a robust foundation for identifying and characterizing promising energy materials.
For instance, the accurate determination of mechanical descriptors, combined with high-throughput phonon calculations, can enable the efficient screening of high thermal conductivity materials or the rational design of strain-tolerant cathode materials for long-lifespan batteries.
Furthermore, our automated extraction of high-order multipoles (particularly quadrupoles and octupoles) provides an atom-resolved pathway to optimize piezoelectric and flexoelectric materials.
By bridging the gap between fundamental material properties and device-level engineering, these developments are crucial for the practical realization of next-generation energy harvesting and electromechanical devices.

\section{Acknowledgments}
We would like to thank Alexander Croy, Jian Han, Lev Kantorovich, Boris Kozinsky, Binbin Liu, Miquel Royo, and Yihan Wu for helpful discussions.
This work is supported by the Sinergia project of the Swiss National Science Foundation (grant No. CRSII5\_189924) and the NCCR MARVEL funded by the Swiss National Science Foundation (grant No. 205602).
S. P. acknowledges support from the Walloon Region in the strategic axe FRFS-WEL-T, the Fonds de la Recherche Scientiﬁque de Belgique (FRS-FNRS) under grants No.~T.0183.23 (PDR) and No.~T.W011.23 (PDRWEAVE).
S.P. is a Research Associate of the FRS-FNRS.
F. Macheda and F. Mauri acknowledge the MORE-TEM ERC-SYN project under grant No. 951215.
The computational time has been provided by the Swiss National Supercomputing Centre (CSCS) under project ID mr33, by the EuroHPC JU award granting access to MareNostrum5 at Barcelona Supercomputing Center (BSC), Spain (Project ID EHPC-EXT-2023E02-050), by PRACE award granting access to Discoverer in SofiaTech, Bulgaria (OptoSpin project ID 2020225411) and the IMX cluster at EPFL.

\section{Data availability}
The data and developed software that support the findings of this article are openly available~\cite{Lin2026}.

\appendix

\section{Invariance and equilibrium conditions on second-order interatomic force constants}\label{inv_conds}

The conservation of the global crystal and angular momenta requires the IFCs to satisfy additional acoustic sum rules (ASRs), and these constraints on the second-order IFCs read~\cite{born1954,leibfried1961,lin2022}:
\begin{align}
\sum_{l\kappa'} \Phi^{l}_{\kappa\alpha,\kappa'\beta} &= 0,\label{eq:ti_2nd} \\
\sum_{l\kappa'} \Phi^{l}_{\kappa\alpha,\kappa'\beta}\tau^{l}_{\kappa\kappa'\gamma} &= \sum_{l\kappa'}  \Phi^{l}_{\kappa\alpha,\kappa'\gamma} \tau^{l}_{\kappa\kappa'\beta}, \label{eq:ri_2nd} \\
\sum_{l\kappa\kappa'}\Phi^{l}_{\kappa\alpha,\kappa'\beta} \tau^{l}_{\kappa\kappa'\gamma} \tau^{l}_{\kappa\kappa'\delta} &= \sum_{l\kappa\kappa'}  \Phi^{l}_{\kappa\gamma,\kappa'\delta}  \tau^{l}_{\kappa\kappa'\alpha} \tau^{l}_{\kappa\kappa'\beta}, \label{eq:eq_conds}
\end{align}
where Eq.~\eqref{eq:ti_2nd} denotes the translational invariance, and Eq.~\eqref{eq:ri_2nd} is the rotational invariance.
If the system is at the equilibrium, the IFCs needs to further fulfill the so-called equilibrium conditions for vanishing external stress, given by Eq.~\eqref{eq:eq_conds}.
This set of constraint equations provide the complete description of the invariance conditions of a harmonic lattice at the equilibrium.
Particularly, the rotational invariance and the equilibrium conditions are found to be important for recovering the physical quadratic dispersions of ZA phonons in LD materials~\cite{lin2022,croy2020}.
Besides, another useful identity derived from translational invariance is
\begin{equation}\label{eq:ti_identity}
\sum_{l\kappa\kappa'}\Phi^{l}_{\kappa\alpha,\kappa'\beta}\tau^{l}_{\kappa\kappa'\gamma}=0,
\end{equation}
which suggests that the lattice remains perfect under any homogeneous deformation, and the resulting forces vanish~\cite{born1954}.
This new identity can be derived by starting from
\begin{equation}\label{eq:ti_pos}
\sum_{l'l}\sum_{\kappa\kappa'}\Phi^{l'l}_{\kappa\alpha,\kappa'\beta}(\tau_{l\kappa'\gamma}-\tau_{l'\kappa\gamma})=0,
\end{equation}
which is valid by considering the translational invariance Eq.~\eqref{eq:ti_2nd}.
One can notice that the summation $\sum_{l,\kappa\kappa'}$ is the same for choosing an arbitrary reference unit cell $l'$, since the lattice is infinite without any surface effect.
Hence, we can set $l'=0$ and obtain Eq.~\eqref{eq:ti_identity}.

\section{Solution of the long-wavelength equations for acoustic vibrations}\label{solvability}

As discussed in the main text, the resulting five coupled equations in terms of phonon momentum $\mathbf{q}$ up to the fourth order, Eqs.~\eqref{eq:eq_0th} to~\eqref{eq:eq_2nd}, \eqref{eq:eq_3rd} and \eqref{eq:eq_4th}, need to be solved iteratively, where the unknowns are the perturbation series of phonon frequencies $\omega(\mathbf{q}) $ and eigenvectors $U_{\kappa\alpha}(\mathbf{q})$ for each phonon mode $\nu$.

The zeroth-order Eq.~\eqref{eq:eq_0th} is independent of $\mathbf{q}$, which contains only the zone-center dynamical matrix $\Phi^{(0)}_{\kappa\alpha,\kappa'\beta}$ given by Eq.~\eqref{eq:D0}.
Owing to the translation invariance, the general solution ansatz is readily obtained as
\begin{equation}\label{eq:solution_0th}
U^{(0)}_{\kappa\alpha}=\overline{U}_\alpha,
\end{equation}
where $\overline{U}_\alpha$ is an arbitrary polarization vector for each acoustic mode, independent of $\kappa$.
It describes the rigid translations of the whole crystal in three Cartesian directions.
The homogeneous zeroth-order equation has an infinite number of solutions, since $\Phi^{(0)}_{\kappa\alpha,\kappa'\beta}$ is singular caused by the constraints of the translational invariance.
Although the zeroth-order solution currently remains arbitrary, it can be further determined in the second-order equation where the linear phonon dispersion is obtained, Eq.~\eqref{eq:linear_disp}.
Using Eqs.~\eqref{eq:D0} and \eqref{eq:solution_0th}, the zeroth-order equation then becomes
\begin{equation}\label{eq:eq_0th_with_0th}
\overline{U}_\beta\sum_{l\kappa'}\Phi^{l}_{\kappa\alpha,\kappa'\beta}=0,
\end{equation}
which holds given the translational invariance Eq.~\eqref{eq:ti_2nd} and thus verifies the zeroth-order solutions.

For the first-order equation, we substitute Eq.~\eqref{eq:solution_0th} into Eq.~\eqref{eq:eq_1st} and arrive at the following inhomogeneous equation:
\begin{equation}\label{eq:eq_1st_with_0th}
\Phi^{(0)}_{\kappa\alpha,\kappa'\beta}U^{(1)}_{\kappa'\beta}(\mathbf{q}) = -q_\gamma\overline{U}_\beta\sum_{\kappa'} \Phi^{(1),\gamma}_{\kappa\alpha,\kappa'\beta}.
\end{equation}
By setting the right-hand side of the above equation to zero, it can be noticed that the homogeneous part of Eq.~\eqref{eq:eq_1st_with_0th} has the same form as the zeroth-order equation.
Since the zeroth-order solution $U^{(0)}_{\kappa\alpha}$ is the orthogonal complement space of the subspace spanned by $\Phi^{(0)}_{\kappa\alpha,\kappa'\beta}$, the sufficient and necessary condition of solvability is that the right-hand side of Eq.~\eqref{eq:eq_1st_with_0th} orthogonalizes to the zeroth-order solution:
\begin{equation}\label{eq:orthogonal_cond}
q_\gamma\overline{U}_\alpha\overline{U}_\beta\sum_{\kappa\kappa'}\Phi^{(1),\gamma}_{\kappa\alpha,\kappa'\beta}=0.
\end{equation}
This solvability condition is fulfilled by substituting $\Phi^{(1),\gamma}_{\kappa\alpha,\kappa'\beta}$ with Eq.~\eqref{eq:D1} and taking into account the identity Eq.~\eqref{eq:ti_identity}:
\begin{equation}\label{eq:verify_orthogonal_cond}
q_\gamma\overline{U}_\alpha\overline{U}_\beta\sum_{l\kappa\kappa'}\Phi^{l}_{\kappa\alpha,\kappa'\beta}\tau^{l}_{\kappa\kappa'\gamma}=0.
\end{equation}
Since $\Phi^{(0)}_{\kappa\alpha,\kappa'\beta}$ is not full-rank due to the translational invariance, there are only $3n_\mathrm{a}-3$ linearly independent first-order equations, and their solutions also become arbitrary.
Without loss of generality, we remove this arbitrariness by imposing the condition of zero displacement for the first atom ($\kappa=1$) in the unit cell: $U^{(1)}_{\kappa=1,\alpha}(\mathbf{q})=0$.
The remaining $(3n_\mathrm{a}-3)\times(3n_\mathrm{a}-3)$ matrix $\Phi^{(0)}_{\kappa\alpha,\kappa'\beta}$, with $\kappa$ and $\kappa'$ both greater than one, is in general non-singular, and the solution of the first-order equation can be obtained by inverting $\Phi^{(0)}_{\kappa\alpha,\kappa'\beta}$.
We introduce the inverse matrix of $\Phi^{(0)}_{\kappa\alpha,\kappa'\beta}$ as
\begin{equation}\label{eq:Gamma}
\Gamma_{\kappa\alpha,\kappa'\beta}=
\begin{cases}
\hat{\Gamma}_{\kappa\alpha,\kappa'\beta} &\kappa,\kappa'\neq1\\
0 & \text{otherwise}
\end{cases},
\end{equation}
where $\hat{\Gamma}_{\kappa\alpha,\kappa'\beta}$ is the inverse of the remaining $\Phi^{(0)}_{\kappa\alpha,\kappa'\beta}$ matrix, and $\Gamma_{\kappa\alpha,\kappa'\beta}$ satisfies the following orthogonal relationship~\cite{born1954}:
\begin{equation}\label{eq:Gamma_orthogonal}
\Gamma_{\kappa\alpha,\kappa'\beta}\Phi^{(0)}_{\kappa'\beta,\kappa''\gamma}=\delta_{\alpha\gamma}\delta_{\kappa\kappa''}.
\end{equation}
Since $\Phi^{(0)}_{\kappa\alpha,\kappa'\beta}$ is a symmetric matrix, its inverse should be symmetric as well, $\Gamma_{\kappa\alpha,\kappa'\beta}=\Gamma_{\kappa'\beta,\kappa\alpha}$.
We now multiply Eq.~\eqref{eq:eq_1st} by $\Gamma_{\kappa''\delta,\kappa\alpha}$, and sum over the indices $\kappa$ and $\alpha$, arriving at the first-order solution by applying Eq.~\eqref{eq:Gamma_orthogonal}:
\begin{equation}\label{eq:solution_1st}
 U^{(1)}_{\kappa\alpha}(\mathbf{q}) = -q_\delta\Gamma_{\kappa\alpha,\kappa'\beta}\Phi^{(1),\delta}_{\kappa'\beta,\kappa''\gamma}U^{(0)}_{\kappa''\gamma},
\end{equation}
with the subscripts relabelled.
It is convenient to introduce the following two notations:
\begin{align}
\Lambda^{\kappa}_{\alpha,\beta\gamma}&=\sum_{\kappa'}\Phi^{(1),\gamma}_{\kappa\alpha,\kappa'\beta}, \label{eq:strain_force_response} \\
\Upsilon^{\kappa}_{\alpha,\beta\gamma}&=\Gamma_{\kappa\alpha,\kappa'\delta}\Lambda^{\kappa'}_{\delta,\beta\gamma}, \label{eq:displace_response}
\end{align}
which serve as the internal-strain force-response tensor and internal-strain displacement-response tensor, respectively~\cite{wu2005,stengel2013}.
The first-order solution Eq.~\eqref{eq:solution_1st} can be then rewritten as
\begin{equation}\label{eq:solution_1st_simple}
 U^{(1)}_{\kappa\alpha}(\mathbf{q}) = -q_\gamma\Upsilon^{\kappa}_{\alpha,\beta\gamma} \overline{U}_\beta.
\end{equation}
Specifically, $\Lambda^{\kappa}_{\alpha,\beta\gamma}$ captures the force acting on the $\kappa$th atom in the unit cell along the direction $\alpha$ due to the internal strain $\varepsilon_{\beta\gamma}$, and $\Upsilon^{\kappa}_{\alpha,\beta\gamma}$ is the corresponding atomic relaxation induced by the strain deformation.

With the obtained zeroth-order solution Eq.~\eqref{eq:solution_0th} and the first-order solution Eq.~\eqref{eq:solution_1st_simple}, the second-order equation~\eqref{eq:eq_2nd} can be recast into
\begin{multline}\label{eq:eq_2nd_long}
\Phi^{(0)}_{\kappa\alpha,\kappa'\beta} U^{(2)}_{\kappa'\beta}(\mathbf{q}) = M_\kappa[\omega^{(1)}(\mathbf{q})]^2\overline{U}_\alpha \\
- \frac{q_\gamma q_\delta}{2}\overline{U}_\beta\sum_{\kappa'}\Phi^{(2),\gamma\delta}_{\kappa\alpha,\kappa'\beta} -q_\gamma q_\delta\overline{U}_\beta\Phi^{(1),\gamma}_{\kappa\alpha,\kappa'\lambda}\Upsilon^{\kappa'}_{\lambda,\beta\delta}.
\end{multline}
To further simplify the above expression, we introduce the type-I strain-gradient force-response tensor~\cite{stengel2013}
\begin{equation}\label{eq:flexo_force_response}
T^{\kappa}_{\alpha\beta,\gamma\delta}=T^{\mathrm{CI},\kappa}_{\alpha\beta,\gamma\delta}+\frac{1}{2}\Big(T^{\mathrm{LM},\kappa}_{\alpha\gamma,\beta\delta}+T^{\mathrm{LM},\kappa}_{\alpha\delta,\beta\gamma}\Big),
\end{equation}
with the clamped-ion and lattice-mediated contributions defined as
\begin{align}
T^{\mathrm{CI},\kappa}_{\alpha\beta,\gamma\delta}&=\frac{1}{2}\sum_{\kappa'}\Phi^{(2),\gamma\delta}_{\kappa\alpha,\kappa'\beta},\label{eq:strain_gradient_frozen} \\
T^{\mathrm{LM},\kappa}_{\alpha\gamma,\beta\delta}&=\Phi^{(1),\gamma}_{\kappa\alpha,\kappa'\lambda}\Upsilon^{\kappa'}_{\lambda,\beta\delta}. \label{eq:strain_gradient_relax}
\end{align}
The symmetrization is performed on the parentheses in Eq.~\eqref{eq:flexo_force_response} with respect to Cartesian components $\gamma$ and $\delta$ (i.e. type-I), because $T^{\mathrm{LM},\kappa}_{\alpha\gamma,\beta\delta}$ itself is a type-II quantity (i.e. invariant under exchange of $\beta$ and $\delta$).
Microscopically, $T^{\mathrm{CI},\kappa}_{\alpha\beta,\gamma\delta}$ describes the force induced by a type-I strain gradient $\nabla_{\delta}\tilde{\varepsilon}_{\beta\gamma}$ on the unit-cell atom $\kappa$ along $\alpha$ direction, while $T^{\mathrm{LM},\kappa}_{\alpha\gamma,\beta\delta}$ accounts for the additional force response resulting from the relaxation of internal coordinates if allowed by point group symmetry.
Following a similar procedure for the first-order solution, the solvability condition for the second-order equation can be obtained as
\begin{equation}\label{eq:solvability_2nd}
\bigg(\sum_{\kappa}m_\kappa[\omega^{(1)}(\mathbf{q})]^2\delta_{\alpha\beta} - q_\gamma q_\delta\sum_{\kappa}T^{\kappa}_{\alpha\beta,\gamma\delta}\bigg)\overline{U}_\beta=0.
\end{equation}
This solvability condition can be further rearranged into the equation for sound wave propagation and yields Eq.~\eqref{eq:eq_acoustic_wave}.
Finally,  the solution of the second-order equation is obtained by multiplying both sides of Eq.~\eqref{eq:eq_2nd_long} by $\Gamma_{\kappa''\lambda,\kappa\alpha}$, summing over the indices $\kappa$ and $\alpha$, and substituting $[\omega^{(1)}(\mathbf{q})]^2$ with Eq.~\eqref{eq:linear_disp}:
\begin{equation}\label{eq:solution_2nd}
 U^{(2)}_{\kappa\alpha}(\mathbf{q}) = q_\gamma q_\delta\Gamma_{\kappa\alpha,\kappa'\lambda}\Big(\frac{M_{\kappa'}}{M}T_{\lambda\beta,\gamma\delta}-T^{\kappa'}_{\lambda\beta,\gamma\delta}\Big)\overline{U}_\beta,
\end{equation}
with the subscripts relabelled and the normalization condition of zeroth-order mode eigenvectors $\overline{U}_\alpha$.
By further introducing the mass-compensated strain-gradient force-response tensor~\cite{stengel2013}
\begin{equation}\label{eq:flexo_force_response_mass}
\hat{T}^{\kappa}_{\alpha\beta,\gamma\delta}=T^{\kappa}_{\alpha\beta,\gamma\delta}-\frac{M_{\kappa}}{M}T_{\alpha\beta,\gamma\delta},
\end{equation}
and the mass-compensated strain-gradient displacement-response tensor (type-I)
\begin{equation}\label{eq:flexo_displace_response_mass}
\hat{\Pi}^{\kappa}_{\alpha\beta,\gamma\delta}=\Gamma_{\kappa\alpha,\kappa'\lambda}\hat{T}^{\kappa'}_{\lambda\beta,\gamma\delta},
\end{equation}
Eq.~\eqref{eq:solution_2nd} can be rewritten as
\begin{equation}\label{eq:solution_2nd_simple}
 U^{(2)}_{\kappa\alpha}(\mathbf{q}) = -q_\gamma q_\delta\hat{\Pi}^{\kappa}_{\alpha\beta,\gamma\delta}\overline{U}_\beta.
\end{equation}
In particular, for the ZA modes in LD materials, the linear dispersion $\omega^{(1)}(\mathbf{q})$ always vanishes, and we can readily drop the mass-compensated part and use the tensors without hat:
\begin{equation}\label{eq:flexo_displace_response}
\Pi^{\kappa}_{\alpha\beta,\gamma\delta}=\Gamma_{\kappa\alpha,\kappa'\lambda}T^{\kappa'}_{\lambda\beta,\gamma\delta},
\end{equation}
which in other words implies the elastic constants $T_{\alpha\beta,\gamma\delta}$ involving the out-of-plane component are completely zero.
In a nutshell, together with Eqs.~\eqref{eq:solution_0th}, \eqref{eq:solution_1st_simple}, and \eqref{eq:solution_2nd_simple} for
$U^{(0)}_{\kappa\alpha}$, $ U^{(1)}_{\kappa\alpha}(\mathbf{q})$ and $ U^{(2)}_{\kappa\alpha}(\mathbf{q})$, respectively, we complete the solutions of the coupled long-wavelength Eqs.~\eqref{eq:eq_0th} to~\eqref{eq:eq_2nd}, which are used for investigating the elastic constants of crystals in this work.

To further obtain the second-order dispersion relation $\omega^{(2)}(\mathbf{q})$ of the bending modes in LD materials,  we continue to solve the third-order and fourth-order long-wavelength equations given by Eqs.~\eqref{eq:eq_3rd} and~\eqref{eq:eq_4th} in the main text.
The terms involving $\omega^{(1)}(\mathbf{q})$ are dropped due to the vanishing linear dispersion of ZA phonons in LD materials~\cite{croy2020,lin2022}.
Analogous to the low-order equations just solved, the solvability condition for the third-order long-wavelength equation is that its inhomogenous part orthogonalizes to the zeroth-order solution, which reads
\begin{multline}\label{eq:solvability_3rd}
q_\gamma U^{(0)}_{\kappa\alpha}\Phi^{(1),\gamma}_{\kappa\alpha,\kappa'\beta} U^{(2)}_{\kappa'\beta}(\mathbf{q}) + \frac{q_\gamma q_\delta}{2}U^{(0)}_{\kappa\alpha}\Phi^{(2),\gamma\delta}_{\kappa\alpha,\kappa'\beta} U^{(1)}_{\kappa'\beta}(\mathbf{q}) \\
+\frac{q_\gamma q_\delta q_\lambda}{6}U^{(0)}_{\kappa\alpha}\Phi^{(3),\gamma\delta\lambda}_{\kappa\alpha,\kappa'\beta}U^{(0)}_{\kappa'\beta} =0,
\end{multline}
with the zero-, first-, and second-order solutions given by Eqs.~\eqref{eq:solution_0th}, \eqref{eq:solution_1st_simple}, and \eqref{eq:solution_2nd_simple}, respectively.
To simplify the third-order solution, we can introduce the type-I in-plane strain-gradient force-response tensor due to a small deflective perturbation as
\begin{equation}\label{eq:shear_force_response}
J^{\kappa}_{\alpha\beta,\gamma\delta\lambda} = J^{\rm{CI},\kappa}_{\alpha\beta,\gamma\delta\lambda} + J^{\rm{LM}1,\kappa}_{\alpha\gamma,\beta\delta\lambda} + J^{\rm{LM}2,\kappa}_{\alpha\delta\lambda,\beta\gamma},
\end{equation}
with the corresponding clamped-ion and lattice-mediated contributions as
\begin{align}
J^{\rm{CI},\kappa}_{\alpha\beta,\gamma\delta\lambda} &=\frac{1}{6} \sum_{\kappa'}\Phi^{(3),\gamma\delta\lambda}_{\kappa\alpha,\kappa'\beta},\label{eq:shear_force_response_CI} \\
J^{\rm{LM1},\kappa}_{\alpha\gamma,\beta\delta\lambda} &=- \Phi^{(1),\gamma}_{\kappa\alpha,\kappa'\mu}\Pi^{\kappa'}_{\mu\beta,\delta\lambda}, \label{eq:shear_force_response_LM1} \\
J^{\rm{LM}2,\kappa}_{\alpha\delta\lambda,\beta\gamma} &= - \frac{1}{2}\Phi^{(2),\delta\lambda}_{\kappa\alpha,\kappa'\mu}\Upsilon^{\kappa'}_{\mu,\beta\gamma} \label{eq:shear_force_response_LM2}.
\end{align}
The physical meaning of those high-order response tensors is that $J^{\rm{CI},\kappa}_{\alpha\beta,\gamma\delta\lambda}$ is the force acting on the atom $\kappa$ in the unit cell along the Cartesian direction $\alpha$ caused by the type-I in-plane strain gradient $\nabla_{\lambda}\tilde{\varepsilon}_{\gamma\delta}$ from the displacements of the atoms $\kappa'$ in the out-of-plane direction $\beta$, while $J^{\rm{LM}1,\kappa}_{\alpha\gamma,\beta\delta\lambda}$ and $J^{\rm{LM}2,\kappa}_{\alpha\delta\lambda,\beta\gamma}$ represent the additional force induced by the relaxation effects of lattice under such strain-gradient perturbation.
Assuming the out-of-plane direction $\beta=z$ with the origin of coordinate system placed at the midsurface of the plate, as discussed in Sec.~\ref{theory_bending_rgd} and according to the Kirchhoff plate theory~\cite{reddy2006,ugural2017,mittelstedt2023}, the in-plane displacements vary linearly in the distance $z$ as $\mathbf{u}^{\parallel}(x,y,z)=-z\nabla w(x,y)$, from which the in-plane strain field is obtained.
In addition, in order to make Eq.~\eqref{eq:solvability_3rd} valid and the third-order long-wavelength equation solvable, we find that the following additional ASRs on the second-order IFCs should hold:
\begin{equation}\label{eq:asr_solvabiliy}
\sum_{l\kappa\kappa'}\Phi^{l}_{\kappa\alpha,\kappa'\beta}\tau^{l}_{\kappa\kappa'\gamma}\tau^{l}_{\kappa\kappa'\delta}\tau^{l}_{\kappa\kappa'\lambda}=0,
\end{equation}
which act as the new constraints on harmonic IFCs along with the invariance conditions in Appendix~\ref{inv_conds}.
Since the left-hand side of Eq.~\eqref{eq:asr_solvabiliy} is a fifth-rank tensor, these additional ASRs contribute 60 constraint equations after considering the tensor symmetry.
More generally, it is noted that the expressions involving a sublattice summation of the odd perturbation term of dynamical matrix should be always vanishing, which gives rise to such ASRs in Eqs.~\eqref{eq:ti_identity} and \eqref{eq:asr_solvabiliy}.
We verify numerically in all the studied 2D materials that the violation of these extra ASRs is small and that their imposition leads to negligible changes in the values of bending rigidities.
Therefore, the solution of the third-order equation can be readily obtained by inverting the matrix $\Phi^{(0)}_{\kappa\alpha,\kappa'\beta}$, i.e. multiplying Eq.~\eqref{eq:eq_3rd} by $\Gamma_{\kappa''\mu,\kappa\alpha}$ with a summation over $\kappa$ and $\alpha$:
\begin{align}\label{eq:solution_3rd}
 U^{(3)}_{\kappa\alpha}(\mathbf{q})&=-q_\gamma q_\delta q_\lambda\Gamma_{\kappa\alpha,\kappa'\mu}J^{\kappa'}_{\mu\beta,\gamma\delta\lambda}\overline{U}_\beta \nonumber \\
&=-q_\gamma q_\delta q_\lambda\Xi^{\kappa}_{\alpha\beta,\gamma\delta\lambda}\overline{U}_\beta,
\end{align}
where $\Xi^{\kappa}_{\alpha\beta,\gamma\delta\lambda}=\Gamma_{\kappa\alpha,\kappa'\mu}J^{\kappa'}_{\mu\beta,\gamma\delta\lambda}$ is the type-I in-plane strain-gradient displacement-response tensor from a small deflective perturbation to the crystalline plate.

Finally, the solvability condition for the fourth-order long-wavelength equation is
\begin{multline}\label{eq:solvability_4th}
\sum_{\kappa} M_\kappa [\omega^{(2)}(\mathbf{q})]^2 U^{(0)}_{\kappa\alpha} = \frac{q_\gamma q_\delta q_\lambda q_\mu}{24}\sum_{\kappa}\Phi^{(4),\gamma\delta\lambda\mu}_{\kappa\alpha,\kappa'\beta}U^{(0)}_{\kappa'\beta}   \\
-q_\gamma\sum_{\kappa}\Phi^{(1),\gamma}_{\kappa\alpha,\kappa'\beta} U^{(3)}_{\kappa'\beta}(\mathbf{q}) + \frac{q_\gamma q_\delta}{2}\sum_{\kappa}\Phi^{(2),\gamma\delta}_{\kappa\alpha,\kappa'\beta} U^{(2)}_{\kappa'\beta}(\mathbf{q}) \\
-\frac{q_\gamma q_\delta q_\lambda}{6}\sum_{\kappa}\Phi^{(3),\gamma\delta\lambda}_{\kappa\alpha,\kappa'\beta} U^{(1)}_{\kappa'\beta}(\mathbf{q}),
\end{multline}
which yields the equations of motion for ZA modes in LD materials with the known quadratic dispersion relation in the long-wavelength limit.
For convenience, we define the following sixth-rank response tensors:
\begingroup
\allowdisplaybreaks
\begin{align}
W^{\kappa}_{\alpha\beta,\gamma\delta,\lambda\mu}=&W^{\rm{CI},\kappa}_{\alpha\beta,\gamma\delta,\lambda\mu}+W^{\rm{LM},\kappa}_{\alpha\beta,\gamma\delta,\lambda\mu},  \label{eq:curvature_force_response} \\
W^{\rm{CI},\kappa}_{\alpha\beta,\gamma\delta,\lambda\mu}=&\frac{1}{24}\sum_{\kappa'}\Phi^{(4),\gamma\delta\lambda\mu}_{\kappa\alpha,\kappa'\beta}, \label{eq:curvature_force_response_CI} \\
W^{\rm{LM},\kappa}_{\alpha\beta,\gamma\delta,\lambda\mu} =& W^{\rm{LM1},\kappa}_{\alpha\beta,\gamma\delta,\lambda\mu} + \frac{W^{\rm{LM2},\kappa}_{\alpha\beta,\gamma,\delta\lambda\mu}+W^{\rm{LM2},\kappa}_{\alpha\beta,\delta,\gamma\lambda\mu}}{2} \nonumber \\
&+\frac{W^{\rm{LM3},\kappa}_{\alpha\beta,\delta\lambda\mu,\gamma}+W^{\rm{LM3},\kappa}_{\alpha\beta,\gamma\lambda\mu,\delta}}{2}, \label{eq:curvature_force_response_LM} \\
W^{\rm{LM1},\kappa}_{\alpha\beta,\gamma\delta,\lambda\mu}=&-\frac{1}{2}\Phi^{(2),\gamma\delta}_{\kappa\alpha,\kappa'\upsilon}\Pi^{\kappa'}_{\upsilon\beta,\lambda\mu}, \label{eq:curvature_force_response_LM1} \\
W^{\rm{LM2},\kappa}_{\alpha\beta,\gamma,\delta\lambda\mu}=&\Phi^{(1),\gamma}_{\kappa\alpha,\kappa'\upsilon}\Xi^{\kappa'}_{\upsilon\beta,\delta\lambda\mu}, \label{eq:curvature_force_response_LM2} \\
W^{\rm{LM3},\kappa}_{\alpha\beta,\delta\lambda\mu,\gamma}=&\frac{1}{6}\Phi^{(3),\delta\lambda\mu}_{\kappa\alpha,\kappa'\upsilon}\Upsilon^{\kappa'}_{\upsilon,\beta\gamma}, \label{eq:curvature_force_response_LM3}
\end{align}
\endgroup
where $W^{\kappa}_{\alpha\beta,\gamma\delta,\lambda\mu}$ is the force-response tensor due to the Laplacian of curvature as a sum of the corresponding clamped-ion and lattice-mediated contributions, respectively.
The symmetrization is performed for the two lattice-mediated tensors, $W^{\rm{LM2},\kappa}_{\alpha\beta,\gamma,\delta\lambda\mu}$ and $W^{\rm{LM3},\kappa}_{\alpha\beta,\delta\lambda\mu,\gamma}$, since they are not invariant by swapping $\gamma$ and $\delta$, while the bending rigidity tensor has such symmetry.
Therefore, using the above defined sixth-rank tensors, the solvability condition in Eq.~\eqref{eq:solvability_4th} can be recast into
\begin{equation}\label{eq:solvability_4th_simple}
\sum_{\kappa}M_{\kappa} [\omega^{(2)}(\mathbf{q})]^2 \overline{U}_{\alpha} = q_\gamma q_\delta q_\lambda q_\mu \!\! \sum_{\kappa}W^{\kappa}_{\alpha\beta,\gamma\delta,\lambda\mu}\overline{U}_{\beta}.
\end{equation}
By multiplying $\Gamma_{\kappa''\eta,\kappa\alpha}$ on both sides of Eq.~\eqref{eq:eq_4th}, summing over the indices $\kappa$ and $\alpha$, and using Eq.~\eqref{eq:dispersion_ZA} to replace $[\omega^{(2)}_{\rm ZA}(\mathbf{q})]^2$, the solution of the fourth-order long-wavelength equation is obtained as
\begin{multline}\label{eq:solution_4th}
 U^{(4)}_{\kappa\alpha}(\mathbf{q}) = q_\gamma q_\delta q_\lambda q_\mu\Gamma_{\kappa\alpha,\kappa'\eta} \\
 \times \Big(\frac{M_{\kappa'}}{M}W_{\eta\beta,\gamma\delta,\lambda\mu}-W^{\kappa'}_{\eta\beta,\gamma\delta,\lambda\mu}\Big)\overline{U}_\beta,
\end{multline}
with the subscripts relabelled.
By further defining the mass-compensated force-response tensor
\begin{equation}\label{eq:curvature_force_response_mass}
\hat{W}^{\kappa}_{\alpha\beta,\gamma\delta,\lambda\mu}=W^{\kappa}_{\alpha\beta,\gamma\delta,\lambda\mu}-\frac{M_{\kappa}}{M}W_{\alpha\beta,\gamma\delta,\lambda\mu},
\end{equation}
and the corresponding mass-compensated displacement-response tensor due to the Laplacian of curvature
\begin{equation}\label{eq:curvature_displace_response_mass}
\hat{\Theta}^{\kappa}_{\alpha\beta,\gamma\delta,\lambda\mu} = \Gamma_{\kappa\alpha,\kappa'\eta}\hat{W}^{\kappa'}_{\eta\beta,\gamma\delta,\lambda\mu},
\end{equation}
Eq.~\eqref{eq:solution_4th} can be rewritten as
\begin{equation}\label{eq:solution_4th_simple}
 U^{(4)}_{\kappa\alpha}(\mathbf{q}) = -q_\gamma q_\delta q_\lambda q_\mu \hat{\Theta}^{\kappa}_{\alpha\beta,\gamma\delta,\lambda\mu}\overline{U}_\beta.
\end{equation}
It is worth mentioning that Stengel~\cite{stengel2016} formulated the theory of strain-gradient elasticity using a similar variational method of lattice dynamics, where the derived strain-gradient elasticity tensor thereof for bulk solids shares many similarities with the bending rigidity tensor of this study.
Our results in Eqs.~\eqref{eq:curvature_force_response} to \eqref{eq:curvature_force_response_LM3} are fully equivalent term-by-term with Eq.~(51) of Ref.~\cite{stengel2016} by summing over $\kappa$ and the symmetrization in Eq.~\eqref{eq:curvature_force_response_LM}.
However, Stengel only considered the purely electronic and lattice-mediated terms in his real calculations, which are Eqs.~\eqref{eq:curvature_force_response_CI} and \eqref{eq:curvature_force_response_LM1} plus \eqref{eq:curvature_force_response_LM2} with only \eqref{eq:shear_force_response_LM1} in $\Xi^{\kappa}_{\alpha\beta,\gamma\delta\lambda}$, respectively.
For the bending rigidities of 2D materials, we find those remaining mixed terms (i.e. both electronic and lattice-mediated in the language of Ref.~\cite{stengel2016}) are non-negligible and crucial to recover the results of the Helfrich Hamiltonian method.

In short, Eqs.~\eqref{eq:solution_3rd} and ~\eqref{eq:solution_4th_simple} constitute the solutions of the third- and fourth-order long-wavelength equations, useful for studying the motion of bending acoustic modes in LD materials.
As demonstrated in the main text, the displacement eigenvectors of ZA modes in 2D materials decouple from the in-plane vibrations in the long-wavelength limit, polarizing purely in the out-of-plane directions.
One can readily set $\alpha$ and $\beta$ in all of those solutions to $z$ with $\overline{U}_z=1$.

\section{Relations of the long-wavelength equations to elastic deformation}\label{relations_longwave_elastic}

We show here how the derived long-wavelength acoustic vibrations in Eqs.~\eqref{eq:eq_0th} to \eqref{eq:eq_2nd} are closely related to the locally homogeneous elastic deformation of a crystal, and a similar discussion can be found in Ref.~\cite{born1954}.
Suppose a sample whose size is smaller than the wavelength of a long acoustic wave.
If one stimulates such a long wave in the sample, it will result in a homogeneous deformation with a uniform stress field and the crystal lattice will remain perfect.
In this situation, the deformation can be obtained as the atomic displacements owing to the zeroth-order waves $U^{(0)}_\alpha(\boldsymbol{\tau})=\overline{U}_\alpha {\rm e}^{ i \mathbf{q}\cdot\boldsymbol{\tau}}$~\cite{born1954} with its derivative to position as the deformation parameter:
\begin{equation}\label{eq:deform_0th}
\tilde{\varepsilon}_{\alpha\beta}=\frac{\partial U^{(0)}_\alpha(\boldsymbol{\tau})}{\partial \tau_\beta}= i  q_{\beta} \overline{U}_\alpha {\rm e}^{ i \mathbf{q}\cdot\boldsymbol{\tau}}.
\end{equation}
Since we focus on a long acoustic wave ($\mathbf{q\to 0}$), the exponential can be regarded as a constant and so do $\tilde{\varepsilon}_{\alpha\beta}$, which indicates that the deformation is indeed homogeneous within the considered situation of sample.
Moreover, by substituting Eqs.~\eqref{eq:D0}, \eqref{eq:D1}, and \eqref{eq:deform_0th} into Eq.~\eqref{eq:eq_1st_with_0th}, the first-order equation can be rewritten as
\begin{equation}\label{eq:deform_1st}
\sum_{l}\Phi^{l}_{\kappa\alpha,\kappa'\beta} U^{(1)}_{\kappa'\beta}(\boldsymbol{\tau})= \sum_{l\kappa'}\Phi^{l}_{\kappa\alpha,\kappa'\beta}\tilde{\varepsilon}_{\beta\gamma}\tau^{l}_{\kappa\kappa'\gamma},
\end{equation}
where $ U^{(1)}_{\kappa'\beta}(\boldsymbol{\tau})= i U^{(1)}_{\kappa'\beta}(\mathbf{q}) {\rm e}^{ i \mathbf{q}\cdot\boldsymbol{\tau}}$ is the first-order wave.
Particularly, the right-hand side of Eq.~\eqref{eq:deform_1st} represents the force acting on the $\kappa$th atom in the unit cell due to the homogeneous strain on the crystal given by Eq.~\eqref{eq:deform_0th}.
Such interatomic forces resulted from the homogeneous elastic deformation create the first-order waves $ U^{(1)}_{\kappa\alpha}(\boldsymbol{\tau})$, which further displace atoms to yield the internal strain that compensates the external strain fields through the left-hand side of Eq.~\eqref{eq:deform_1st}, ultimately reaching a balanced strained state.
Moreover, in order to further analyze the second-order waves $ U^{(2)}_{\kappa\alpha}(\boldsymbol{\tau})$, which is closely related to the strain gradient field
\begin{equation}\label{eq:strain_field_deform}
\nabla_\gamma\tilde{\varepsilon}_{\alpha\beta}=\frac{\partial^2 U^{(0)}_\alpha(\boldsymbol{\tau})}{\partial \tau_\beta\partial \tau_\gamma}=-q_\beta q_\gamma \overline{U}_\alpha {\rm e}^{ i \mathbf{q}\cdot\boldsymbol{\tau}},
\end{equation}
one can rewrite the second-order long-wavelength equation by substituting Eqs.~\eqref{eq:flexo_force_response_mass} and \eqref{eq:strain_field_deform} into Eq.~\eqref{eq:eq_2nd_long}:
\begin{equation}\label{eq:deform_2nd}
\sum_{l}\Phi^{l}_{\kappa\alpha,\kappa'\beta} U^{(2)}_{\kappa'\beta}(\boldsymbol{\tau})=\hat{T}^{\kappa}_{\alpha\beta,\gamma\delta}\nabla_\delta\tilde{\varepsilon}_{\beta\gamma},
\end{equation}
with $ U^{(2)}_{\kappa'\beta}(\boldsymbol{\tau})= U^{(2)}_{\kappa'\beta}(\mathbf{q}) {\rm e}^{ i \mathbf{q}\cdot\boldsymbol{\tau}}$ defined as the second-order waves.
It is easy to notice that the above equation also has the unit of forces, which corresponds to the internal force response due to the induced strain gradient field.
Particularly, $\hat{T}^{\kappa}_{\alpha\beta,\gamma\delta}$ in Eq.~\eqref{eq:flexo_force_response_mass} can be decomposed into the clamped-ion and lattice-mediated contributions to the internal force response due to the strain gradient and strain fields created by the zeroth-order and first-order waves, respectively.
In the case of a long-wavelength deformation, the induced strain gradient field can be also assumed to be homogeneous, indicating the strain field varies linearly throughout the sample material.
As can be seem from Eq.~\eqref{eq:deform_2nd}, the atomic displacements perturbed by such a constant strain gradient field generate the second-order waves $ U^{(2)}_{\kappa\alpha}(\boldsymbol{\tau})$.
When it comes to the bending waves in 2D crystals, one needs to further solve the third- and fourth-order long-wavelength equations, as shown in Appendix~\ref{solvability}.

In a nutshell, the zeroth-order waves first subject the whole crystal to a homogeneous deformation and generate the external strain field.
This leads to the creation of the first-order waves through a rigid displacement of all atoms in the crystal, and the induced forces acting on the atoms due to the external stress must be balanced by the forces of the internal stress arising from the first-order waves, as shown in Eq.~\eqref{eq:deform_1st}.
In the same manner, the variation of the strain field from the zeroth-order waves and the atomic displacements from the first-order waves together generate the second-order waves as described by Eq.~\eqref{eq:deform_2nd}, and the induced forces are further compensated by the atomic displacements from the second-order waves.
The dispersions of longitudinal and transverse elastic waves are then given by the solvability condition of second-order Eq.~\eqref{eq:eq_acoustic_wave}.
Therefore, the long-wavelength equations for acoustic waves are intimately linked to the homogeneous elastic deformation, and can be utilized to determine the elastic tensor of crystals.

\section{Cluster expansion model for multipole and dielectric tensors}\label{multipole_cluster_expansion}

The dynamical multipole and dielectric tensors of crystalline solids should respect the fundamental symmetry relations indicated by the underlying space group symmetry as well as the derivative commutativity.
To explicitly consider the symmetry requirements on multipole and dielectric tensors, one can use the cluster expansion model to represent the Taylor expansions of charge density response and dielectric screening function in Eqs.~\eqref{eq:charge_density} and \eqref{eq:screening_func} as
\begin{align}
\overline{\rho}_{\boldsymbol{\kappa}}(\mathbf{q}) &= \frac{e}{\Omega} \frac{(- i )^{|\alpha|}}{\alpha!}\mathcal{Q}_{\boldsymbol{\kappa}}^{I}(\alpha)q^{}_{I}, \label{eq:charge_density_cluster} \\
\xi(\mathbf{q}) &= \epsilon^{}_{I}q^{}_I, \label{eq:epsilon_cluster}
\end{align}
where $\boldsymbol{\kappa}\equiv\{\kappa, i\}$ is a composite index for both the atom site $\kappa$ and the corresponding Cartesian direction $i$, $\mathcal{Q}_{\boldsymbol{\kappa}}^{I}(\alpha)$ is a flattened vector of the generalized multipole tensor of each cluster used in the expansion, $q^{}_I\equiv q^{}_{i_1}q^{}_{i_2}\cdots q^{}_{i_n}$ is a combined wave vector, and we have used the multi-index notation introduced in Ref.~\cite{zhou2019} for cluster expansion: $\alpha$ is the so-called cluster comprised of $n$ lattice sites $\{ \kappa_1, \kappa_2,\ldots,\kappa_n\}$ of the primitive unit cell and $I\equiv\{i_1,i_2,\ldots,i_n\}$ is a collection of Cartesian components.
The order of a cluster is also the number of atom sites within it.
For each cluster $\alpha$, its absolute value and factorial are defined as
\begin{align}
|\alpha|&\equiv\sum_\kappa\alpha_\kappa, \\
\alpha!&\equiv\prod_\kappa\alpha_\kappa!,
\end{align}
where $\alpha_\kappa$ is the number of the atom site $\kappa$ in the cluster.
It should be noted that a cluster can contain duplicate atom sites, and such a cluster is called an \emph{improper} cluster, otherwise being a \emph{proper} cluster.
As can be seen from the left-hand side of Eq.~\eqref{eq:charge_density_cluster}, the cluster expansion for charge density response to a phonon perturbation only contains a single atom site $\kappa$, and such an expansion is thus just an on-site cluster expansion with only the improper clusters at different orders, i.e. the cluster $\alpha$ comprising only the repeated atom sites $\kappa$ in the charge density response.
For a given atom site $\kappa$, we should also stress that the cluster $\alpha$ only defines the superscript components $I$ of a multipole tensor, which actually represent the electric field response, and the order of multipole tensor is hence $n+1$ given an $n$-order cluster.
Besides, for the expansion of dielectric screening function in Eq.~\eqref{eq:epsilon_cluster}, it is actually not a cluster expansion since no atom site is involved, and we also use the introduced multi-index notation for simplicity.

The symmetry-independent irreducible components of the multipole tensors at each order can be obtained by analyzing the symmetries of the corresponding clusters.
The symmetries of each cluster consist of two parts: space-group symmetry and permutation symmetry, where the permutation only exists for an improper cluster as the permutation of a proper cluster is included already in the space-group symmetry.
In addition, a sum rule should be further imposed on the lowest-order clusters with the non-vanishing tensor components to fulfill the charge neutrality conditions of the system.

\begin{table}[t!]
\centering
\caption{The dynamical Born effective charge $Z$ [$e$], quadrupole $Q$ [$e~\mathrm{bohr}$] and octupole $O$ [$e~\mathrm{bohr}^2$], clamped-ion dielectric permittivity $\boldsymbol{\epsilon}^{\infty}$ and dielectric dispersion tensors $\boldsymbol{\epsilon}^{(4)}$ [bohr$^2$] of BaTiO$_3$ calculated with the same lattice structure and computational parameters as in Ref.~\cite{royo2020}, via a least-squares fit to the Taylor expansion of unscreened charge density response in Eq.~\eqref{eq:charge_density} and dielectric screening function in Eq.~\eqref{eq:screening_func}.
In the parentheses, the components of Born effective charge and dielectric permittivity from a DFPT calculation and quadrupole, octupole and dielectric dispersion tensors from Ref.~\citep{royo2020} are listed out for comparison.
Only symmetry-independent components are shown.
}
\label{table:multipole_Royo_BTO}
\begin{ruledtabular}
\footnotesize
\begin{tabular}{lccc}
 & Ba & Ti & O \\
\hline
$Z_{x}^{x}$ & 2.786 (2.786)  & 6.500 (6.500) & -2.043 (-2.043) \\
$Z_{x}^{y}$ & -0.106 (-0.106)  & -0.226 (-0.226) & -0.006 (-0.006) \\
$Z_{x}^{z}$ & -0.106 (-0.106)  & -0.226 (-0.226) & 0.168 (0.168) \\
$Z_{z}^{x}$ & -0.106 (-0.106) & -0.226 (-0.226) & 0.074 (0.074) \\
$Z_{z}^{z}$ & 2.786 (2.786) & 6.500 (6.500) & -5.200 (-5.200) \\
\hline
$Q_{x}^{xx}$ & -1.060 (-1.060) & 1.561 (1.560) & -0.843 (-0.843) \\
$Q_{x}^{xy}$ & 0.468 (0.468) & -0.316 (-0.317) & 0.020 (0.020) \\
$Q_{x}^{xz}$ & 0.468 (0.468) & -0.316 (-0.317) & -0.120 (-0.120) \\
$Q_{x}^{yy}$ & -0.102 (-0.102) & 1.565 (1.566) & -0.069 (-0.069) \\
$Q_{x}^{yz}$ & 0.004 (0.004) & -0.060  (-0.059)& -0.044 (-0.044) \\
$Q_{x}^{zz}$ & -0.102 (-0.102) & 1.565 (1.566) & 1.166 (1.167) \\
$Q_{z}^{xx}$ & -0.102 (-0.102) & 1.565 (1.566) & -1.559 (-1.561) \\
$Q_{z}^{xy}$ & 0.004 (0.004) & -0.060 (-0.059) & 0.011 (0.010) \\
$Q_{z}^{xz}$ & 0.468 (0.468) & -0.316 (-0.317) & 1.419 (1.420) \\
$Q_{z}^{zz}$ & -1.060 (-1.060) & 1.561 (1.560) & -2.176 (-2.175) \\
\hline
$O_{x}^{xxx}$ & -283.510 (-285.161) & 91.416 (94.437) & -106.071 (-106.134) \\
$O_{x}^{xxy}$ & 3.739 (3.699) & -2.481 (-2.380) & 0.858 (0.849) \\
$O_{x}^{xxz}$ & 3.739 (3.699) & -2.481 (-2.380) & 2.278 (2.294) \\
$O_{x}^{xyy}$ & -84.392 (-84.574) & -39.283 (-38.987) & -35.358 (-35.231) \\
$O_{x}^{xyz}$ & 1.284 (1.273) & 1.661 (1.720) & 0.146 (0.165) \\
$O_{x}^{xzz}$ & -84.392 (-84.574) & -39.283 (-38.987) & -38.334 (-38.317) \\
$O_{x}^{yyy}$ & -0.668 (-0.650) & -8.291 (-8.638) & -0.036 (-0.106) \\
$O_{x}^{yyz}$ & -0.113 (0.006) & -0.720 (-0.848) & 0.715 (0.718) \\
$O_{x}^{yzz}$ & -0.113 (0.006) & -0.720 (-0.848) & -0.001 (0.004) \\
$O_{x}^{zzz}$ & -0.668 (-0.650) & -8.291 (-8.638) & 4.244 (4.220) \\
$O_{z}^{xxx}$ & -0.668 (-0.650) & -8.291 (-8.638) & 7.943 (8.095) \\
$O_{z}^{xxy}$ & -0.113 (0.006) & -0.720 (-0.848) & 0.587 (0.755) \\
$O_{z}^{xxz}$ & -84.392 (-84.574) & -39.283 (-38.987) & -69.926 (-70.265) \\
$O_{z}^{xyz}$ & 1.284 (1.273) & 1.661 (1.720) & 0.690 (0.642) \\
$O_{z}^{xzz}$ & 3.739 (3.699) & -2.481 (-2.380) & 5.943 (5.982) \\
$O_{z}^{zzz}$ & -283.510 (-285.161) & 91.416 (94.437) & -325.458 (-328.941) \\
\hline
$\epsilon^\infty_{xx}$ & \multicolumn{3}{c}{6.083 (6.083)} \\
$\epsilon^\infty_{xy}$ & \multicolumn{3}{c}{-0.111 (-0.111)} \\
$\epsilon^{(4)}_{xxxx}$ & \multicolumn{3}{c}{-29.705 (-30.159)} \\
$\epsilon^{(4)}_{xxxy}$ & \multicolumn{3}{c}{0.567 (0.561)} \\
$\epsilon^{(4)}_{xxyy}$ & \multicolumn{3}{c}{-7.066 (-7.104)} \\
$\epsilon^{(4)}_{xxyz}$ & \multicolumn{3}{c}{0.075 (0.088)} \\
\end{tabular}
\end{ruledtabular}
\end{table}

\emph{Space-group symmetry.} The charge density response $\overline{\rho}_{\boldsymbol{\kappa}}(\mathbf{q})$ to a phonon perturbation transforms covariantly in its atom site $\kappa$ under the operations in the crystal space group $\mathbb{G}_S$.
As a result, for each symmetry operation $\hat{s}\in\mathbb{G}_S$, one can have the following transformation:
\begin{equation}\label{eq:multipole_sym_transform}
\mathcal{Q}_{\boldsymbol{\kappa}'}^{I}(\hat{s}\alpha)=\mathfrak{R}_{IJ}(\hat{s})\mathcal{Q}_{\boldsymbol{\kappa}}^{J}(\alpha),
\end{equation}
where $\mathfrak{R}_{IJ}(\hat{s})=\mathfrak{r}_{i_1 j_1}\cdots\mathfrak{r}_{i_{n+1} j_{n+1}}$ is a $3^{n+1}\times 3^{n+1}$ matrix, which corresponds to an operation $\hat{s}$ consisting of an orthogonal transformation with a 3$\times$3 rotation matrix $\mathfrak{r}$ followed by a translation $\mathfrak{t}$, i.e. $\hat{s}\boldsymbol{\tau}_{\kappa}=\mathfrak{r}\cdot\boldsymbol{\tau}_\kappa+\mathfrak{t}$, and $n$ is the order of cluster $\alpha$.
It is important to mention that the order of atom sites in a cluster may be changed by the symmetry operation $\hat{s}$.
However, the clusters for representing multipole tensors are all improper and the order of atom sites within the cluster thus becomes trivial.
Furthermore, with the help of space-group symmetry, all the clusters of lattice sites used in the expansion can be grouped into the so-called \textit{orbits} which only include the symmetry-equivalent clusters of $\alpha$ under the operations of space group, denoted as $\mathbb{G}_{S\alpha}\equiv\{ \hat{s}\alpha|\hat{s}\in\mathbb{G}_S \}$.
In other words, the symmetry operations within a space group can be divided into two subgroups: one for mapping a cluster $\alpha$ to all possible clusters in its orbit $\mathbb{G}_{S\alpha}$, and the other for building symmetry constraints on the cluster $\alpha$ itself.
The latter is further defined as the \textit{isotropy group} $\mathbb{G}_\alpha$ of a cluster $\alpha$, which represents a set of transformations making $\alpha$ invariant: $\mathbb{G}_\alpha\equiv\{ \hat{s}\in\mathbb{G}_S|\hat{s}\alpha=\alpha \}$.
Therefore, the cluster expansion of charge density response can be recast using only the representative clusters of the symmetry-distinct atom sites as
\begin{equation}\label{eq:charge_density_cluster_reduced}
\overline{\rho}_{\boldsymbol{\kappa}'}(\mathbf{q}) = \frac{e}{\Omega}\frac{(- i )^{|\alpha|}}{\alpha!}\sum_{\hat{s}\alpha\in\mathbb{G}_{S\alpha}}\mathfrak{R}_{IJ}(\hat{s})\mathcal{Q}_{\boldsymbol{\kappa}}^{J}(\alpha)q^{}_{I}.
\end{equation}
Then, the symmetry constraints from the space group of a crystal on the multipole tensors can be built using those operations in the isotropy group of the corresponding clusters as
\begin{equation}\label{eq:sym_const_isotropy}
\mathcal{Q}_{\boldsymbol{\kappa}}^{I}(\alpha)=\mathfrak{R}_{IJ}(\hat{s})\mathcal{Q}_{\boldsymbol{\kappa}}^{J}(\alpha),\qquad\forall\hat{s}\in\mathbb{G}_{\alpha}.
\end{equation}

\emph{Permutation symmetry.} Since the multipole tensors are defined as the partial derivatives of the charge density response of each atom to the phonon momentum, they should be commutative with respect to the order of doing the partial derivatives, which reads
\begin{equation}\label{eq:multipole_sym_permute}
\mathcal{Q}_{\boldsymbol{\kappa}}^{I}(\alpha)=\mathfrak{P}_{IJ}(\alpha)\mathcal{Q}_{\boldsymbol{\kappa}}^{J}(\alpha),
\end{equation}
where $\mathfrak{P}_{IJ}(\alpha)$ is a $3^{n+1}\times 3^{n+1}$ permutation matrix for a $n$-order cluster $\alpha$.
One should be careful that the permutation is only allowed for the superscripted Cartesian components $I$, but not the one in the composite index $\boldsymbol{\kappa}$ which denotes the direction of phonon perturbation.

\emph{Charge neutrality conditions.} The summation of the effective charge of each atom in the unit cell should vanishing~\cite{gonze1997,giannozzi1991}.
Such a sum rule is imposed on the second-rank Born effective charge tensor as $\sum_{\kappa}Z_{\kappa\alpha}^{\beta}=0$.
In the case of systems with vanishing Born effective charges, the charge neutrality condition should be fulfilled on the first non-vanishing multipole tensors (i.e. quadrupole tensors) for infrared-inactive solids~\cite{vogl1976} as $\sum_{\kappa}Q_{\kappa\alpha}^{\beta\gamma}=0$.
For dielectric materials, the Born effective charge and quadrupole tensors will never vanish together.
Finally, using only the representative clusters, the above two charge neutrality conditions can be merged into a single formula as
\begin{equation}\label{eq:charge_asr}
\sum_{\alpha\in\mathbb{A}/\mathbb{G}_S} \sum_{\hat{s}\alpha\in\mathbb{G}_{S\alpha}} \mathfrak{R}_{IJ}(\hat{s})\mathcal{Q}_{\boldsymbol{\kappa}}^{J}(\alpha)=0,
\end{equation}
where $\mathbb{A}$ stands for a set of all possible clusters, and $\mathbb{A}/\mathbb{G}_S$ is the \textit{orbit space} that the representative clusters live in.
Unlike the DFPT or finite-difference scheme adopted in most \textit{ab initio} codes where the sum rule of charge neutrality conditions is fulfilled as a post-processing step~\cite{gonze1997,giannozzi1991}, we explicitly consider it as an additional set of linear constraints when determining the number of irreducible parameters of multipole tensors.
We expect this can further improve the accuracy of our fitted multipole tensors which will automatically satisfy the charge neutrality conditions.

In the same manner, the symmetry-independent components of dielectric tensors can be found by imposing the space-group (or point-group) and permutation symmetries:
\begin{align}
\epsilon_{I} &= \mathfrak{R}_{IJ}(\hat{s}) \epsilon_{J},\quad\forall\hat{s}\in\mathbb{G}_S, \\
\epsilon_{I} &= \mathfrak{P}_{IJ} \epsilon_{J},
\end{align}
where $\mathfrak{P}_{IJ}$ now is simply a $3^n\times 3^n$ permutation matrix independent of clusters for each even $n$-order dielectric tensor.
Once all symmetry constraints on the multipole and dielectric tensors are built, we calculate the corresponding null space $\mathbb{N}$ by performing a maximal-pivot Gauss-Jordan elimination, which  is used as a basis set to expand the irreducible components.
Finally, the cluster expansion of multipole in Eq.~\eqref{eq:charge_density_cluster_reduced} can be rewritten into a matrix-vector form as a linear regression problem:
\begin{equation}\label{eq:multipole_linear_equation}
\overline{\boldsymbol{\rho}}=\mathbb{B}\cdot\boldsymbol{\mathcal{Q}}=\mathbb{B}\cdot\mathbb{N}\cdot\boldsymbol{\mathcal{Q}}'=\mathbb{B}'\cdot\boldsymbol{\mathcal{Q}}',
\end{equation}
where $\mathbb{B}'=\mathbb{B}\cdot\mathbb{N}$ is the correlation matrix, and $\boldsymbol{\mathcal{Q}}'=\mathbb{N}^{-1}\cdot\boldsymbol{\mathcal{Q}}$ is a vector of unknown parameters to be solved which denotes the irreducible parameters of all multipole tensors.
Explicitly, the matrix elements of $\mathbb{B}$ take the following form
\begin{equation}
\mathbb{B}(\boldsymbol{\kappa}',\alpha I) =  \frac{e}{\Omega} \frac{(- i )^{|\alpha|}}{\alpha!} \sum_{\hat{s}\alpha\in\mathbb{G}_{S\alpha}} \mathfrak{R}_{JI}(\hat{s})q^{}_{J},
\end{equation}
where the composite row index $\boldsymbol{\kappa}'$ is contained in the cluster $\hat{s}\alpha$ on the right-hand side, i.e. the first atom of each cluster.
Also, the null space $\mathbb{N}$ has the shape of $N_{\boldsymbol{\mathcal{Q}}}\times N_{\boldsymbol{\mathcal{Q}}'}$, where $N_{\boldsymbol{\mathcal{Q}}}$ and $N_{\boldsymbol{\mathcal{Q}}'}$ are the number of total and irreducible parameters of multipole tensors, respectively.
Then, an efficient least-squares regression can be adopted to solve the linear problem in Eq.~\eqref{eq:multipole_linear_equation}, and the solvability condition is that $\mathbb{B}'$ needs to be full-rank, i.e.  an overdetermined linear system.
A simple estimation of the minimum number of configurations of phonon perturbation can be done via $N^\mathrm{min}_\mathbf{q}=N_{\boldsymbol{\mathcal{Q}}'}/(6n_\mathrm{a})$, where the factor of 6 comes from the fact that each atom has three Cartesian components and the charge density response is a complex number.
Even though we are using a set of random small $\mathbf{q}\to\mathbf{0}$ perturbations, it is still possible that some rows of $\mathbb{B}'$ are linearly correlated, and hence we recommend to generate the number of $\mathbf{q}$ perturbations that is of 2 to 3 folds of $N^\mathrm{min}_\mathbf{q}$, in order to obtain the reliable and accurate solution of multipole tensors.
Last but not least, the linear regression problem of dielectric screening function can be regarded as a special case of charge density response (i.e. for a single atom along one specific direction), and here we are not going to discuss it further.

\begin{table}[t!]
\centering
\caption{The dynamical Born effective charge $Z$ [$e$], quadrupole $Q$ [$e~\mathrm{bohr}$], octupole $O$ [$e~\mathrm{bohr}^2$], clamped-ion dielectric permittivity $\boldsymbol{\epsilon}^{\infty}$ and dielectric dispersion tensors $\boldsymbol{\epsilon}^{(4)}$ [bohr$^2$] of BaTiO$_3$, calculated by a least-squares fit to the Taylor expansion of unscreened (or transverse) charge density response and dielectric screening function in Eqs.~\eqref{eq:charge_density} and \eqref{eq:screening_func}, respectively.
We report without parentheses the results obtained with the PC prescription, while the corresponding LOC ones are provided in parentheses.
Only symmetry-independent components are shown.
Evidently, the relevant differences that cannot be attributed to numerical error arise only in the case of octupoles.
}
\label{table:multipole_BTO}
\begin{ruledtabular}
\footnotesize
\begin{tabular}{lccc}
 & Ba & Ti & O \\
\hline
$Z_{x}^{x}$ & 2.785 (2.785)  & 6.836 (6.836) & -2.056 (-2.056) \\
$Z_{y}^{y}$ & 2.785 (2.785)  & 6.836 (6.836) & -4.358 (-4.358) \\
$Z_{z}^{z}$ & 2.759 (2.759)  & 6.249 (6.249) & -3.003 (-3.003) \\
$Z_{y}^{z}$ & --  & -- & -1.530 (-1.530) \\
$Z_{z}^{x}$ & --  & -- & -1.416 (-1.416) \\
\hline
$Q_{x}^{xy}$ & 0.716 (0.716) & -0.242 (-0.242) & 0.258 (0.258) \\
$Q_{x}^{xz}$ & -0.261 (-0.262) & -0.197 (-0.196) & -0.478 (-0.477) \\
$Q_{y}^{xx}$ & 0.716 (0.716) & -0.242 (-0.242) & -0.893 (-0.892) \\
$Q_{y}^{yy}$ & -0.716 (-0.716) & 0.242 (0.242) & -3.530 (-3.529) \\
$Q_{y}^{yz}$ & -0.261 (-0.262) & -0.197 (-0.196) & -0.217 (-0.217) \\
$Q_{y}^{zz}$ & -- & -- & 0.079 (0.079) \\
$Q_{z}^{xx}$ & -1.207 (-1.206) & 3.078 (3.074) & -1.423 (-1.424) \\
$Q_{z}^{yy}$ & -1.207 (-1.206) & 3.078 (3.074) & -1.343 (-1.343) \\
$Q_{z}^{yz}$ & -- & -- & 0.994 (0.994) \\
$Q_{z}^{zz}$ & 0.315 (0.315) & 1.771 (1.771) & 0.039 (0.040) \\
\hline
$O_{x}^{xxx}$ & -278.832 (-201.308) & 1.749 (46.030) & -108.226 (-89.640) \\
$O_{x}^{xyz}$ & 7.469 (7.444) & -52.598 (-52.648) & -1.146 (-1.170) \\
$O_{x}^{yyx}$ & -92.944 (-67.103) & 0.583 (15.343) & -39.549 (-33.355) \\
$O_{x}^{zzx}$ & -93.259 (-68.748) & 35.030 (49.666) & -36.715 (-30.661) \\
$O_{y}^{xxy}$ & -92.944 (-67.103) & 0.583 (15.343) & -64.445 (-58.171) \\
$O_{y}^{xxz}$ & 7.469 (7.444) & -52.598 (-52.648) & -14.674 (-14.725) \\
$O_{y}^{yyy}$ & -278.832 (-201.308) & 1.749 (46.030) & -248.156 (-229.579) \\
$O_{y}^{yyz}$ & -7.469 (-7.444) & 52.598 (-52.648) & -54.832 (-54.874) \\
$O_{y}^{zzy}$ & -- & -- & -85.060 (-79.210) \\
$O_{y}^{zzz}$ & -- & -- & -63.598 (-64.204) \\
$O_{z}^{xxy}$ & 8.704 (8.707) & -52.083 (-52.280) & -17.457 (-17.436) \\
$O_{z}^{xxz}$ & -98.099 (-72.369) & 27.880 (42.800) & -46.312 (-40.126) \\
$O_{z}^{yyz}$ & -98.099 (-72.369) & 27.880 (42.800) & -73.016 (-66.854) \\
$O_{z}^{yyy}$ & -8.704 (8.707) & 52.083 (52.280) & -89.099 (-89.051) \\
$O_{z}^{zzy}$ & -- & -- & -36.250 (-36.401) \\
$O_{z}^{zzz}$ & -261.776 (-188.048) & -53.260 (-8.289) & -151.880 (-134.310) \\
\hline
$\epsilon^\infty_{xx}$ & \multicolumn{3}{c}{6.295 (6.295)} \\
$\epsilon^\infty_{zz}$ & \multicolumn{3}{c}{6.001 (6.001)} \\
$\epsilon^{(4)}_{xxxx}$ & \multicolumn{3}{c}{-27.855 (-27.859)} \\
$\epsilon^{(4)}_{zzzz}$ & \multicolumn{3}{c}{-24.082 (-24.081)} \\
$\epsilon^{(4)}_{yyyz}$ & \multicolumn{3}{c}{-2.142 (-2.142)} \\
$\epsilon^{(4)}_{xxzz}$ & \multicolumn{3}{c}{-10.183 (-10.181)} \\
\end{tabular}
\end{ruledtabular}
\end{table}

To validate our approach for extracting multipole and dielectric tensors from a least-squares linear regression of the cluster expansion model proposed here, we take the rhombohedral BaTiO$_3$ as an example since it is the only material for which octupole interactions were reported in the literature~\cite{royo2020}.
For this comparison, we use their structure for BaTiO$_3$ and keep every computational parameter the same as reported in Ref.~\cite{royo2020}.
A set of 20 $\mathbf{q}$ perturbations is randomly sampled over a small region around the zone center in arbitrary directions, with the magnitude of $\mathbf{q}$ ranging from 0.01 to 0.05 in reciprocal lattice unit.
Then, the calculations of charge density response and dielectric function for those $\mathbf{q}$ points are then performed based on a recent implementation within the DFPT framework~\cite{macheda2022,macheda2024}, with the unscreened charge response obtained by switching off the macroscopic potential.
Our results for the multipole and dielectric tensors of BaTiO$_3$ are summarized in Table~\ref{table:multipole_Royo_BTO} and are in excellent agreement with Ref.~\cite{royo2020}.
Also, we note that the choice of the rhombohedral cell in Ref.~\cite{royo2020} is not standard and gives rise to different symmetry-independent components, compared to the results shown in Table~\ref{table:multipole_BTO} (i.e. the one used in our elastic constant calculations).
In general, when any linear regression technique is applied to get the expansion coefficient of a Taylor series, it is difficult to remove the crosstalk among different orders, even if the system-dependent magnitude of correlation matrix is carefully chosen.
There is a \textit{renormalization effect} on the fitted quadrupole and octupole tensors from the higher-order terms that are not included in the expansion.
Since a higher even order can only affect the lower even order (similarly for odd order), the obtained quadrupoles and octupoles should inherently contain the influence from hexadecapoles ($n=4$) and triacontadipoles ($n=5$), respectively.
In order to obtain the unscreened and exact quadrupole and octupole tensors without those higher-order effects, we recommend to include those next-leading terms in the expansion, although this increases the number of DFPT calculations.
Nevertheless, the exact values of multipole tensors are not necessary for accurate phonon dispersion or short-circuit elastic constant calculations.
In Table~\ref{table:multipole_Royo_BTO}, we have performed the multipole expansion up to the hexadecapole term to ensure the obtained quadrupole tensor is unscreened and exact, which can be readily compared with those values from a direct linear response calculation in Ref.~\cite{royo2020}.

\begin{table*}[t!]
\centering
\caption{The contribution of dipole-octupole interactions to the elastic constants of BaTiO$_3$ at different levels of multipolar treatment based on the real-space IFCs of a $14\times14\times14$ supercell.
For each component of the elastic tensor, left and right columns correspond to the results using the multipole and dielectric tensors fitted from the response functions with the PC and LOC representations of ionic charges, respectively.
It is evident that the PC and LOC prescriptions give rise to the equal contribution to the elastic constants.
}
\label{table:elastic_compare}
\begin{ruledtabular}
\begin{tabular}{ccccccccccccc}
\multirow{2}{*}{Levels} & \multicolumn{2}{c}{$C_{11}$} & \multicolumn{2}{c}{$C_{12}$} & \multicolumn{2}{c}{$C_{13}$} & \multicolumn{2}{c}{$C_{14}$} & \multicolumn{2}{c}{$C_{33}$} & \multicolumn{2}{c}{$C_{44}$} \\
 & \multicolumn{2}{c}{$\text{PC} \qquad \text{LOC}$} & \multicolumn{2}{c}{$\text{PC} \qquad \text{LOC}$} & \multicolumn{2}{c}{$\text{PC} \qquad \text{LOC}$} & \multicolumn{2}{c}{$\text{PC} \qquad \text{LOC}$} & \multicolumn{2}{c}{$\text{PC} \qquad \text{LOC}$} & \multicolumn{2}{c}{$\text{PC} \qquad \text{LOC}$} \\
\hline
DD & 242.90 & 242.90 & 81.37 & 81.37 & 48.29 & 48.29 & 70.87 & 70.87 & 293.84 & 293.84 & 25.43 & 25.43 \\
DD+DO & 251.16 & 251.23 & 94.66 & 94.76 & 41.38 & 41.48 & 70.45 & 70.47 & 284.21 & 284.32 & 23.80 & 23.76 \\
DO contribution & 8.26 & 8.33 & 13.29 & 13.39 & -6.91 & -6.81 & -0.42 & -0.40 & -9.63 & -9.52 & -1.63 & -1.67 \\
\hline
DD+DQ & 295.42 & 295.40 & 86.02 & 86.01 & 60.53 & 60.55 & 44.32 & 44.31 & 299.67 & 299.70 & 56.87 & 56.88 \\
DD+DQ+DO & 303.73 & 303.79 & 99.27 & 99.37 & 53.64 & 53.75 & 43.87 & 43.89 & 290.06 & 290.19 & 55.26 & 55.24 \\
DO contribution & 8.31 & 8.39 & 13.25 & 13.36 & -6.89 & -6.80 & -0.45 & -0.42 & -9.61 & -9.51 & -1.61 & -1.64 \\
\hline
DD+DQ+QQ & 294.68 & 294.67 & 86.31 & 86.31 & 60.79 & 60.80 & 44.72 & 44.72 & 299.96 & 299.98 & 56.67 & 56.69 \\
DD+DQ+QQ+DO & 303.00 & 303.06 & 99.57 & 99.67 & 53.89 & 54.00 & 44.28 & 44.30 & 290.34 & 290.48 & 55.06 & 55.04 \\
DO contribution & 8.32 & 8.39 & 13.26 & 13.36 & -6.90 & -6.80 & -0.44 & -0.43 & -9.62 & -9.50 & -1.61 & -1.65 \\
\hline
DD+DQ+QQ+D$\epsilon$D & 292.31 & 292.30 & 80.44 & 80.44 & 63.38 & 63.39 & 44.59 & 44.58 & 303.27 & 303.29 & 56.45 & 56.46 \\
DD+DQ+QQ+DO+D$\epsilon$D & 300.63 & 300.71 & 93.72 & 93.81 & 56.44 & 56.59 & 44.14 & 44.14 & 293.61 & 293.76 & 54.86 & 54.84 \\
DO contribution & 8.32 & 8.41 & 13.28 & 13.37 & -6.94 & -6.80 & -0.45 & -0.44 & -9.66 & -9.53 & -1.59 & -1.62 \\
\end{tabular}
\end{ruledtabular}
\end{table*}

\section{Influence of the choice of ionic charge representation in pseudopotentials}\label{ionic_pot}
Within an all-electron formulation, the displacement of an atom $\kappa$ corresponds to the movement of a point charge, with a charge equal to the atomic number $\mathcal{Z}_\kappa$.
When dealing with pseudopotential (PP) calculations, the choice of the representation of the ionic charge is not unique.
As a matter of fact, one possible choice is to describe the motion of the ion as a point charge with the pseudo charge $\mathcal{Z}^{\text{PP}}_\kappa$, where $\mathcal{Z}^{\text{PP}}_\kappa$ is determined by how many electrons have been used as \textit{valence} electrons in the construction of the pseudopotential, and we refer to this choice as ``PC'':
\begin{equation}
\rho^{\textrm{ion,PC}}_{\kappa\alpha}(\mathbf{q})=-i\frac{|e|}{\Omega}q_{\alpha}\mathcal{Z}^{\textrm{PP}}_{\kappa}.
\label{eq:rhoionPC}
\end{equation}
Another possibility, for example, is to describe the displacement of the ion as the motion of the pseudo charge cloud distribution associated to the local part of the pseudopotential, and we refer to this choice as ``LOC'':
\begin{equation}
\rho^{\textrm{ion,LOC}}_{\kappa\alpha}(\mathbf{q})=\frac{e}{v(\mathbf{q})} V^{\textrm{ion,LOC}}_{\kappa\alpha}(\mathbf{q}),
\label{eq:marchese}
\end{equation}
where $v(\mathbf{q})=4\pi e^2/|\mathbf{q}|^2$ and $V^{\textrm{ion,LOC}}_{\kappa\alpha}(\mathbf{q})$ is the local part of the pseudopotential.
Clearly, the PC and LOC representations of ionic charges are inequivalent choices, having an impact on the determination of the multipolar expansion of charge density response.
This problem has been discussed in Appendix~B3 of Ref.~\cite{macheda2024}, where Eqs.~\eqref{eq:rhoionPC} and \eqref{eq:marchese} are Eqs. (B29) and (B28), respectively. We report here, following the arguments of the cited reference, the proof that in semiconductors the choice of PC or LOC leads to different values for octupole tensors, while dipoles and quadrupoles should not be affected by the choice of the description of the ionic displacement.
%

According to Eq.~(B34) of Ref.~\cite{macheda2024}, the bare and screened total charge density responses are linked by the relationship
\begin{align}
\epsilon^{-1}_{\textrm{L,PP}}(\mathbf{q}) \overline{\rho}^{\textrm{PP}}_{\kappa\alpha}(\mathbf{q})=  \rho^{\textrm{PP}}_{\kappa\alpha}(\mathbf{q}),
\label{eq:epsm1pprhorhobar}
\end{align}
where $\epsilon^{-1}_{\textrm{L,PP}}(\mathbf{q})$ is the inverse dielectric function computed in presence of a pseudopotential, and the subscript ``$\textrm{L}$'' is used to distinguish it from the response function which relates the external to the full self-consistent potential.
We indicate with $\overline{\rho}^{\textrm{PP,PC}}_{\kappa\alpha}(\mathbf{q})$ and $\overline{\rho}^{\textrm{PP,LOC}}_{\kappa\alpha}(\mathbf{q})$ the unscreened total charge density variations, where the bare atomic displacement is added to the self-consistently determined induced density via Eqs.~\eqref{eq:rhoionPC} and \eqref{eq:marchese}, respectively, and the same notation is used for the screened responses.
For simplicity, we consider the case of a cubic semiconductor, where the following relation holds:
\begin{align}
\epsilon^{-1}_{\textrm{L,PP}}(\mathbf{q})=\epsilon^{\infty,-1}+\mathcal{O}(q^2),
\end{align}
and the demonstration here can be easily extended to any crystal symmetry.
The difference between the two choices of ionic charge representation can be then calculated as
\begin{align}
&\overline{\rho}^{\textrm{PP,PC}}_{\kappa\alpha}(\mathbf{q})-\overline{\rho}^{\textrm{PP,LOC}}_{\kappa\alpha}(\mathbf{q})=\frac{ \rho^{\textrm{PP,PC}}_{\kappa\alpha}(\mathbf{q})-\rho^{\textrm{PP,LOC}}_{\kappa\alpha}(\mathbf{q})}{\epsilon^{-1}_{\textrm{L,PP}}(\mathbf{q})} \nonumber \\
&=\Big[\rho^{\textrm{ion,PC}}_{\kappa\alpha}(\mathbf{q})-\rho^{\textrm{ion,LOC}}_{\kappa\alpha}(\mathbf{q})\Big] \left[\epsilon^{\infty,-1}+\mathcal{O}(q^2)\right] \nonumber \\
&=\frac{e}{v(\mathbf{q})}\Big[V^{\textrm{ion,PC}}_{\kappa\alpha}(\mathbf{q})- V^{\textrm{ion,LOC}}_{\kappa\alpha}(\mathbf{q})\Big]\left[\epsilon^{\infty,-1}+\mathcal{O}(q^2)\right],
\end{align}
where we have defined $\rho^{\textrm{ion,PC}}_{\kappa\alpha}(\mathbf{q})=e V^{\textrm{ion,PC}}_{\kappa\alpha}(\mathbf{q})/v(\mathbf{q})$.
As discussed in Appendix~C of Ref.~\cite{royo2019}, one can demonstrate
\begin{align}
V^{\textrm{ion,PC}}_{\kappa\alpha}(\mathbf{q})- V^{\textrm{ion,LOC}}_{\kappa\alpha}(\mathbf{q})=\mathcal{O}(q),
\end{align}
and we then have
\begin{align}
\overline{\rho}^{\textrm{PP,PC}}_{\kappa\alpha}(\mathbf{q})-\overline{\rho}^{\textrm{PP,LOC}}_{\kappa\alpha}(\mathbf{q})=\mathcal{O}{(q^3)},
\end{align}
meaning that differences between the Taylor expansions of the two densities starts to arise at the octupolar level.
The dependence of the values of the dynamical octupoles on the choice of the pseudopotential is also discussed in the context of flexoelectricity~\cite{hong2013,hong2011,stengel2013}.
We show the numerical evidence of the above theoretical argument in Table~\ref{table:multipole_BTO}, where the multipole expansion for BaTiO$_3$ with the crystal structure adopted in this work are presented.
It can be noticed that the dipole and quadrupole values in the PC and LOC prescriptions are equal up to numerical error, while the octupoles show large differences.
The choice implemented in this work is PC, which is consistent with the one in the \textsc{abinit}~\cite{royo2019,royo2021,gonze2016,gonze2020}, as demonstrated in Table~\ref{table:multipole_Royo_BTO}.

We now investigate if the PC and LOC prescriptions would lead to different values for the elastic constants.
To achieve our goal, we first report in Table~\ref{table:elastic_compare} the values of the elastic constants obtained for a given truncation level of the multipole expansion of the long-range dynamical matrix as in Eq.~\eqref{eq:Phi_expand}.
Immediately below, we report the DO contribution to a given truncation level, as the choice of PC or LOC results in different octupole values.
Since all the operations that we perform to obtain the elastic constants depend linearly on the terms appearing in Eq.~\eqref{eq:Phi_expand}, we expect the DO contribution to the elastic constants to be independent of the truncation level.
This is respected to a very high precision for both the PC and LOC calculations.
Importantly, we notice that the difference between the PC and LOC cases in the DO contribution to elastic constants is much lower than the contribution itself.
This is a strong hint that the choice between PC and LOC has low or no impact on the evaluation of elastic constants.

\section{Computational details}\label{comp_details}

First-principles calculations are performed using the \textsc{Quantum ESPRESSO} package~\cite{giannozzi2009,giannozzi2017}, with the norm-conserving pseudopotentials from \textsc{PseudoDojo}~\cite{van2018}.
We employ the Perdew, Burke, and Ernzerhof's (PBE)~\cite{perdew1996} parametrization of the generalized gradient approximation (GGA) for silicon, NaCl and graphene, and Perdew-Wang's parametrization of the local density approximation (LDA)~\cite{perdew1992} for BaTiO$_3$, to describe the exchange-correlation effect of electrons.
In the case of GaAs, the revision of PBE functional for solid (PBEsol)~\cite{perdew2008} is adopted to overcome the infamous problem of metallic-like screening with the standard PBE and have better lattice parameters compared to experimental values.
For silicon, NaCl, and GaAs, the electronic $\mathbf{k}$ grid is chosen to be 12$\times$12$\times$12, while a smaller 10$\times$10$\times$10 grid is used for BaTiO$_3$, and their plane-wave energy cutoffs are 40, 160, 90, and 100 Ry, respectively.
To simulate graphene as a 2D system, the Coulomb cutoff technique~\cite{sohier2017-2} is used to remove the long-range electrostatic interactions along the vacuum direction, using the Marzari-Vanderbilt smearing~\cite{marzari1999} with a smearing of 0.02 Ry, where the electronic $\mathbf{k}$ grid and plane-wave energy cutoff are set to 48$\times$48$\times$1 and 100 Ry, respectively.
For other 2D materials including \emph{h}-BN, MoS$_2$ and InSe, we use the same primitive cell structure and computational parameters as described in Ref.~\cite{ponce2023}.
All of structural optimizations are carried out based on the convergence thresholds as the pressure and forces smaller than $10^{-3}$ kbar and $10^{-4}$ Ry/bohr.
Then, the DFPT simulations are performed to calculate the dynamical matrices on a set of phonon grids, from which the IFCs used in this work are obtained by taking the inverse Fourier transform.
We modified the \textsc{matdyn} code of the \textsc{Quantum ESPRESSO} software~\cite{giannozzi2009,giannozzi2017} and implemented Eq.~\eqref{eq:elastic_tensor} to obtain the elastic constants and Eq.~\eqref{eq:bend_rgd_sym} for the bending rigidities, where the macroscopic long-range electrostatics has been removed.
For the bulk materials considered in this study, we perform a least-squares fit to the Taylor expansion of the unscreened (or transverse) charge density response and dielectric screening function to a phonon perturbation, to extract the multipole and dielectric tensors up to the third and fourth orders as in Eqs.~\eqref{eq:charge_density} and \eqref{eq:screening_func}, respectively (see Appendix~\ref{multipole_cluster_expansion} for details).
We generate a list of $\mathbf{q}$ points (five to ten depending on crystal symmetry) along random directions with the magnitude ranging from 0.01 to 0.05 of the reciprocal lattice constant, and the unscreened charge density response after removing the macroscopic potential as well as the dielectric function for those $\mathbf{q}$ points are calculated within the DFPT framework based on recent developments~\cite{macheda2024,macheda2023,marchese2023,macheda2022}.
The electronic grid for GaAs is further increased to 30$\times$30$\times$30 to accurately predict the dielectric properties of the system.
The dynamical quadrupole tensors of 2D \emph{h}-BN, MoS$_2$, and InSe are directly taken from Ref.~\cite{ponce2023} which were computed using the \textsc{abinit} software~\cite{gonze2016,gonze2020}.

For the finite-difference calculations of elastic constants of bulk solids,  we use the strain-stress relation in Eq.~\eqref{eq:hooke_law} as implemented in the \textsc{thermo\_pw}~\cite{dal2016} code, which externally relies on \textsc{Quantum ESPRESSO} subroutines.
A few independent strains with a magnitude of 0.005 are applied to deform the system through the specific directions, according to its Laue class and crystal symmetry.
Then, several self-consistent field calculations are performed to obtain the corresponding stress tensor of those deformations, with the allowed relaxation of ionic positions to the strain perturbation.
The elastic stiffness tensor as the first-order derivative of stress with respect to strain is calculated numerically by finite difference.

\bibliography{Bibliography}

\onecolumngrid

\ifarXiv
    

\section*{Supplementary material}

# Supplemental Material: Elastic Constants and Bending Rigidities from Long-Wavelength Perturbation Expansions

Changpeng Lin,[1, *] Samuel Poncé,[2, 3] Francesco Macheda,[4, 5] Francesco Mauri,[4, 5] and Nicola Marzari[1, 6]

[1] *Theory and Simulation of Materials (THEOS), and National Centre for Computational Design and Discovery of Novel Materials (MARVEL), École Polytechnique Fédérale de Lausanne, 1015 Lausanne, Switzerland*
[2] *European Theoretical Spectroscopy Facility, Institute of Condensed Matter and Nanosciences, Université catholique de Louvain, Chemin des Étoiles 8, B-1348 Louvain-la-Neuve, Belgium*
[3] *WEL Research Institute, Avenue Pasteur 6, 1300 Wavre, Belgium*
[4] *Dipartimento di Fisica, Università di Roma La Sapienza, Piazzale Aldo Moro 5, 00185 Roma, Italy*
[5] *Istituto Italiano di Tecnologia, Graphene Labs, Via Morego 30, 16163 Genova, Italy*
[6] *Laboratory for Materials Simulations, Paul Scherrer Institut, 5232 Villigen PSI, Switzerland*
(Dated: December 24, 2024)

## I. ADDITIONAL INFORMATION IN BULK CRYSTALS AND MONOLAYERS

In Fig. S1, we show the phonon dispersions and independent components of elastic tensor for NaCl as a function of supercell size, where the multipolar interactions are teated at DD+DO+D$\epsilon$D level. In Fig. S2, we present the phonon dispersions and independent components of elastic tensor for GaAs as a function of supercell size, and the results are calculated at the levels of DD and DD+DQ+QQ interactions. In Fig. S3, the phonon dispersions of BaTiO$_3$ calculated using the DD+DQ+QQ multipolar interpolation are shown. In Fig. S4, we present the phonon dispersions and the independent components of elastic constants and bending rigidities of MoS$_2$ monolayer. In Fig. S5, the independent components of elastic constants and bending rigidities of single-layer InSe are shown, where no macroscopic long-range interactions are removed. In Table S1, we show the multipolar and dielectric properties of 2D h-BN, MoS$_2$ and InSe.

## II. EFFECTIVE ELASTIC MODULI AND ANISOTROPY OF MEDIUM

The mechanical and elastic properties of real materials depend explicitly on their constituent grains and phases, among which the stress and strain tensors can vary. Different polycrystalline average schemes [1] have been proposed to estimate the macroscopic effective elastic properties of materials, based on the knowledge of elastic tensor. The Voigt scheme [2] assumes that the strain field is uniform throughout the sample, known as the isostrain average and acting as the upper bound of elastic moduli. In contrast, the Reuss scheme [3] assumes uniform stress fields, also known as the isostress average, which gives rise to the lower bound of elastic moduli. A

* changpeng.lin@epfl.ch

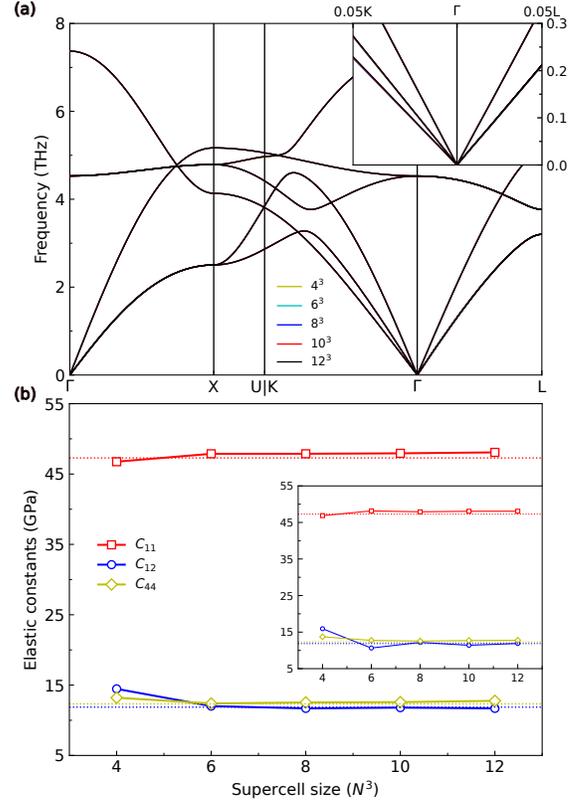

FIG. S1. (a) Phonon dispersion and (b) independent components of elastic tensor for NaCl as a function of supercell size, where the results are obtained from the multipolar (DD+DO+D$\epsilon$D) subtraction. The inset of the panel (a) is the zoom-in acoustic dispersion around the $\Gamma$ point along the K and L directions In the panel (b), the inset shows the elastic constants of NaCl without long-range removal and the dotted lines are the reference values from the THERMO_PW finite-difference calculations.

further arithmetic mean of Voigt and Reuss bounds that agrees better with experimental moduli was later noticed by Hill [4] and is called the Voigt-Reuss-Hill (VRH) average.

Before getting the expressions of elastic moduli under

TABLE S1. Dynamical Born effective charge $Z$ [$e$], quadrupole $Q$ [$e \cdot$ bohr], clamped-ion dielectric permittivity $\epsilon^\infty$ and polarizability tensors $\alpha$ [bohr] of 2D h-BN, MoS$_2$ and InSe, calculated using density functional perturbation theory. The quadrupole values are taken from Ref. [5]. Only symmetry-independent components are shown.

| Elements | $Z_x^x$ | $Z_z^z$ | $Q_x^{xy}$ | $Q_x^{xz}$ | $Q_z^{xx}$ | $Q_z^{zz}$ | $\epsilon_{xx}^\infty$ | $\epsilon_{zz}^\infty$ | $\alpha_{xx}$ | $\alpha_{zz}$ |
|---|---|---|---|---|---|---|---|---|---|---|
| B | 2.685 | 0.246 | 4.261 | – | – | – | 1.591 | 1.098 | 1.881 | 0.310 |
| N | -2.685 | -0.246 | 0.384 | – | – | – | | | | |
| Mo | -0.988 | -0.070 | -5.533 | – | – | – | 6.105 | 1.299 | 13.050 | 0.765 |
| S | 0.494 | 0.035 | -0.391 | -0.174 | 7.878 | -0.297 | | | | |
| In | 2.444 | 0.169 | -7.136 | -1.040 | -7.090 | -0.369 | 3.772 | 1.342 | 8.823 | 1.089 |
| Se | -2.444 | -0.169 | -1.566 | 0.451 | -1.192 | 0.340 | | | | |

Voigt, Reuss and VRH average schemes, we first simplify the notation of elastic tensor by introducing the Voigt notation, where the fourth-rank elastic tensor is transformed into a matrix. Using the transformation rule: $xx \to 1$, $yy \to 2$, $zz \to 3$, $yz, zy \to 4$, $xz, zx \to 5$ and $xy, yx \to 6$, the stress-strain relations in Eqs. (1) and (5) of the main text can be recast into the matrix form as

$$\sigma_i = C_{ij}\varepsilon_j, \quad \varepsilon_i = S_{ij}\sigma_j, \tag{S1}$$

with $i$ and $j$ ranging from 1 to 6. It is emphasized that the strain $\varepsilon_i$ in Voigt notation is the engineering strain, where a factor of 2 is included for the shear strains, i.e. $\varepsilon_4 = 2\varepsilon_{yz}$, $\varepsilon_5 = 2\varepsilon_{xz}$ and $\varepsilon_6 = 2\varepsilon_{xy}$. Also, we should reiterate here that the strain, stress and elastic tensors in Voigt notation are actually non-tensorial, which do not obey the laws of tensor transformation or preserve the norm of the fourth-rank tensor notation.

### A. 3D case

In Voigt notation, the bulk ($K$) and shear moduli ($G$) of bulk materials under the Voigt, Reuss and VRH average schemes can be expressed as [4]

$$K^V = \frac{1}{9}[C_{11} + C_{22} + C_{33} + 2(C_{12} + C_{13} + C_{23})] \tag{S2}$$

$$G^V = \frac{1}{15}[C_{11} + C_{22} + C_{33} - C_{12} - C_{13} - C_{23} \\ + 3(C_{44} + C_{55} + C_{66})] \tag{S3}$$

$$K^R = [S_{11} + S_{22} + S_{33} + 2(S_{12} + S_{13} + S_{23})]^{-1} \tag{S4}$$

$$G^R = [\frac{4}{15}(S_{11} + S_{22} + S_{33} - S_{12} - S_{13} - S_{23}) \\ + \frac{1}{5}(S_{44} + S_{55} + S_{66})]^{-1}, \tag{S5}$$

$$K^{VRH} = \frac{K^V + K^R}{2} \tag{S6}$$

$$G^{VRH} = \frac{G^V + G^R}{2}, \tag{S7}$$

where the superscripts V and R represent the Voigt and Reuss bounds, respectively. For a homogeneous isotropic linear elastic medium, its stress-strain relation and elastic properties can be wholly determined by any two of elastic moduli. Based on the obtained bulk and shear moduli, the Young's modulus ($E$) and Poisson's ratio ($\nu$) for the isotropic linear elastic materials of each polycrystalline average scheme are calculated by the following expressions [1]:

$$E = \frac{9KG}{3K + G}, \tag{S8}$$

$$\nu = \frac{3K - 2G}{2(3K + G)}. \tag{S9}$$

From the knowledge of elastic moduli, diverse useful elastic and wave properties are readily obtained, including P-wave modulus, Lamé's first parameter, Pugh's ratio [6], Cauchy pressure [7], Kleinman parameter [8], sound velocities [9], Debye [10] and melting temperatures [11], and Vickers hardness [12], as are summarized in Section III A.

Besides, another important quantity for characterizing the directional-dependent elastic response of materials to external strain fields is elastic anisotropy, which arises from the anisotropy of their chemical bondings and depends on the specific crystal class. In order to provide a universal measure for the extent of anisotropy of any crystal symmetry, Ranganathan and Ostoja-Starzewski [16] proposed the universal anisotropy index $A^U$ by considering the full tensorial nature of elasic tensor as

$$A^U = 5\frac{G^V}{G^R} + \frac{K^V}{K^R} - 6, \tag{S10}$$

which represents a fractional difference between the elastic moduli of Voigt and Reuss bounds, and is valid for all kinds of crystal systems. This anisotropy measure has the advantage of taking into account the anisotropic bulk contributions through the tensorial nature of elastic tensor. It vanishes for the locally isotropic single crystals, with any departure from zero quantifying the extent of elastic anisotropy. Although $A^U$ is a universal anisotropy measure for all crystal symmetries, it is still a relative measure with respect to the perfect isotropic system, the increase of which does not explicitly suggest the enhanced elastic anisotropy by that extent. To overcome this deficiency, Kube [17] has derived the log-Euclidean anisotropy index

$$A^L = \sqrt{5\left[\ln\left(\frac{G^V}{G^R}\right)\right]^2 + \left[\ln\left(\frac{K^V}{K^R}\right)\right]^2}, \tag{S11}$$

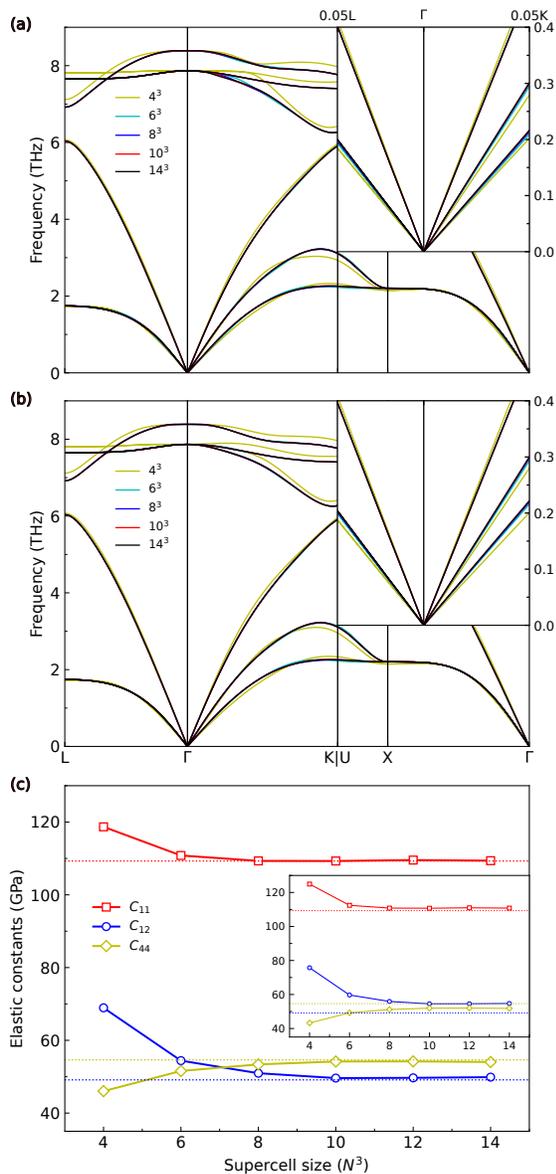

FIG. S2. Phonon dispersions and independent components of elastic tensor for GaAs as a function of supercell size. The panels (a) and (b) are the phonon dispersions obtained from the standard (DD) and multipolar (DD+DQ+QQ) interpolations, respectively; the inset shows the zoom-in acoustic branches. In the panel (c), the elastic constants are calculated after removing the contributions of DD, DQ and QQ terms in its IFCs, while the results shown in the inset are obtained from the IFCs with the removal of only DD interactions; the dotted lines there denote the reference values of elastic constants from the finite-difference calculations in THERMO_PW code.

which measures the elastic anisotropy for all crystal types based on a distance of elastic tensor in the log-Euclidean space. Similarly, a null distance for $A^{\mathrm{L}}$ indicates the case of perfect elastic isotropy, and it provide an absolute measure of the anisotropy when larger than zero. Thus, one can apply $A^{\mathrm{L}}$ to have an unambiguous and definite comparison of elastic anisotropy among different crystals with different symmetries.

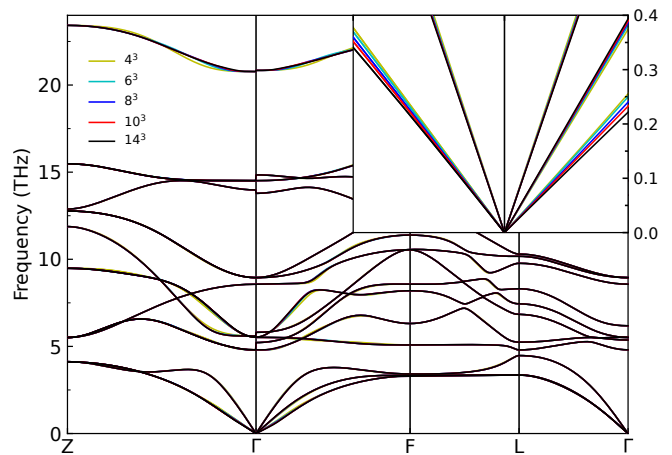

FIG. S3. Phonon dispersion of BaTiO$_3$ as a function of supercell size using the multipolar interpolation at the DD+DQ+QQ level, where the inset shows a zoom-in result of the acoustic branches.

### B. 2D case

For a 2D system, the Voigt- and Reuss-averaged layer ($K$) and shear moduli ($G$) are obtained as [18]

$$K^{\mathrm{V}} = \frac{C_{11} + C_{22} + 2C_{12}}{4}, \tag{S12}$$

$$G^{\mathrm{V}} = \frac{C_{11} + C_{22} - 2C_{12} + 4C_{66}}{8}, \tag{S13}$$

$$K^{\mathrm{R}} = \frac{1}{S_{11} + S_{22} + 2S_{12}}, \tag{S14}$$

$$G^{\mathrm{R}} = \frac{2}{S_{11} + S_{22} - 2S_{12} + S_{66}}, \tag{S15}$$

while the VRH average scheme remains the same as the 3D case shown in Eqs. (S6) and (S7). Similar to the bulk modulus in the 3D case, $K$ becomes the so-called layer (or area) modulus for 2D materials. In particular, if the 2D crystal is elastically isotropic, its shear modulus $G$ is just $C_{66}$. Under the homogeneous and isotropic assumptions of the linear elastic medium, the 2D Young's modulus ($E$) and Poisson's ratio $\nu$ of each polycrystalline average scheme can be calculated as [1]

$$E = \frac{4KG}{K+G}, \tag{S16}$$

$$\nu = \frac{K-G}{K+G}. \tag{S17}$$

It is worth pointing out that the Poisson's ratio in 2D system is allowed to be larger than 0.5, ranging from -1.0 to 1.0, which is a striking character due to the reduced dimensionality. Furthermore, in some literature, one can

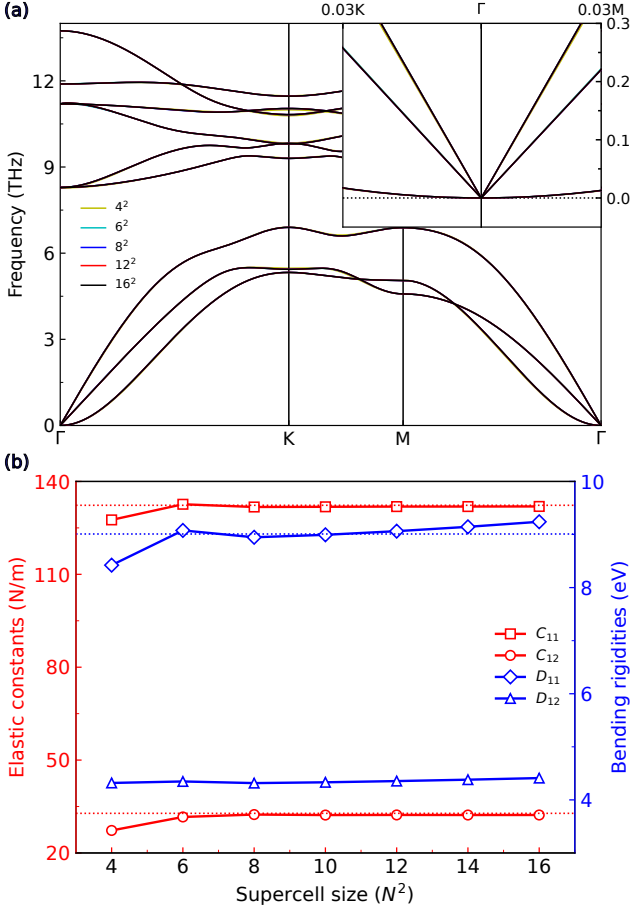

FIG. S4. (a) Phonon dispersion and (b) independent components of the elastic and bending rigidity tensors in $MoS_2$ as a function of supercell size. The multipolar interactions up to the quadrupole terms are considered for the long-range effect, and the inset of panel (a) shows the zoom-in acoustic branches around the $\Gamma$ point along the K and M directions. In the panel (b), the short-circuit elastic constants and bending rigidities of $MoS_2$ are calculated at the level of removing DD+DQ+QQ; the dotted lines represent the theoretical elastic constants $C_{11}/C_{12}$ and bending rigidity $D_{11}$ taken from Ref. [13] and Ref. [14], respectively.

find that the 2D Young's modulus and Poisson's ratio along the two Cartesian components $x$ and $y$ are [19]

$$E_x = \frac{C_{11}C_{22} - C_{12}^2}{C_{22}}, \tag{S18}$$

$$E_y = \frac{C_{11}C_{22} - C_{12}^2}{C_{11}}, \tag{S19}$$

$$\nu_x = \frac{C_{12}}{C_{22}}, \tag{S20}$$

$$\nu_y = \frac{C_{12}}{C_{11}}. \tag{S21}$$

A complete set of macroscopic elastic moduli and wave properties that can be derived based on the elastic constants and bending rigidities of 2D materials are further summarized in Sec. III B.

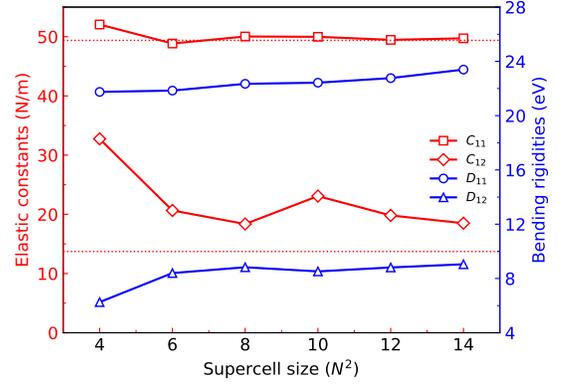

FIG. S5. Independent components of elastic and bending rigidity tensors in InSe as a function of supercell size, where no long-range interactions are subtracted. The dotted lines represent the theoretical elastic constants $C_{11}$ and $C_{12}$ taken from Ref. [15].

In order to quantitatively characterize the degree of elastic anisotropy in 2D crystals, the 2D elastic anisotropy index has been proposed by Li et al. recently [18]:

$$A^{SU} = \sqrt{2\left(\frac{G^V}{G^R} - 1\right)^2 + \left(\frac{K^V}{K^R} - 1\right)^2}, \tag{S22}$$

which provides a measure of the Euclidean distance between the normalized elastic stiffness averaged by Voigt and Reuss schemes. It is easy to find that such a measure has a minimum value of zero which is exactly the case of perfect isotropy, and any non-vanishing value corresponds to the extent of elastic anisotropy. In addition, they also extended the universal anisotropy index $A^U$ [16] and the log-Euclidean anisotropy index $A^L$ [17] which are originally proposed for bulk materials and obtained the corresponding 2D forms as [18]

$$A^U = 2\frac{G^V}{G^R} + \frac{K^V}{K^R} - 3, \tag{S23}$$

$$A^L = \sqrt{2\left[\ln\left(\frac{G^V}{G^R}\right)\right]^2 + \left[\ln\left(\frac{K^V}{K^R}\right)\right]^2}. \tag{S24}$$

As already discussed in the situation of bulk system (see Sec. II A for details), $A^U$ is just a relative measure that accounts for the deviation from the perfectly isotropic case, while $A^L$ is able to provide the absolute measure of the degree of anisotropy. Furthermore, according to 2D crystallography, the 2D crystals can be classified into four catalogues based on their rotational symmetries: hexagonal (six-fold), square (four-fold), (centered) rectangular (two-fold) and oblique (without rotational symmetry) lattices. All of these three elastic anisotropy indices are applicable to the four types of 2D crystals, regardless of their symmetries, and only the hexagonal lattice is perfectly isotropic.


## III. SUMMARY OF THE DERIVED MACROSCOPIC ELASTIC AND WAVE PROPERTIES OF MATERIALS

In this Appendix, we present the elastic and wave properties of bulk and 2D materials that can be calculated based on the knowledge of the full elastic tensor and the elastic moduli under the Voigt, Reuss or VRH average scheme from Sec. II, as listed alphabetically below. All of these calculations have been implemented into the MATDYN code of QUANTUM ESPRESSO distribution [20, 21].

### A. 3D case

The Cauchy pressure ($P^\mathrm{C}$) [7] is computed by

$$P_a^\mathrm{C} = C_{12} - C_{44}, \tag{S25a}$$

$$P_a^\mathrm{C} = C_{13} - C_{44},\ P_b^\mathrm{C} = C_{12} - C_{66}, \tag{S25b}$$

$$P_a^\mathrm{C} = C_{23} - C_{44},\ P_b^\mathrm{C} = C_{13} - C_{55},\ P_c^\mathrm{C} = C_{12} - C_{66}, \tag{S25c}$$

$$P_a^\mathrm{C} = C_{23} - C_{44},\ P_b^\mathrm{C} = C_{13} - C_{55},$$
$$P_c^\mathrm{C} = C_{12} - C_{66},\ P_d^\mathrm{C} = C_{25} - C_{46}, \tag{S25d}$$

$$P_a^\mathrm{C} = C_{23} - C_{44},\ P_b^\mathrm{C} = C_{13} - C_{55},\ P_c^\mathrm{C} = C_{12} - C_{66},$$
$$P_d^\mathrm{C} = C_{14} - C_{56},\ P_e^\mathrm{C} = C_{25} - C_{46},\ P_f^\mathrm{C} = C_{36} - C_{45}, \tag{S25e}$$

where Eq. (S25a) is for cubic system, Eq. (S25c) is for orthorhombic system, Eq. (S25d) is for monoclinic system, Eq. (S25e) is for triclinic system, and Eq. (S25b) is for hexagonal, rhombohedral (trigonal) and tetragonal systems. The Pettifor's Cauchy pressure describes the angular character of chemical bonding of solids; $P^\mathrm{C} < 0$ suggests covalent bonding, whereas $P^\mathrm{C} > 0$ indicates metallic bonding. Since decent ductile materials are often formed through the metallic bonding with the largely delocalized valance states, $P^\mathrm{C}$ also gives a good measure for the ductility-brittleness of materials [22, 23].

The Kleinman's parameter [8] is obtained as

$$\zeta = \frac{C_{11} + 8C_{12}}{7C_{11} - 2C_{12}}, \tag{S26}$$

which characterizes the stability of materials against the stretching and bending loads, also known as the internal strain parameter. It is dimensionless with values generally in the range of 0 to 1, and quantifies the relative ease of bond bending to stretching. Under external stress fields, the lower limit $\zeta = 0$ implies that the system will be governed by bond stretching, while the upper limit $\zeta = 1$ indicates that bond bending will dominate the system.

The Lamé's first parameter is calculated as

$$\lambda = K - \frac{2}{3}G, \tag{S27}$$

which formulates the 3D Hooke's law of homogeneous isotropic elastic linear materials together with the Lamé's second parameter $\mu = G$. They are of great significance as the eigenelastic constants [24] or the principal elasticities [25] of materials. It is noted that $\lambda$ is often regarded to give a measure of incompressibility, which in fact is not the reciprocal of the compressibility $\beta = 1/K$ [1].

The melting point can be evaluated by [11]

$$T^\mathrm{melt} = 607 + 9.3K, \tag{S28}$$

which is an empirical formula to estimate the melting temperature of solids based on their bulk moduli. The units of $T^\mathrm{melt}$ and $K$ are kelvin and gigapascal, respectively.

The Pugh's ratio [6]:

$$\frac{K}{G} \tag{S29}$$

is defined as the ratio of the bulk modulus to the shear modulus of a given material, which empirically describes the ductile-brittle nature of solids. The empirical critical value of Pugh's ratio for the ductile-to-brittle transition is 1.75 [6]. Together with Pettifor's Cauchy pressure $P^\mathrm{C}$ and Poisson's ratio $\nu$, they provide the criteria for determining the ductile-to-brittle transition of materials. The materials with $K/G > 1.75$, $P^\mathrm{C} > 0$ and $\nu > 0.25$ are thought to be ductile in nature, otherwise behaving brittle [6, 22, 26].

The pressure (P)-wave modulus:

$$M = K + \frac{4G}{3}, \tag{S30}$$

is defined as the ratio of the axial stress to the axial strain in an uniaxial strain state [1], also known as the longitudinal modulus.

The sound (elastic wave) velocities can be obtained as [9, 10]

$$v_\mathrm{l} = \sqrt{\frac{M}{\rho}} = \sqrt{\frac{3K + 4G}{3\rho}}, \tag{S31}$$

$$v_\mathrm{t} = \sqrt{\frac{G}{\rho}}, \tag{S32}$$

$$v_\mathrm{v} = \sqrt{\frac{K}{\rho}}, \tag{S33}$$

$$\frac{1}{v_\mathrm{m}} = \sqrt[3]{\frac{1}{3}\left(\frac{1}{v_\mathrm{l}^3} + \frac{2}{v_\mathrm{t}^3}\right)}, \tag{S34}$$

with the subscripts denoting the longitudinal (or compressional, l), transverse (or shear, t), volumetric (v) and mean (m) sound velocities, respectively. In the long-wavelength limit, the group velocities of acoustic phonon modes coincide with the phase velocities of sounds (elastic waves). The Debye temperature ($\theta_\mathrm{D}$) of bulk solids

can be further predicted from the mean wave velocity ($v_{\rm m}$) as [10]

$$\theta_{\rm D} = \frac{hv_{\rm m}}{k_{\rm B}} \sqrt[3]{\frac{3n_{\rm a}\rho N_{\rm A}}{4\pi M_{\rm mol}}}, \quad (S35)$$

where $h$ is the Planck constant, $k_{\rm B}$ is the Boltzmann constant, $n_{\rm a}$ is the number of atoms in the unit cell, $N_{\rm A}$ is Avogadro constant, and $M_{\rm mol}$ is the molecular weight of a given crystal.

The Vickers hardness [27, 28] is given as

$$H_{1a} = 0.0963K, \quad (S36a)$$
$$H_{1b} = 0.1475G, \quad (S36b)$$
$$H_{1c} = 0.0607E, \quad (S36c)$$
$$H_2 = 0.1769G - 2.899, \quad (S36d)$$
$$H_3 = 0.0635E, \quad (S36e)$$
$$H_4 = \frac{(1-2\nu)K}{6(1+\nu)}, \quad (S36f)$$
$$H_5 = 2\left(\frac{G^3}{K^2}\right)^{0.585} - 3, \quad (S36g)$$

which are semi-empirical formulae adopted to estimate the hardness from the Vickers test [12] of a solid. The calculation of Vickers hardness is made based on its relation to the elastic moduli, whose accuracy in fact varies dramatically when going from a specific class of materials to another. Considering the influence of band gap and crystal system, Singh et al. [27] proposed the hardness recommendation model by comparing the calculated hardness of fifty different materials, where the experimental data are available. This hardness recommendation model is summarized in Table S2.

### B. 2D case

The 2D Lamé's first parameter is calculated as [1]

$$\lambda = K - G, \quad (S37)$$

while the 2D Lamé's second parameter is just $\mu = G$. They together parameterize the elastic moduli for homogeneous isotropic media, and the Hooke's law can be recast as $\sigma_{ij} = \lambda \delta_{ij}\varepsilon_{kk} + 2\mu\varepsilon_{ij}$. Moreover, the 2D pressure-wave modulus is obtained as

$$M = K + G. \quad (S38)$$

The different types of speed of sound in 2D materials can readily be calculated from Eqs. (S31) to (S33) with $\rho$ replaced by the 2D mass density $\rho_{\rm 2D}$, which now stand for the longitudinal ($v_{\rm l}$), transverse ($v_{\rm t}$) and area ($v_{\rm a}$) sound velocities, respectively. Particularly, the pressure-wave modulus used in the calculation of longitudinal sound velocity should be calculated by Eq. (S38). The mean sound velocity averaged between the longitudinal and transverse sound velocities is computed by

$$\frac{1}{v_{\rm m}} = \sqrt{\frac{1}{2}\left(\frac{1}{v_{\rm l}^2} + \frac{1}{v_{\rm t}^2}\right)}. \quad (S39)$$

With the knowledge of mean sound velocity, the Debye temperature $\theta_{\rm D}$ of 2D crystals can be predicted based on the 2D Debye model as [29]

$$\theta_{\rm D} = \frac{hv_{\rm m}}{k_{\rm B}}\sqrt{\frac{n_{\rm a}}{\pi A}}. \quad (S40)$$

The strain energy density $\mathcal{U}(\boldsymbol{\kappa})$ of 2D materials under small deflection perturbations is a quadratic function of curvature tensor $\boldsymbol{\kappa}$ as [30]

$$\mathcal{U}(\boldsymbol{\kappa}) = \frac{1}{2}D_{\alpha\beta,\gamma\delta}\kappa_{\alpha\beta}\kappa_{\gamma\delta} = \frac{1}{2}D_{ij}\kappa_i\kappa_j, \quad (S41)$$

where we have adopted a similar Voigt notation to rewrite the fourth-rank bending rigidity tensor $D_{\alpha\beta\gamma\delta}$ and the second-rank curvature tensor $\kappa_{\alpha\beta}$ into a matrix and vector form, respectively. The linear term $\mathcal{M}_i\kappa_j$ vanishes in Eq. (S41) for a symmetric 2D plate or membrane (i.e. the existence of out-of-plane reflection symmetry) [31], since the strain energy density is an even function in this case with $\mathcal{U}(\boldsymbol{\kappa}) = \mathcal{U}(-\boldsymbol{\kappa})$. For determining the effective bending rigidities of 2D crystals, we should make an assumption of orthotropic materials, whose anisotropy only displays in two mutually perpendicular directions. Consequently, the bending rigidity tensor is reduced to having only four independent components:

$$\mathbb{D} = \begin{bmatrix} D_{11} & D_{12} & 0 \\ D_{12} & D_{22} & 0 \\ 0 & 0 & D_{66} \end{bmatrix}, \quad (S42)$$

and the strain energy density can be explicitly written as

$$\mathcal{U}(\boldsymbol{\kappa}) = \frac{1}{2}\left(D_{11}\kappa_1^2 + D_{22}\kappa_2^2 + 2D_{12}\kappa_1\kappa_2 + 4D_{66}\kappa_6^2\right). \quad (S43)$$

According to the crystal symmetry of four catalogues of 2D lattces, the bending rigidity tensor of hexagonal, square and rectangular lattices is all in the form of orthotropic materials, except for the oblique one which has the complete six independent components. The principal curvatures at a given point of the plate are the two curvatures perpendicular to each other with the maximum and minimum values, also known as the maximum and minimum curvatures, denoted as $\kappa_a$ and $\kappa_b$, respectively. When $\kappa_1$ and $\kappa_2$ coincide with the principal curvatures, the two principal bending rigidities are just $D_{11}$ and $D_{22}$. In our implementation, we simply assume the in-plane Cartesian directions coincide with the principal axis directions, and hence please be careful with the situations when they are not. For the hexagonal lattice which are perfectly isotropic in all directions, the principal bending rigidity $D_{\rm P}$ is simply $D_{11}$. In such isotropic case,

TABLE S2. Hardness recommendation model for Eqs. (S36) [27] based on the electronic band gap ($E_g$) in eV and crystal system of materials.

| Type | Cubic | Hexagonal | Orthorhombic | Rhombohedral | General |
|---|---|---|---|---|---|
| Insulator ($E_g > 2$) | $H_2$ | $H_{1c}$ | $H_2$ | $H_2$ | $H_2$ |
| Semiconductor ($0 < E_g < 2$) | $H_5$ | $H_{1c}$, $H_3$ | – | $H_2$ | $H_5$ |
| Metal ($E_g = 0$) | $H_{1b}$ | $H_4$ | $H_4$ | $H_4$ | $H_4$ |

one can further calculate the Gaussian curvature modulus (or Gaussian rigidity), which relates to the Gaussian curvature defined as the product of two principal curvatures, i.e. $\kappa_\mathrm{G} = \kappa_a \kappa_b = \kappa_1 \kappa_2 - \kappa_6^2$. In a homogeneous and isotropic plate, the components of bending rigidity tensor is no longer independent, and the relation $D_{11} = D_{22} = (D_{12} + D_{21} + 4D_{66})/2$ holds. The strain energy density in Eq. (S43) can be then rewritten as

$$\begin{aligned}\mathcal{U}(\boldsymbol{\kappa}) &= \frac{1}{2}\left[D_{11}\left(\kappa_1 + \kappa_2\right)^2 - 4D_{66}\left(\kappa_1\kappa_2 - \kappa_6^2\right)\right] \\ &= \frac{1}{2}\left[D^\mathrm{P}\left(\kappa_a + \kappa_b\right)^2 + D^\mathrm{G}\kappa_\mathrm{G}\right],\end{aligned} \quad (S44)$$

with the corresponding Gaussian curvature modulus calculated as $D^\mathrm{G} = -4D_{66}$, where the sum of $\kappa_1 + \kappa_2 = \kappa_a + \kappa_b$ is an invariant quantity as a result of the rotation of the coordinate axes on the Mohr's circle of curvature [32]. It should be noted in some literature the prefactor $1/2$ is absorbed into the definition of Gaussian rigidities [33], i.e. $D^\mathrm{G} = -2D_{66}$, and we choose to use this definition in this study. In simple models such as the classical Kirchhoff plate, $D^\mathrm{G}$ is usually negative which is necessary to stabilize saddle distortions of an initially flat thin plate [34].


\fi

\twocolumngrid

\end{document}